\documentclass[twocolumn]{aastex63}

\usepackage{bm}
\usepackage{amsmath}
\usepackage{comment}

\newcommand{\kepler}{\textit{Kepler}}
\newcommand{\teff}{T_\mathrm{eff}}
\newcommand{\feh}{\mathrm{[Fe/H]}}
\newcommand{\ks}{{K_\mathrm{s}}}
\newcommand{\plx}{\varpi}
\newcommand{\mass}{M_\star}

\newcommand{\age}{t_\star}
\newcommand{\eep}{e}
\newcommand{\masuda}{\citet{2022ApJ...933..195M}}
\newcommand{\masudap}{\citep{2022ApJ...933..195M}}
\newcommand{\prot}{P_\mathrm{rot}}
\newcommand{\tauc}{\tau_\mathrm{c}}
\newcommand{\ro}{\mathrm{Ro}}
\newcommand{\amp}{R_\mathrm{per}}

\received{Jun 7, 2022}
\revised{Aug 17, 2022}
\accepted{Aug 27, 2022}
\shorttitle{Ages of the CKS Stars (Understanding the McQuillan Sample II)}
\shortauthors{Masuda}
\graphicspath{{./}{figures/}}

\begin{document}

\title{
Detectability of Rotational Modulation in \textit{Kepler} Sun-like Stars as a Function of Age
}

\correspondingauthor{Kento Masuda}
\email{kmasuda@ess.sci.osaka-u.ac.jp}

\author[0000-0003-1298-9699]{Kento Masuda}
\affiliation{Department of Earth and Space Science, Osaka University, Osaka 560-0043, Japan}



\begin{abstract}

We examine how the fraction $f$ of stars for which rotational modulation has been detected in \kepler\ light curves depends on the stellar mass $\mass$ and age $\age$.
Our sample consists of $\approx 850$ FGK stars hosting transiting planet candidates detected from the prime \kepler\ mission. For these stars, atmospheric parameters have been derived using high-resolution spectra from the California-\kepler\ survey, and rotational modulation has been searched in \kepler\ light curves homogeneously. We fit stellar models to the atmospheric parameters, Gaia parallax, and 2MASS magnitude of these stars and obtain samples drawn from the posterior probability distributions for their masses and ages under a given, uninformative prior. We combine them with the result of rotational modulation search to simultaneously infer the mass--age distribution of the sample as well as $f(\mass, \age)$,
in a manner that fully takes into account mass and age uncertainties of individual stars.
We find that $f$ remains near unity up to $\age \sim 3\,\mathrm{Gyr}$ and drops to almost zero by $\age \sim 5\,\mathrm{Gyr}$, although the trend is less clearly detected for stars with $\lesssim 0.9\,M_\odot$ due to weaker age constraints. This finding is consistent with a view that the detection of rotational modulation is limited by photometric precision to younger stars that exhibit higher-amplitude modulation, and suggests that the detectability of rotational modulation in \kepler\ light curves is insensitive to metallicity and activity cycles for stars younger than the Sun.

\end{abstract}



\section{Introduction}

High precision, continuous photometry made available by the NASA \kepler\ mission \citep{2010Sci...327..977B, 2010ApJ...713L..79K} enabled detection of rotational brightness variations for tens of thousands of FGKM stars \citep[e.g.][]{2013A&A...557L..10N, 2013A&A...560A...4R, 2014ApJS..211...24M, 2014A&A...572A..34G, 2019ApJS..244...21S, 2020A&A...635A..43R, 2021ApJS..255...17S}. While these stars have been studied extensively, 
less attention has been paid to stars without detected variations \citep[but see][]{2012MNRAS.423.2966J},
limiting our understanding of
why some stars exhibit detectable rotational modulation while others do not.
Obviously age must be an important factor, 
because a star's activity weakens as it ages and spins down, and so does the amplitude of photometric rotational modulation.
It is unclear, though, whether age is the only important  
parameter.
Long-term activity cycles may cause some young/old stars to become occasionally invisible/visible in their rotational modulation. 
Metal-rich stars may have enhanced stellar variabilities at a given rotation period because of their deeper convective envelopes \citep[e.g.,][]{2020A&A...634L...9W,2020MNRAS.499.3481A,2021ApJ...912..127S}.

It has been known that the distribution of rotation periods $\prot$ derived from photometric modulation for \kepler\ stars exhibits a rather sharp upper edge as a function of stellar effective temperature $\teff$ \citep[e.g.,][]{2014ApJS..211...24M, 2021ApJS..255...17S}. \citet{2014ApJS..211...24M} noted that the edge lies roughly on a gyrochrone of the solar age.
\citet{2019ApJ...872..128V} pointed out that the longest detected periods for stars with different spectral types scale with their convective turnover timescales $\tauc$: the edge lies around a 
Rossby number $\ro=\prot/\tauc$ close to the solar value.
\citet{2019ApJ...872..128V} discussed two possible interpretations that are not mutually exclusive: (i) it is a detection edge associated with modulation amplitudes decreasing with increasing rotation periods, where they hypothesized that the edge might have been sharpened by a sudden drop in the amplitude due to a change in the stellar spottedness, or (ii) stellar spin down stalls around the solar Rossby number \citep{2015MNRAS.450.1787A, 2016Natur.529..181V}, causing older stars to masquerade as young in their rotational appearance (the weakened magnetic braking hypothesis).
\citet{2019ApJ...872..128V} was prudent in deciding which interpretation is more likely, although they noted that longest-period stars around the upper edge do not show similar variability amplitudes and argued that this feature is at odds with the simple detection edge as posited in scenario (i).

More recently, \masuda\ presented evidence for the magnitude-dependent detection threshold in the \citet{2014ApJS..211...24M} sample, which was shown to agree with the location of the observed upper edge for main-sequence stars: this is scenario (i) in \citet{2019ApJ...872..128V} but does not require a discontinuous drop in the photometric modulation amplitude.
The crux of this argument is that the modulation amplitude decreases very rapidly with increasing $\ro$ in a manner roughly independent of $\teff$ for solar-type main-sequence stars, and that stars with different magnitudes have different detection thresholds for rotational modulation. 
Since the \kepler\ sample is dominated by the faintest stars, the combination of the two imprints a sharp $\ro$ cutoff in the sample of detected rotational modulation; here the cooler stars have slightly higher thresholds for the detectable amplitude because they are fainter, which explains why the upper $\prot$--$\teff$ edge does not correspond to a constant variability amplitude.
If this view is correct, the stars should exhibit detectable rotational modulation if and only if a star is younger than a certain age threshold that is primarily determined by its $\teff$ or mass (and depends weakly on its visual magnitude): the threshold Rossby number translates into the threshold rotation period via the dependence of $\tau_c$ on $\teff$, which then translates into the threshold age via gyrochronal relations.\footnote{
This prediction is insensitive to the weakened magnetic braking, because most stars cross down the detection thresholds before its likely onset \citep[see][]{2022ApJ...933..195M}.
}
This prediction is in contrast to what we expect assuming that the edge is due to weakened magnetic braking (scenario ii), in which case stars with detected rotational modulation should have a broad age distribution at a fixed mass.


In this paper, we estimate the fraction of \kepler\ stars with detected rotation periods as a function of mass and age to better understand the detection bias of photometric rotational modulation,
and to test the above prediction.
This requires ages of the stars both {\it with and without} detected rotational modulation. We also need a sample of stars for which rotational modulation has been uniformly searched. 
For these reasons, we focus on the stars studied in the California-\kepler\ Survey \citep[CKS;][]{2017AJ....154..107P}. This is a sample of stars with transiting planet candidates detected in the \kepler\ data, for which high-resolution spectra have been obtained with Keck/HIRES and stellar atmospheric parameters have been uniformly derived. The information can be combined with precise parallaxes from {\it Gaia} EDR3 \citep{2021A&A...649A...1G} and ground-based photometry to derive isochrone-based ages.
Most of these stars have also been searched for rotational modulation by \citet{2015ApJ...801....3M} using the same autocorrelation function (ACF) based method as adopted in \citet{2014ApJS..211...24M}. Planetary transits were masked in the search, and in any case affect only a small portion of the light curve. 
Rather, we consider the selection conditioned on the presence of transiting planets to be advantageous for the present study: 
various surveys have shown that stars with close-in planets are roughly 10 times less likely to host binary companions within $\sim 10\,\mathrm{au}$ \citep[see][for a summary]{2021MNRAS.507.3593M}, and so 
these stars are unlikely to be tight binaries where tidal interactions may have significantly affected the stellar rotation.\footnote{Massive, close-in planets may also tidally affect the stellar rotation \citep{2021ApJ...919..138T}. However, such ``hot Jupiters" exist only around $\sim 1\%$ of Sun-like stars \citep{2012ApJ...753..160W} and so are negligible statistically.} 

As is well known, the age from isochrone fitting is often highly degenerate with mass and so is uncertain \citep[e.g.,][]{2010ARA&A..48..581S}; this remains to be the case even in our analysis leveraging the {\it Gaia} parallax and focusing on solar-mass main-sequence stars. Nevertheless, we will show that such degenerate solutions, if properly taken into account, still provide useful information on the underlying age distribution for a large number of stars.
In Section \ref{sec:isochrone}, we describe our isochrone fitting method, and validate it using simulated and asteroseismic data. Here we show typical constraints we can obtain for mass and age with the current data, and discuss limitations and caveats in interpreting such probabilistic constraints as obtained from isochrone fitting. In Section \ref{sec:cks}, we describe our sample stars and apply the fitting method to obtain samples drawn from the posterior probability distributions for their physical parameters including ages, also incorporating gyrochronology information where available. 
We also present initial observations on how the amplitudes of rotational modulation as well as their detectability depend on stellar mass, age, and metallicity.
Section \ref{sec:hbayes} describes our hierarchical Bayesian framework to infer the fraction of stars with detected rotational modulation fully taking into account the mass and age uncertainties. 
In Section \ref{sec:results}, we apply the method to our sample to derive how the fraction of stars with detected modulation depends on stellar mass and age, and show that the result is consistent with the view that the longest detected periods are determined by a simple detection edge.
In Section \ref{sec:discussion} we summarize and conclude the paper.

\section{Isochrone Fitting}\label{sec:isochrone}

Here we describe our method of isochrone fitting (Section~\ref{ssec:isochrone_method}), and perform internal (Section~\ref{ssec:isochrone_recovery}) and external (Section~\ref{ssec:isochrone_seismic}) tests to validate the procedure. 
These tests are also used to illustrate limitations and caveats of single-value estimates based on the Bayesian posterior probability distribution, which motivates the hierarchical treatment in Section \ref{sec:hbayes}.

\subsection{The Method}\label{ssec:isochrone_method}

Essentially, we perform the fitting on the color-magnitude diagram 
interpolating the MIST models \citep{2011ApJS..192....3P,2013ApJS..208....4P, 2015ApJS..220...15P, 2016ApJS..222....8D, 2016ApJ...823..102C}. 
The physical parameters (effective temperature, mass, radius etc.) and the magnitudes in different photometric bands are derived by linearly interpolating model grids 
for a given set of age $\age$, iron metallicity $\feh$, and the equivalent evolutionary phase \citep[EEP;][]{2016ApJS..222....8D} $\eep$.
This approach is  
also adopted in the {\tt isochrones} package \citep{2015ascl.soft03010M}.

In this paper, we feed the measurements of effective temperature $\teff$, iron metallicity $\feh$, $\ks$-band magnitude $\ks$, and parallax $\plx$ as the data $D$,
and infer the probability density function (PDF) for the set of parameters $\bm{\theta}=(\age, \feh, \eep, d)$ given the data $D$, where $d$ is the distance to a star.
The measurements of $\teff$ and $\feh$ come from high-resolution spectroscopy; $\ks$ is from the Two Micron All Sky Survey \citep[2MASS;][]{2006AJ....131.1163S}; and the parallax is from {\it Gaia} EDR3. 
We do not use spectroscopic surface gravity $\log g$, because the above data usually provide much more stringent constraints on stellar radii than $\log g$. We do not use magnitudes in other photometric bands following \citet{2018AJ....156..264F}: they are redundant given the spectroscopic $\teff$ and could make the results more sensitive to interstellar extinction.

The inference requires the likelihood function $\mathcal{L}(D|\theta)$, the probability to obtain $D$ for a given set of parameter values $\bm{\theta}$.
This is computed as a product of independent Gaussians for each measured parameter:\footnote{Here we ignore the possible correlation between $\teff$ and $\feh$ that may exist when the two parameters are derived from the same spectra. This is not a fundamental limitation to the method.}
given the stellar model, the observable $\bm{y}=(\teff, \feh, \ks, \varpi)$ is computed as a deterministic function of $\bm{\theta}$ as described above (and $d=1/\varpi$), and is used to compute the likelihood 
\begin{equation}
\mathcal{L}(D|\bm{\theta}) = \prod_i {1\over\sqrt{2\pi(\sigma_i^{\rm obs})^2}}\exp\left[-{1\over 2}\left({y_i^{\rm obs}-y_i(\bm{\theta}})\over \sigma_i^{\rm obs}\right)^2\right]
\end{equation}
where $y_i^{\rm obs}$ and $\sigma_i^{\rm obs}$ denote the ``measured value" and ``error bar" of the parameter $y_i$ (i.e., $\teff$, $\feh$, $\ks$, $\plx$), respectively.
We then sample from the following joint posterior PDF of age $\age$, metallicity $\feh$, EEP $\eep$, and distance $d$:
\begin{align}
    \label{eq:posterior}
    p(\bm{\theta}|D) \propto \mathcal{L}(D|\bm{\theta})\,\pi(\bm{\theta})
\end{align}
adopting a certain prior PDF $\pi(\bm{\theta})$.
The prior PDF $\pi$ is assumed to be separable as $\pi(\bm{\theta})=\pi_0(\age,\feh,\eep)\,\pi_1(d)$, and $\pi_0$ is chosen so that   $\pi_0(\age, \feh, \eep)$ is proportional to the Jacobian $|\partial(\age,\feh,\mass)/\partial(\age,\feh,\eep)|$, i.e.,
the probability density is constant in the $(\age, \feh, \mass)$ space where valid models exist.\footnote{This means that the marginalized prior PDFs for individual parameters are not necessarily uniform.} 
The age, [Fe/H], and EEP were bounded to be within (0.1, 13.8)Gyr, ($-0.5, 0.5$), and (0, 600), respectively.  
For the distance prior $\pi_1(d)$, we adopt 
an exponentially decreasing volume density prior with a length scale of 1.35~kpc \citep{2015PASP..127..994B,2016ApJ...832..137A}. In practice, the choice of $\pi_1$ does not play a significant role for our sample stars whose parallaxes are well constrained.
Isochrone fitting based on a Bayesian framework  
is not without precedent: earlier works include \citet{2004MNRAS.351..487P, 2005A&A...436..127J, 2007ApJS..168..297T, 2015ascl.soft03010M}.

The code was implemented using {\tt JAX} \citep{jax2018github}. The sampling was performed using Hamiltonian Monte Carlo and the No-U-Turn Sampler \citep{DUANE1987216, 2017arXiv170102434B} as implemented in {\tt NumPyro} \citep{bingham2018pyro, phan2019composable}. We typically draw 20,000 samples, after which the resulting chains had the split $\hat{R}<1.02$ \citep{BB13945229} for all the parameters in most cases.
The tool is available through GitHub.\footnote{\url{https://github.com/kemasuda/jaxstar}}

\subsection{Injection and Recovery Tests}\label{ssec:isochrone_recovery}

\begin{figure*}
\epsscale{1.05}
\plottwo{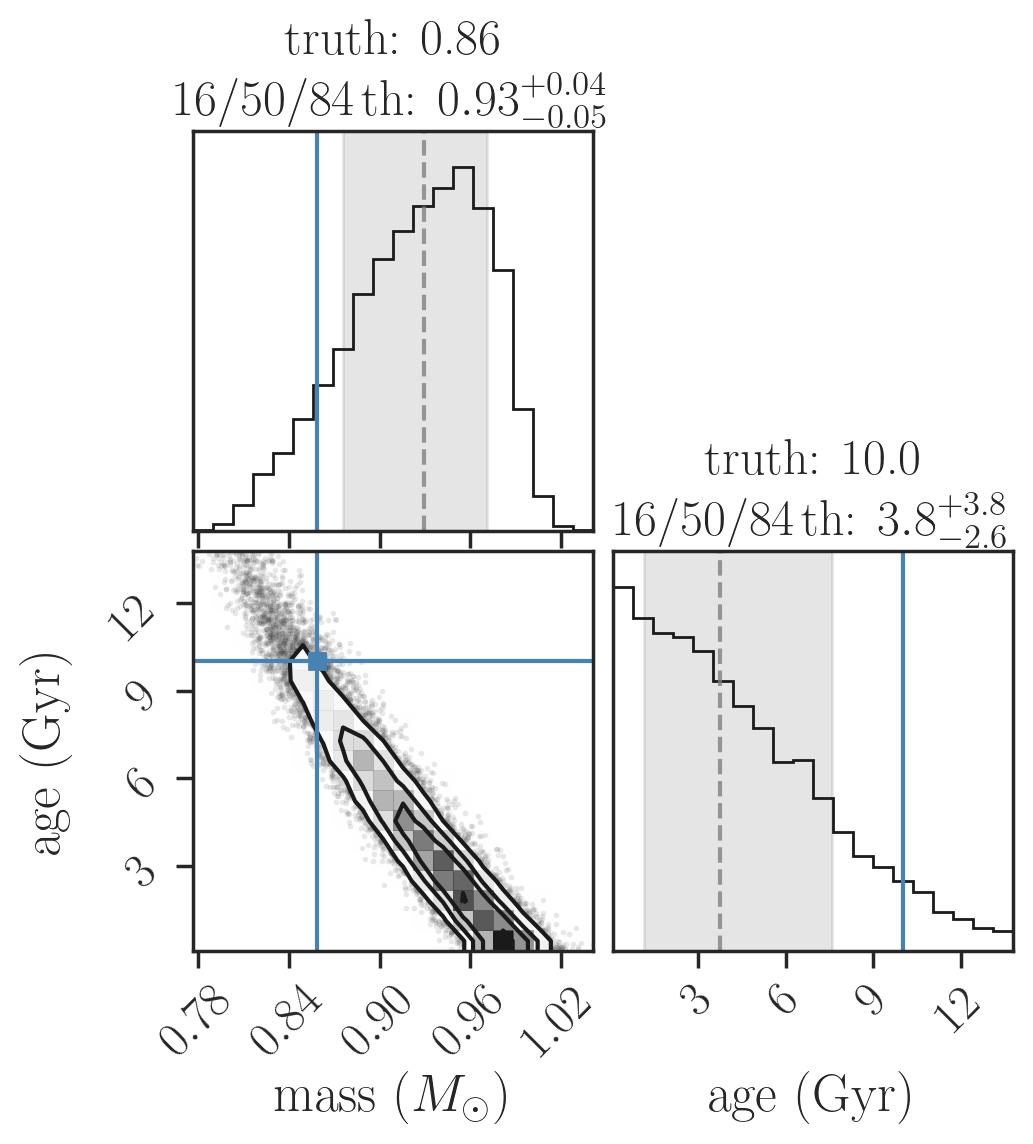}{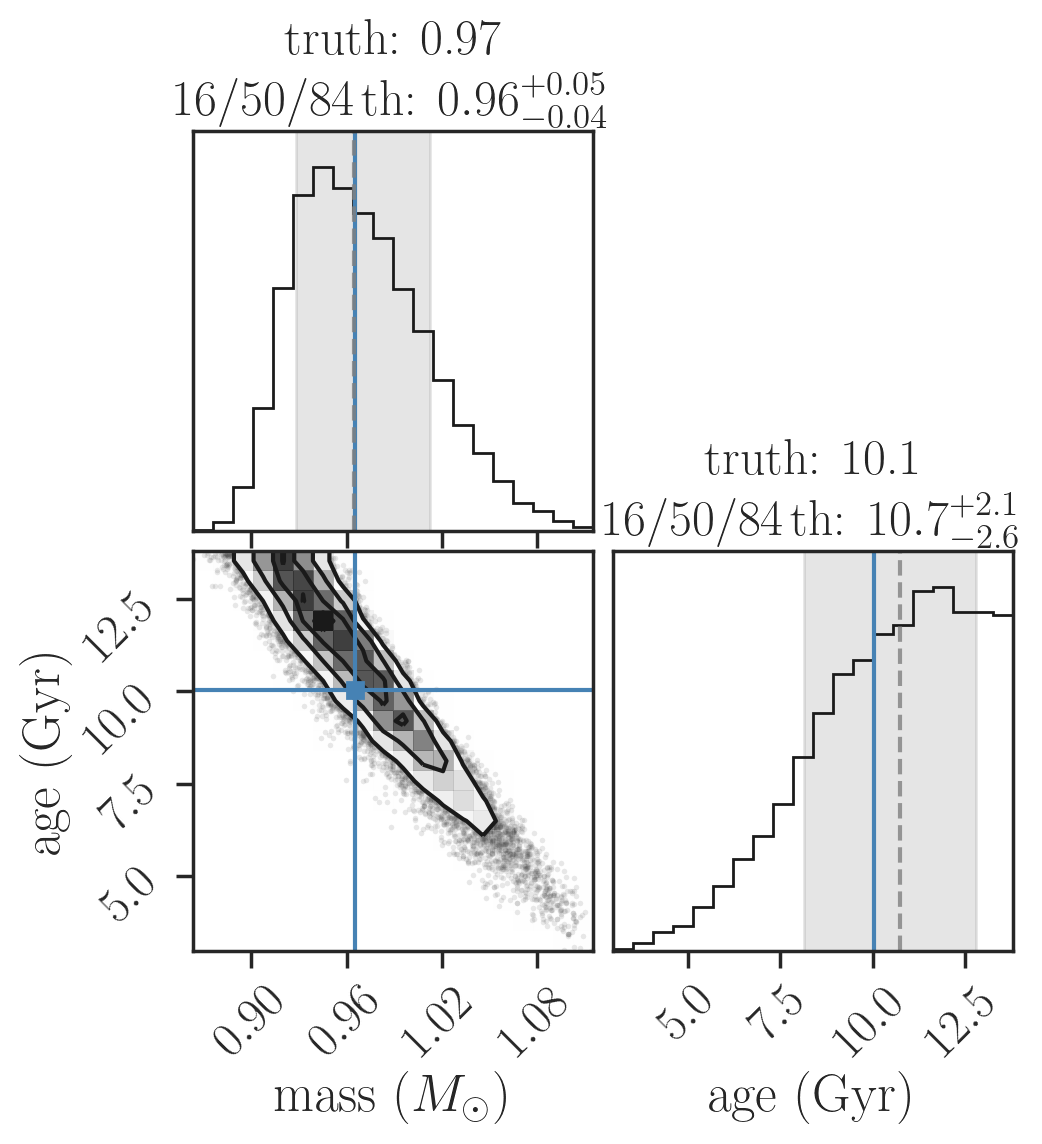}
\plottwo{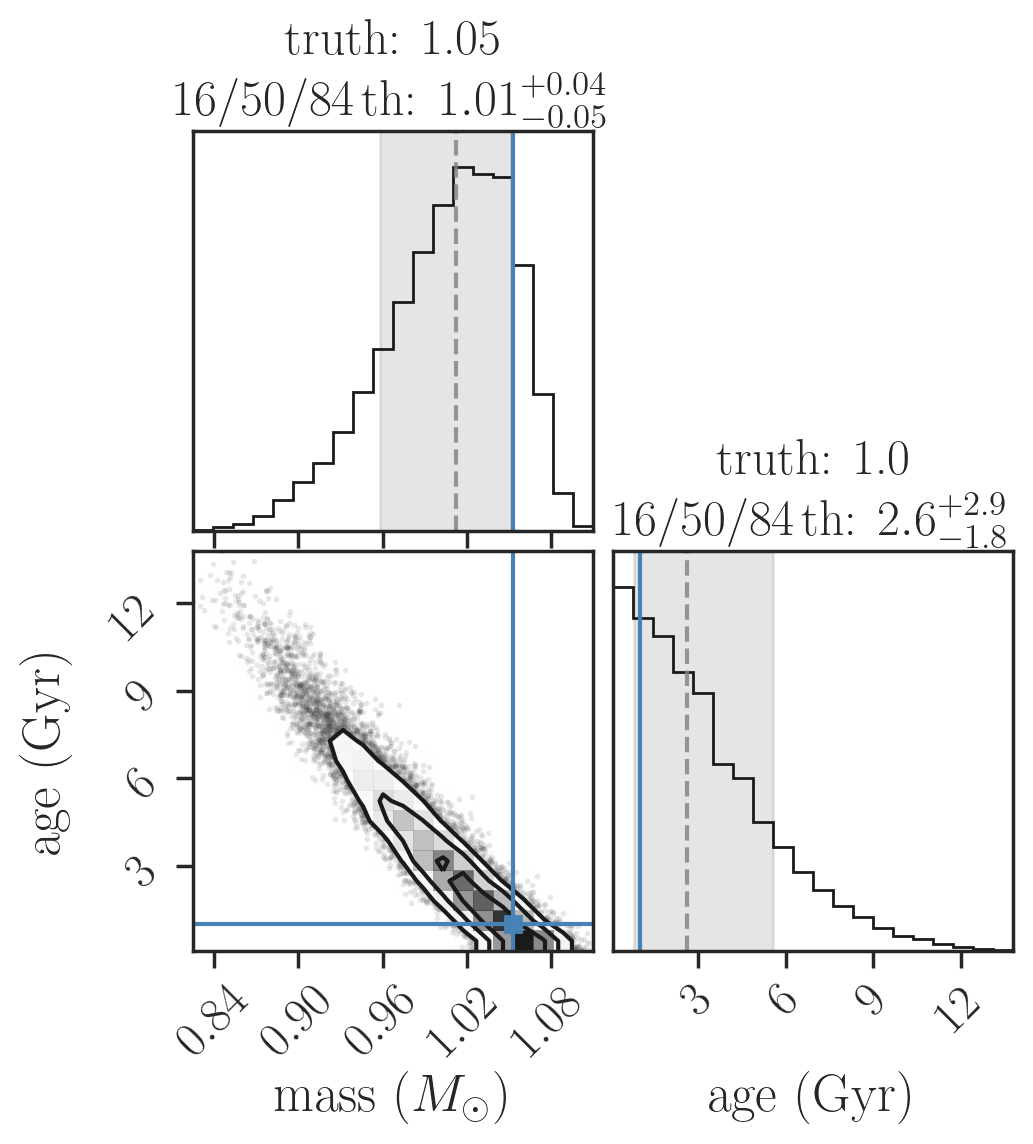}{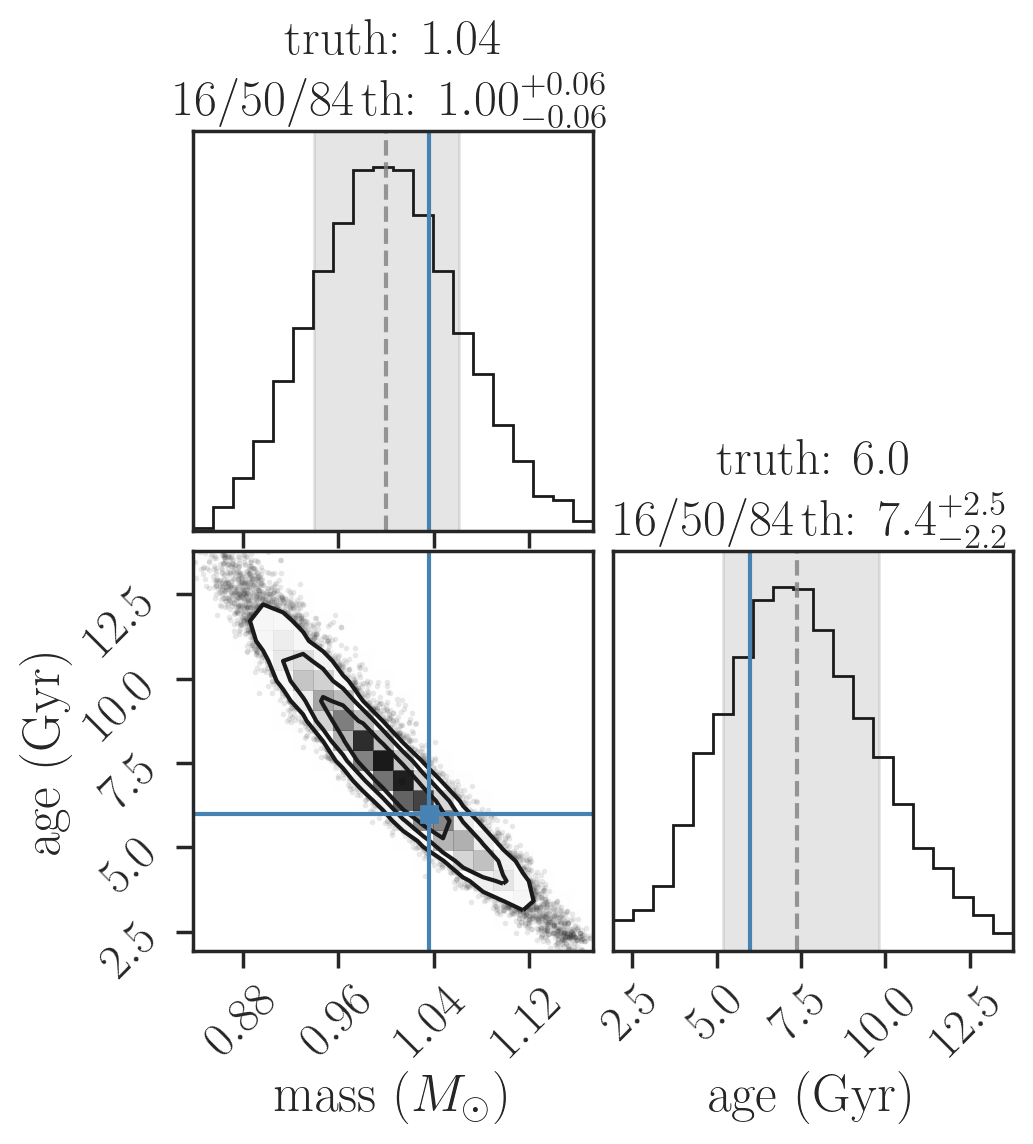}
\caption{Results of injection-and-recovery tests for individual stars. Typically masses are well determined, and age posteriors are not inconsistent with the truths, but simple summary statistics such as median often miss the mark; see also Figure \ref{fig:bias_maps}, and Section~\ref{ssec:isochrone_recovery} for details.}
\label{fig:examples}
\end{figure*}

As we will see, given the current measurement precision and non-linear nature of the function $\bm{y}(\bm{\theta})$, the posterior PDF (\ref{eq:posterior}) is usually non-Gaussian and broad, and is sometimes even multi-modal. For main-sequence stars as we will mostly focus on, the mass and age exhibit a negative correlation so that they result in the same luminosity. For these reasons, the resulting constraints are often not adequately summarized by a single representative value (e.g., ``best-fit" and ``error"), as illustrated in Figure \ref{fig:examples}. One of the main focuses of this paper is how to appropriately interpret such probabilistic constraints. 

To illustrate what kind of constraints we typically obtain, as well as to test the validity of our fitting procedure, we perform injection and recovery tests using simulated data. We computed $(\teff, \feh, \ks, \plx)$ for 5,000 stars whose physical parameters were randomly drawn from the distributions in Table \ref{tab:simulated}, which are based on the CKS sample analyzed in the following sections. The simulated observable parameters were perturbed by the errors shown in the table, and we feed those ``measured" values and the assumed errors into the code. 

\begin{deluxetable}{l@{\hspace{.1cm}}@{\hspace{.15cm}}cc}[!ht]
\tablecaption{Parameters of the Simulated Stars\label{tab:simulated}}
\tablehead{
\colhead{Parameter} & \colhead{Distribution} & \colhead{Assumed Error}
}
\startdata
stellar mass $\mass$ $(M_\odot)$ & $\mathcal{U}(0.7, 1.3)$ & \nodata\\
age $\age$ (Gyr) & $\mathcal{U}(0, 13.8)$ & \nodata\\
$\feh$ & $\mathcal{N}(0.03, 0.18)$ & $0.1$\\
$\log_{10}\plx$ (mas) & $\mathcal{N}(0.17, 0.26)$ & $\plx/100$\\
$\ks$ (mag)& \nodata & $0.023$\\
$\teff$ (K) & \nodata & $110$
\enddata
\tablecomments{The parameters without specified distributions were computed in a deterministic way from the other parameters. Assumed errors show standard deviations of Gaussians. EEP was truncated at 600. $\mathcal{N}(\mu, \sigma)$ means the normal distribution centered on $\mu$ and with variance $\sigma^2$. $\mathcal{U} (a,b)$ is the uniform probability density function between $a$ and $b$.
}
\end{deluxetable}

Figure \ref{fig:examples} shows examples of the resulting posterior samples for four simulated stars along with the ground truths. We find that masses are typically recovered within $\sim 0.05\,M_\odot$. 
Ages are not inconsistent with the truths in the sense that the resulting posteriors have significant probability masses around the truths. However, the marginal distribution for the age is wide and skewed for stars in the main-sequence. 
This makes simple summary statistics such as the median and symmetric $68\%$ interval less useful than for the mass, because different statistics capture different aspects of the distribution and may give very different values.
The fact that the posterior distribution is wide also suggests that such statistics, as well as the entire posterior distribution, are sensitive to the adopted prior.

\begin{figure*}
    \epsscale{1.15}
    \plottwo{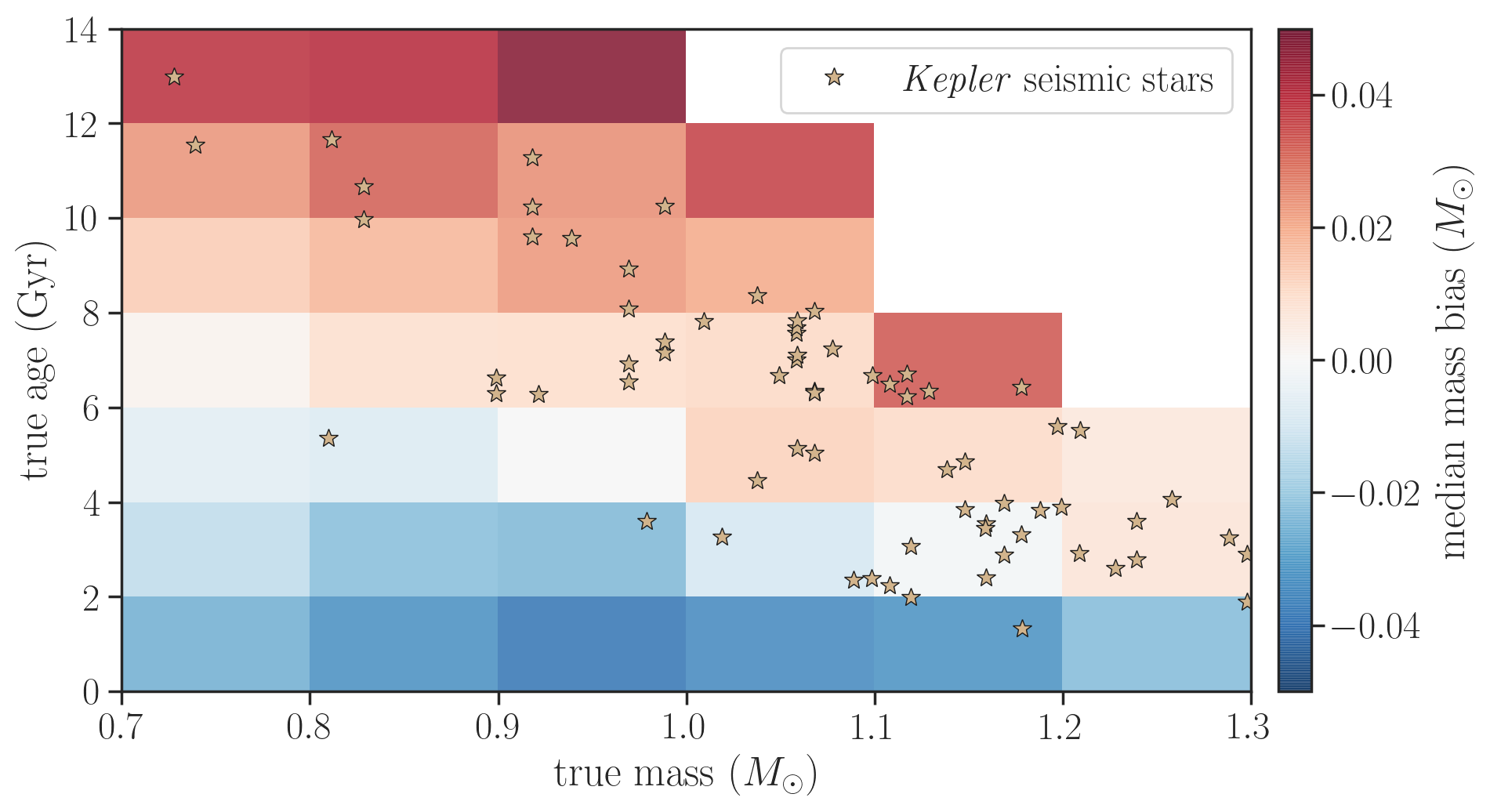}{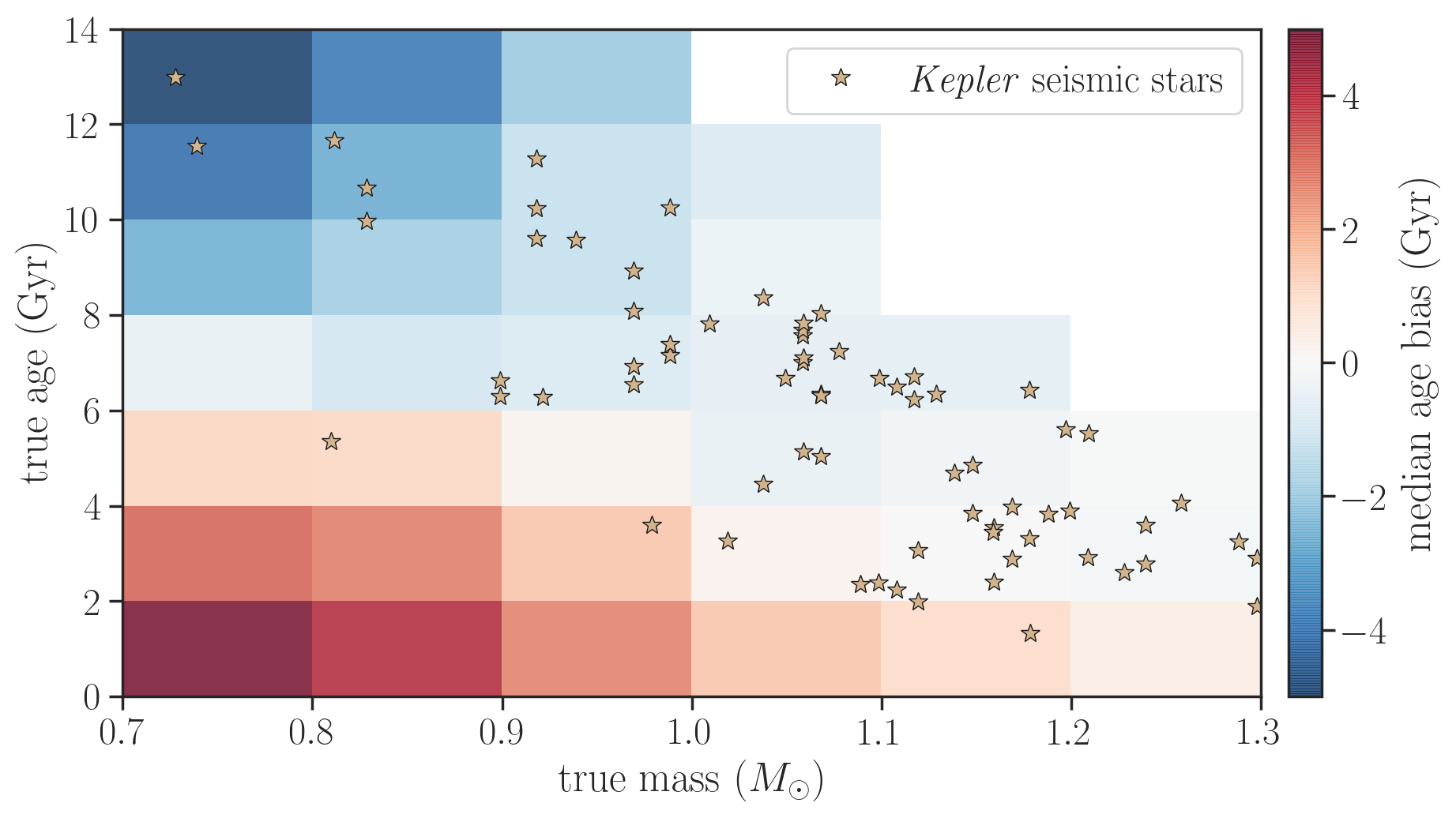}
    \caption{Results of injection-and-recovery tests for 5,000 simulated stars. The plots show the biases in the recovered mass {\it (Left)} and age {\it (Right)} as a function of true stellar mass and age, where the median of the marginal posterior PDF is adopted as the ``recovered" value.
    Each cell contains many stars with different metallicities and parallaxes. In each cell, we compute the median of the differences between the recovered and true masses/ages and show the value with different colors indicated in the right scales. 
    See Section~\ref{ssec:isochrone_recovery} for details.
    The star symbols show the ages and masses of \kepler\ asteroseismic stars from the LEGACY and KEGAS projects; see Section~\ref{ssec:isochrone_seismic}.}
    \label{fig:bias_maps}
    \vspace{0.5cm}
    \epsscale{1.15}
    \plotone{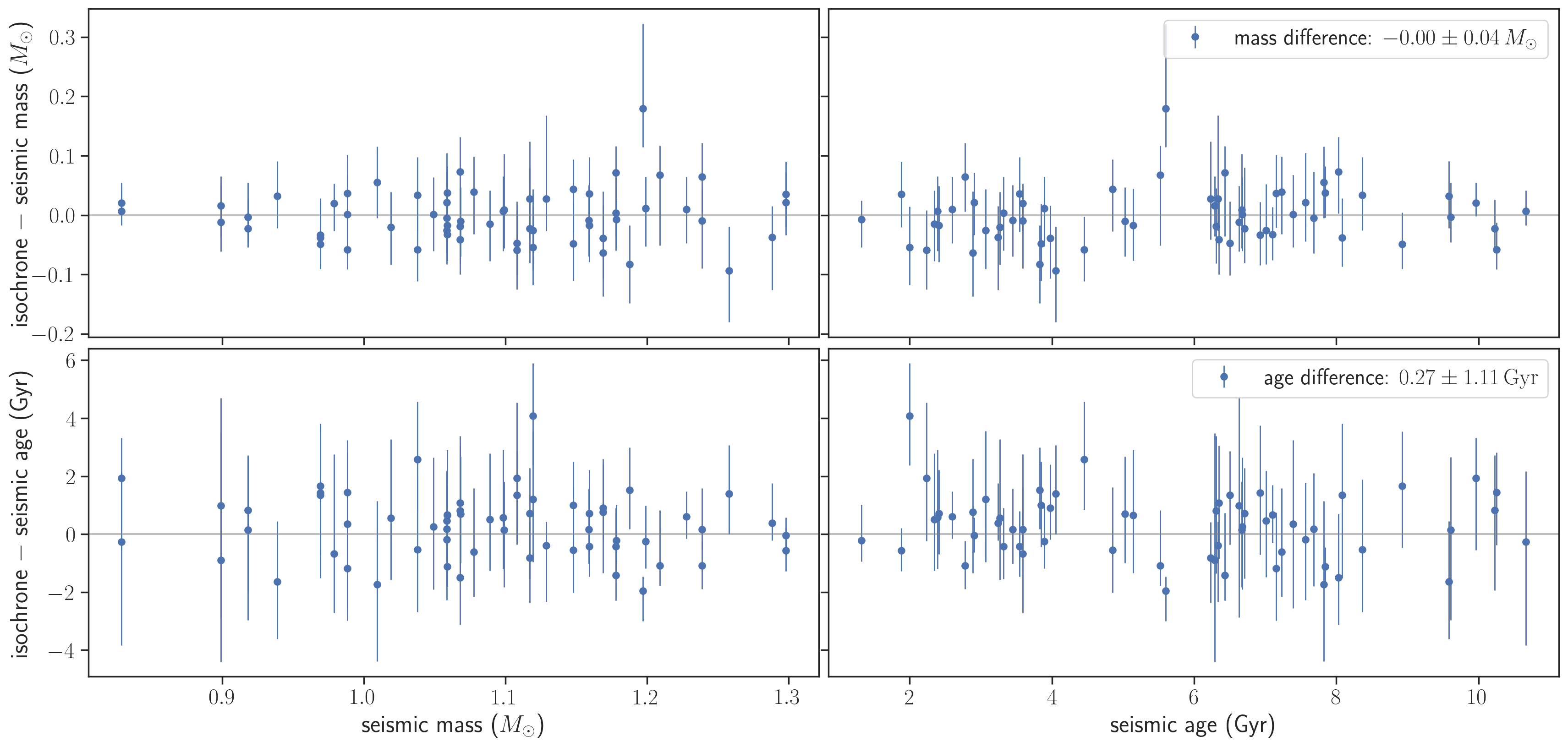}
    \caption{Comparison between our isochrone parameters and seismic parameters for Kepler seismic stars. They happen to have masses and ages for which isochrone fitting works particularly well (see also Figure \ref{fig:bias_maps}). The vertical axes show the difference between the median of the posterior of the isochrone fit from the seismic value. The error bars only show the $68\%$ interval of the isochrone fit.}
    \label{fig:iso_vs_astero}
\end{figure*}

How well does the posterior median (not) work statistically, depending on the true stellar mass and age?
Figure \ref{fig:bias_maps} summarizes how the recovered mass and age are biased or not as a function of the input (true) mass and age. Each cell contains stars with different $\feh$ and $\plx$, and the color corresponds to the median of the differences between the medians of the recovered distribution from the truths. For the whole sample, we found $\mass^{\rm med}-\mass^{\rm true} = 0.00 \pm 0.05 \,M_\odot$ and $\age^{\rm med}-\age^{\rm true} = -0.1 \pm 2.9\,\mathrm{Gyr}$ (mean and standard deviation); so overall the results are not systematically biased.\footnote{Here the simulated stars follow the same uniform mass--age distribution as the adopted prior, so this agreement is not surprising.
The agreement would have been poorer if the parameter distribution of the simulated stars was far from the adopted prior.} 
However, the accuracy depends on the true mass and age as shown in the figure.
The ages are well estimated for super-solar mass stars in the latter halves of their main-sequence lives, as expected; but for less massive stars, the median posterior estimates for individual stars are systematically biased, for the youngest and oldest stars in each mass range. Here the posterior PDFs are not very informative and tend to be close to the prior PDF, and so their medians tend to be around the middle of the prior range. This large uncertainty in age, on the other hand, results in smaller changes in mass. 
We do see a sysetmatic trend in the mass bias that is inversely correlated with the age bias, but the mass bias is typically less than $\sim 0.05\,M_\odot$. 

To summarize, we show that a simple summary (such as median) of the age posterior is of limited use for individual sub-solar mass stars even with the Gaia parallax, while that for mass is good to $\sim 0.05\,M_\odot$. In Section \ref{sec:hbayes}, we will show that such weak constraints, if properly handled, are still useful for inferring the underlying age distribution.

\subsection{Tests using Kepler Seismic Stars}\label{ssec:isochrone_seismic}

Also shown with the star symbols in Figure \ref{fig:bias_maps} are the masses and ages of the \kepler\ seismic stars from the LEGACY \citep{2017ApJ...835..173S} and KEGAS \citep{2015MNRAS.452.2127S} projects; for stars in both, the LEGACY parameters are used. 
Here we test our isochrone fitting method further by applying it to these stars with precise and accurate parameter constraints from asteroseismology.
Interestingly, they are in the regions where the median posterior estimates are supposed to perform relatively well;
the right panel of Figure~\ref{fig:bias_maps}, for example, suggests that isochronal ages for individual stars would be good to $\sim 1\,\mathrm{Gyr}$.

We applied our fitting code to these 94 stars adopting the spectroscopic $\teff$ and $\feh$ as used in the semimic analyses \citep{2015MNRAS.452.2127S, 2017ApJ...835..173S}, 2MASS $\ks$, and Gaia EDR3 parallaxes. 
We assumed a common uncertainty of 110K for $\teff$ and 0.1 for $\feh$ considering the systematics in the absolute scales. The extinction in the $\ks$ band was corrected using {\tt Bayestar17} \citep{2018MNRAS.478..651G}, although the effect was found to be very minor for $\ks$ magnitudes of those nearby stars.
We found the corresponding Gaia sources using their 2MASS IDs, corrected parallax zero points following the recipe in \citet{2021A&A...649A...4L}, and inflated the parallax error using the fitting function of the {\it Gaia} magnitude $G$ derived by \citet{2021MNRAS.506.2269E}.
We did not use $\log g$ from asteroseismic modeling so that the information used here is independent from the seismic analyses. 
We excluded stars 
with their inferred $(\teff, \feh, \ks, \plx)$ (medians of the marginal posteriors) differing from the input values by more than two standard deviations,
which indicates that 
an isochrone model consistent with the observed parameters was not found.
We also excluded stars with $\mathrm{|[Fe/H]|}>0.4$ or with seimic mass $>1.3\,M_\odot$; for those most metal-poor or metal-rich stars we typically found larger mismatches between the models and observables.

The results of our isochrone fitting are compared with the seismic values in Figure \ref{fig:iso_vs_astero}, where we show the difference between the posterior medians and the reported seismic values. The error bars are those for isochrone fitting alone, and show the 16th/84th percentiles of the marginal posteriors.
We find the mean differences for mass and age to be $0.00\,M_\odot$ and $0.3\,\mathrm{Gyr}$, and their standard deviations to be $0.04\,M_\odot$ and $1.1\,\mathrm{Gyr}$.\footnote{For more massive stars, the isochrone-based masses were found to agree less well. We do not deal with such stars in this paper.} 
This good agreement --- including that the amount of scatter is as expected from our injection-and-recovery simulations in Section~\ref{ssec:isochrone_recovery} --- further validates our isochrone fitting procedure.
We reiterate, though, that these seismic stars happened to be the stars for which isochrone fitting works best;
age precision, in particular, would be poorer for younger, lower-mass stars
as shown in Figure \ref{fig:bias_maps}.

\subsection{Note on the Summary Statistics}

In this section, we used the median and symmetric $68\%$ interval of the marginal posterior PDF as summary statistics. Since the PDF is not Gaussian (see Figure~\ref{fig:examples}), the two values inherently miss some information in the PDF. 
We chose them for illustration here, simply because they are widely used. One could adopt other metrics such as mean and standard deviation, or maximum a posteriori and highest probability density intervals. We tried them and found that none of them is obviously closer to the truths than the others; some work better in a certain region of the parameter space but less well in other parts.
The results as shown in Figure \ref{fig:bias_maps} depend on such choices and that is exactly what we mean by simple summaries are of limited use. The following analyses in this paper will not be based on any such summaries but on the whole likelihood functions, i.e., the information on how well different sets of parameters to be inferred fit the observed parameters (data) for each star.

\begin{figure*}
    \epsscale{1.12}
    \plottwo{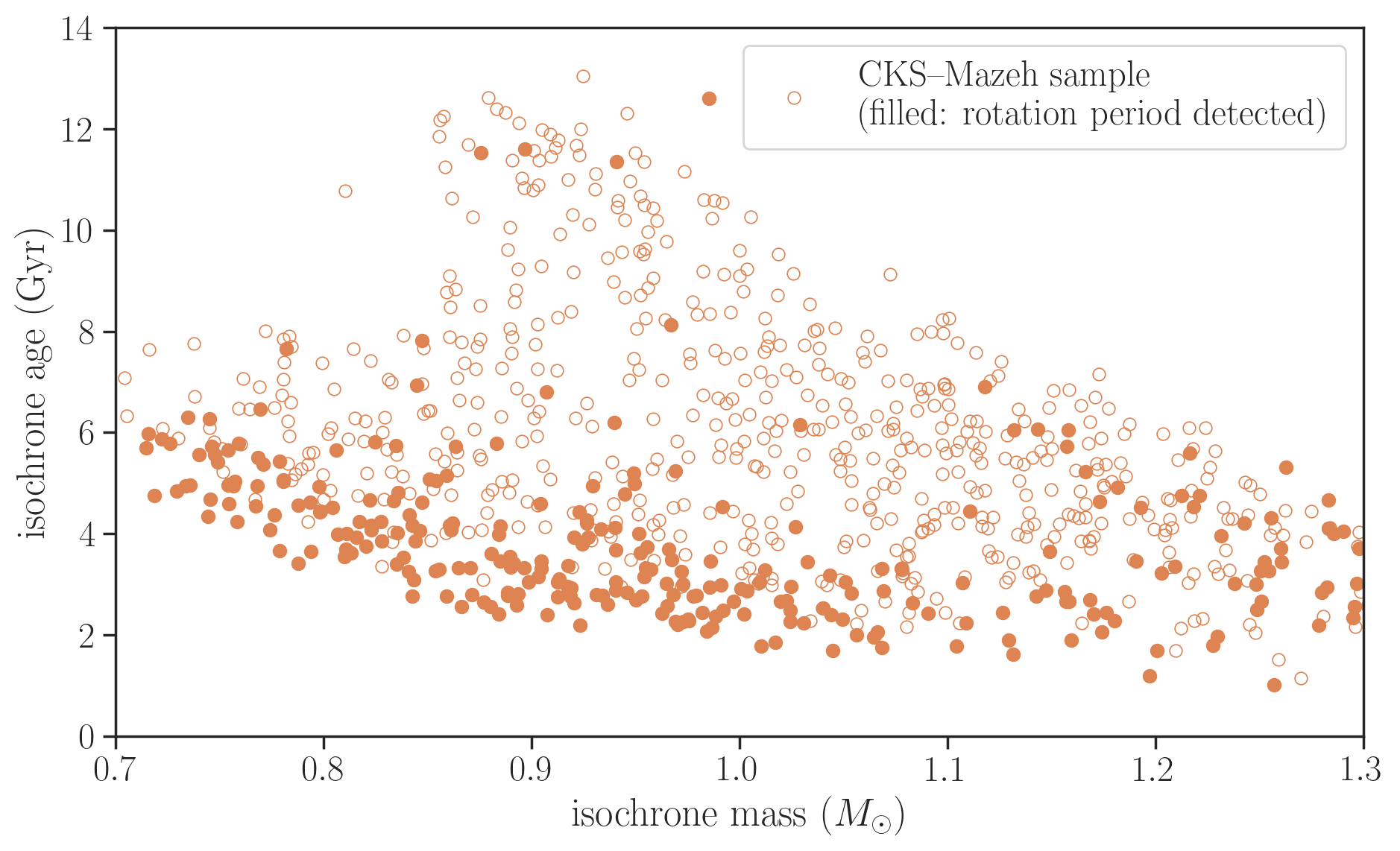}{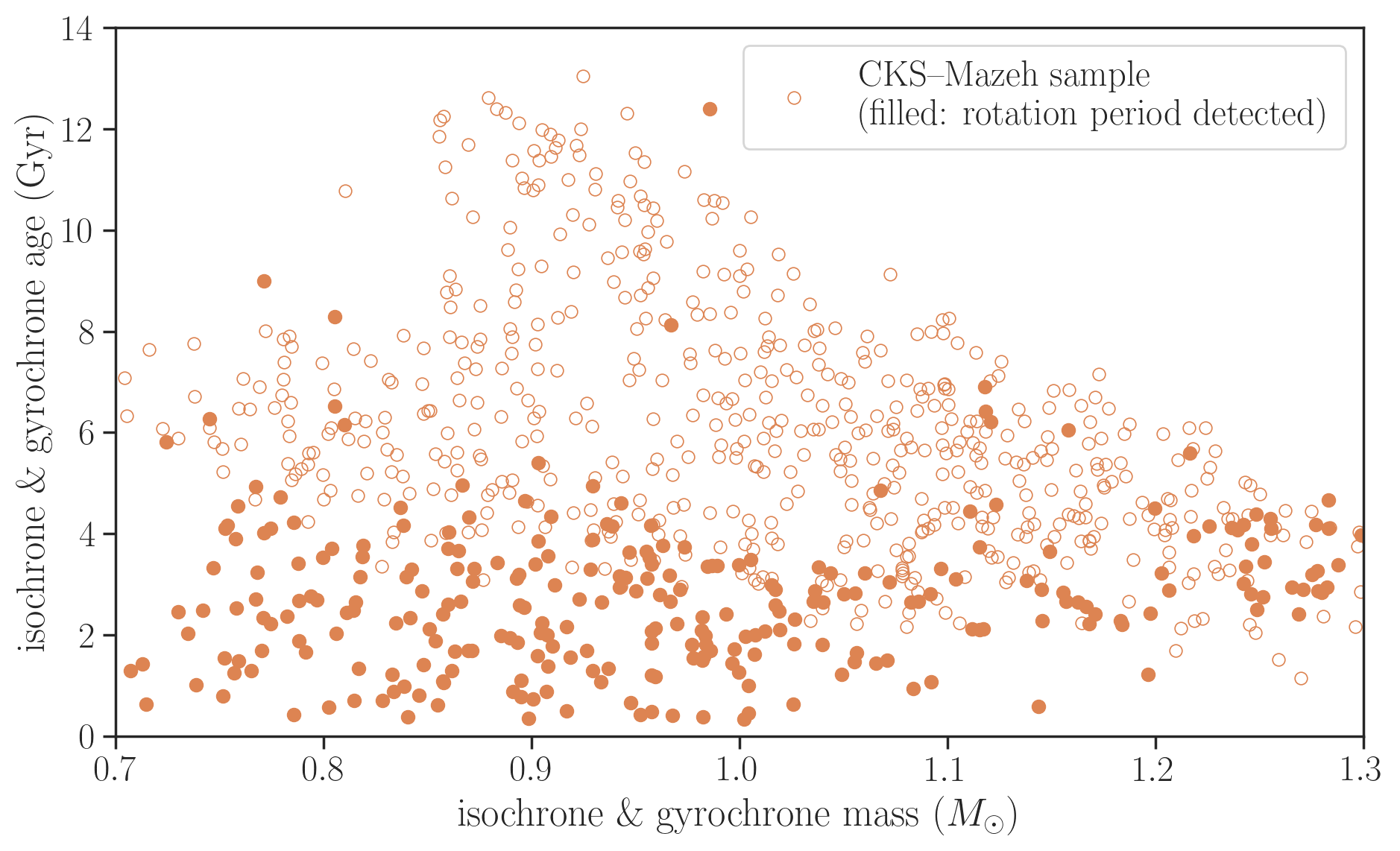}
    \caption{Ages and masses of the CKS stars derived from {\it (Left)} isochrone fitting only and from {\it (Right)} isochrone \& gyrochrone in Section~\ref{sec:cks}. Filled circles show stars for which rotational modualtion has been detected by \citet{2015ApJ...801....3M}.
    These plots show the medians of the marginal posteriors, which are significantly biased in a large part of the parameter space (see Figure \ref{fig:bias_maps}). For example, although no star appears to have ages $\lesssim 2\,\mathrm{Gyr}$ in the left panel, many of the stars around the lower edge in fact have large age uncertainties and are (also) consistent with ages $\lesssim 2\,\mathrm{Gyr}$. Therefore these plots are only for illustration; a more careful analysis fully taking into account the age and mass uncertainties will be performed in Section \ref{sec:results} to derive the fraction of stars with robust rotation periods as a function of mass and age.}
    \label{fig:age_vs_mass}
\end{figure*}

\section{Isochrone Modeling of the CKS Stars}\label{sec:cks}

The California-\kepler\ Survey \citep[CKS;][]{2017AJ....154..107P, 2017AJ....154..108J} 
provided high-resolution ($R\sim 55,000$) Keck/HIRES spectra as well as the spectroscopic parameters for $>1,000$ FGK \kepler\ stars with known (candidate) transiting planets. 
For generic \kepler\ stars with candidate transiting planets, \citet{2015ApJ...801....3M} performed a homogeneous search for stellar rotation periods using the auto-ACF method \citep{2014ApJS..211...24M} and published their search results {\it both for stars with and without detected rotation periods.} In this paper, we focus on the intersection of the two samples, for which rotational modulation has been uniformly searched and atmospheric parameters have been homogeneously derived from high-resolution spectra.

We use $\teff$ and $\feh$ obtained from the {\tt SpecMatch} pipeline for 1305 CKS stars assuming Gaussian errors of 110K and 0.1 respectively \citep{2017AJ....154..107P}. The stars were cross matched with Gaia EDR3 using their 2MASS IDs. We chose the stars within $10^{-3}$~arcsec and $|G-Kp|<0.2$
and found 1202 sources with measured parallaxes.\footnote{We omitted KIC 3957082 for which CKS parameters were missing.}
We corrected the parallax zero point for each star according to the recipe given by \citet{2021A&A...649A...4L} and inflated the parallax error using the fitting function of $G$ derived by \citet{2021MNRAS.506.2269E}.
We took $\ks$ magnitudes from 2MASS and corrected for extinction using the dust map by \citet{2019ApJ...887...93G}. 
When the errors of the $K_s$ magnitudes were missing, they were replaced by the median error of the sample. Among them, there are 1054 stars for which rotation periods have been searched by \citet{2015ApJ...801....3M}. 

We perform isochrone fitting for these 1,054 stars as described below, with and without incorporating gyrochronal age constraints from the photometrically measured rotation periods when available. The results are used to define two sets of samples that will be analyzed separately in Section~\ref{sec:results}.

\subsection{Isochrone-only Sample}\label{ssec:cks_iso}

First, we performed the isochrone fitting as described in Section \ref{sec:isochrone} and obtained posterior samples for the physical parameters of the stars including mass and age. As we did in Section \ref{ssec:isochrone_seismic}, we excluded stars whose fitted 
$(\teff, \feh, \ks, \plx)$ differ from the input values by more than two standard deviations.
We also excluded 29 stars with $\mathrm{|[Fe/H]|>0.4}$ considering the test results using seismic stars in Section~\ref{ssec:isochrone_seismic}, as well as those with 
potentially problematic Gaia astrometry and/or potential binary contaminants from the sample following \citet{2022MNRAS.510.5623M}.
These criteria left us with 948 stars. We then focus on its subset whose isochrone masses (as specified by medians of the marginal posteriors) are between $0.7$--$1.3\,M_\odot$. The final sample consists of 855 stars.\footnote{The summary table is available through GitHub: \url{https://github.com/kemasuda/acheron/tree/main/cks_frot}.}

Our results are consistent with those in \citet{2022AJ....163..179P}, who derived physical parameters of the CKS stars in almost the same way, within typical uncertainties. All of the 855 stars have been analyzed in \citet{2022AJ....163..179P}, and the differences of our posterior medians from their values (mean offset and standard deviation) are $0.00\pm0.02\,M_\odot$ for the mass, $0.01\pm0.03\,R_\odot$ for the radius, and $0.33\pm1.33\,\mathrm{Gyr}$ for the age. The origin of the slight offset in the median posterior age is unclear. It could be attributed to the difference in the priors adopted in the isochrone fitting. Since the following analyses require the entire samples from the posterior PDF, we will use the results from our own isochrone fitting.

\subsection{Joint Isochrone \& Gyrochrone Sample}\label{ssec:cks_joint}

Among the 1,054 stars, robust detection of rotational modulation has been reported by \citet{2015ApJ...801....3M} for 359 stars ($34\%$ of the sample). Here ``robust" means that the detected period is consistent in different quarters (their flag M1), the signal is strong enough to be reliable (M2), and has passed visual examination (R). 
For those stars, we also performed isochrone fitting additionally incorporating gyrochrone information calibrated to the Praesepe cluster and the Sun, following the method in \citet{2019AJ....158..173A}. We found that the resulting age constraints were typically dominated by the information from rotation periods.
Using the outputs from this joint fitting when available, the same cut as in Section~\ref{ssec:cks_iso}
left us with 855 stars, in which 278 stars ($33\%$) have robustly detected rotation periods. 

\subsection{Masses, Ages, and Photometric Modulation Amplitudes of Stars With and Without Robustly Detected Rotation Periods}\label{ssec:cks_result}

In Figure \ref{fig:age_vs_mass}, we show ``point estimates" for ages and masses from the above analyses with and without gyrochronal constraints, using the medians of the marginal posterior PDFs.
The filled circles show stars for which robust detection of rotational modulation (and hence rotation period) has been reported 
in \citet{2015ApJ...801....3M}. In the right panel, gyrochronal information was used in addition to the isochrone likelihood (Section~\ref{ssec:cks_joint}) for those stars.  
Both results show that rotational modulation has been detected for the youngest stars, and that the fraction of such stars rapidly decreases to almost zero as they become older (but see also Section~\ref{ssec:hbayes_movitation} for caveats in interpreting these results based on such point estimates and for why the two results appear to be different).

\begin{figure*}
    \epsscale{1.12}
    \plottwo{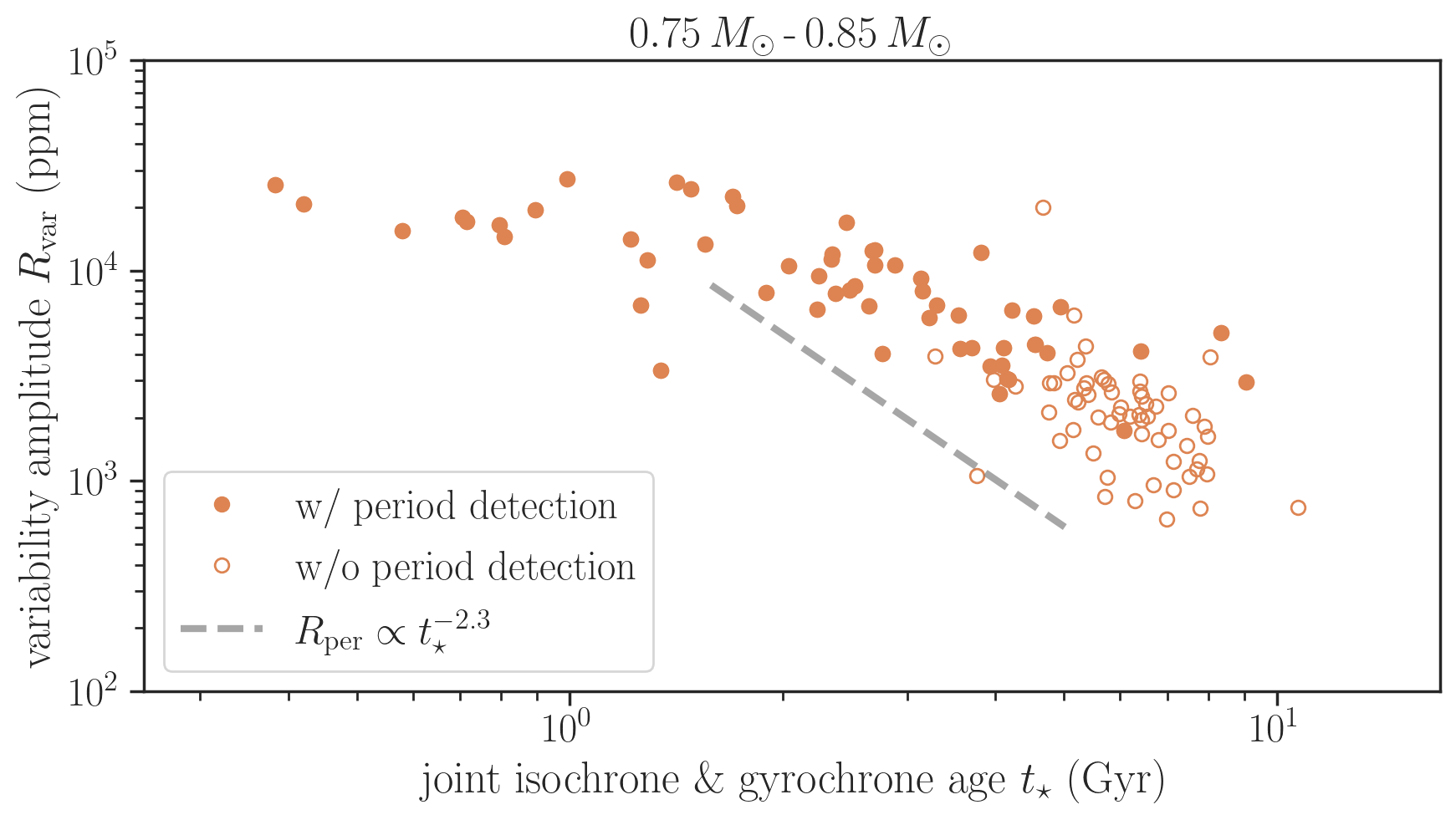}{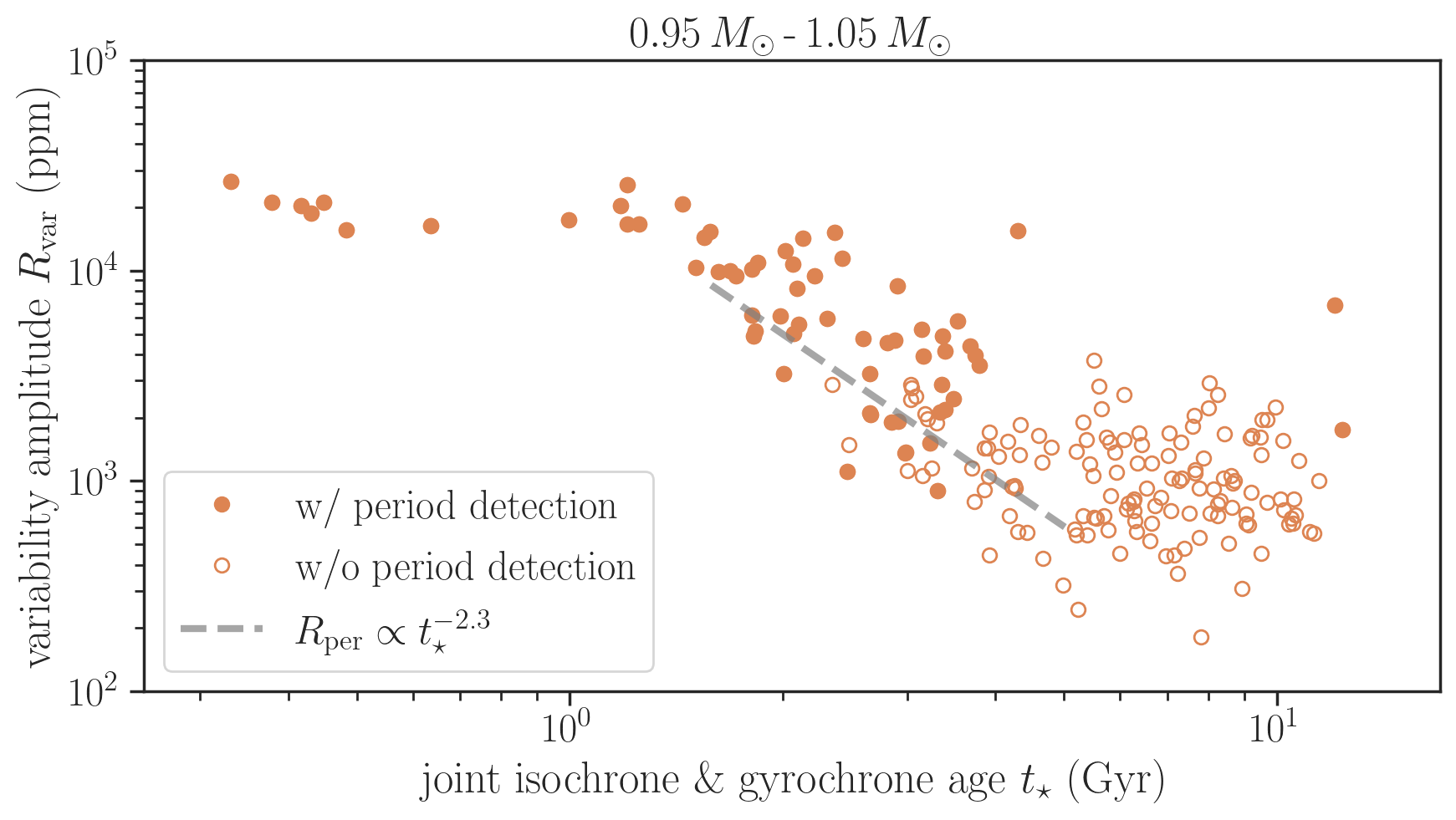}
    \plottwo{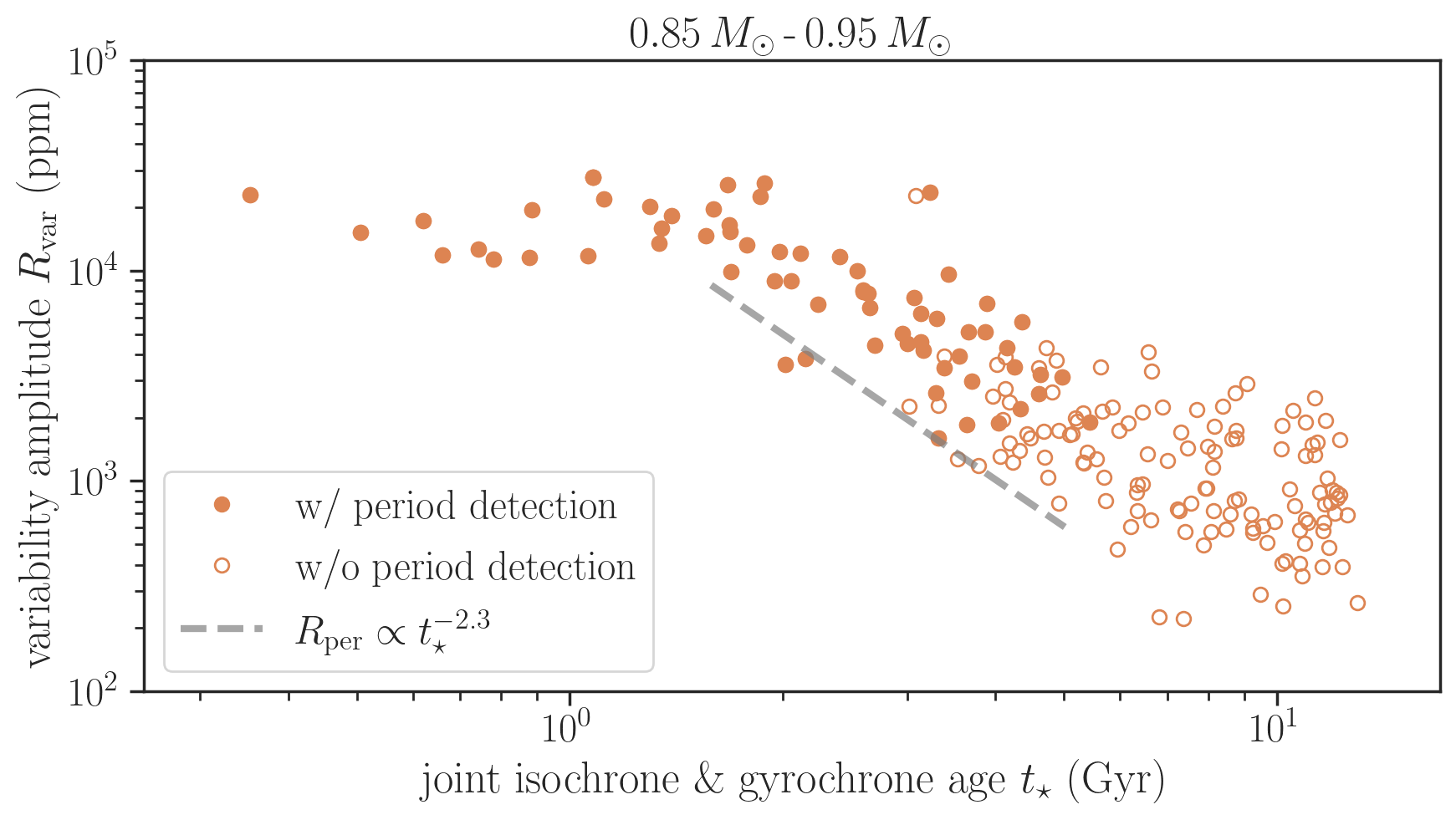}{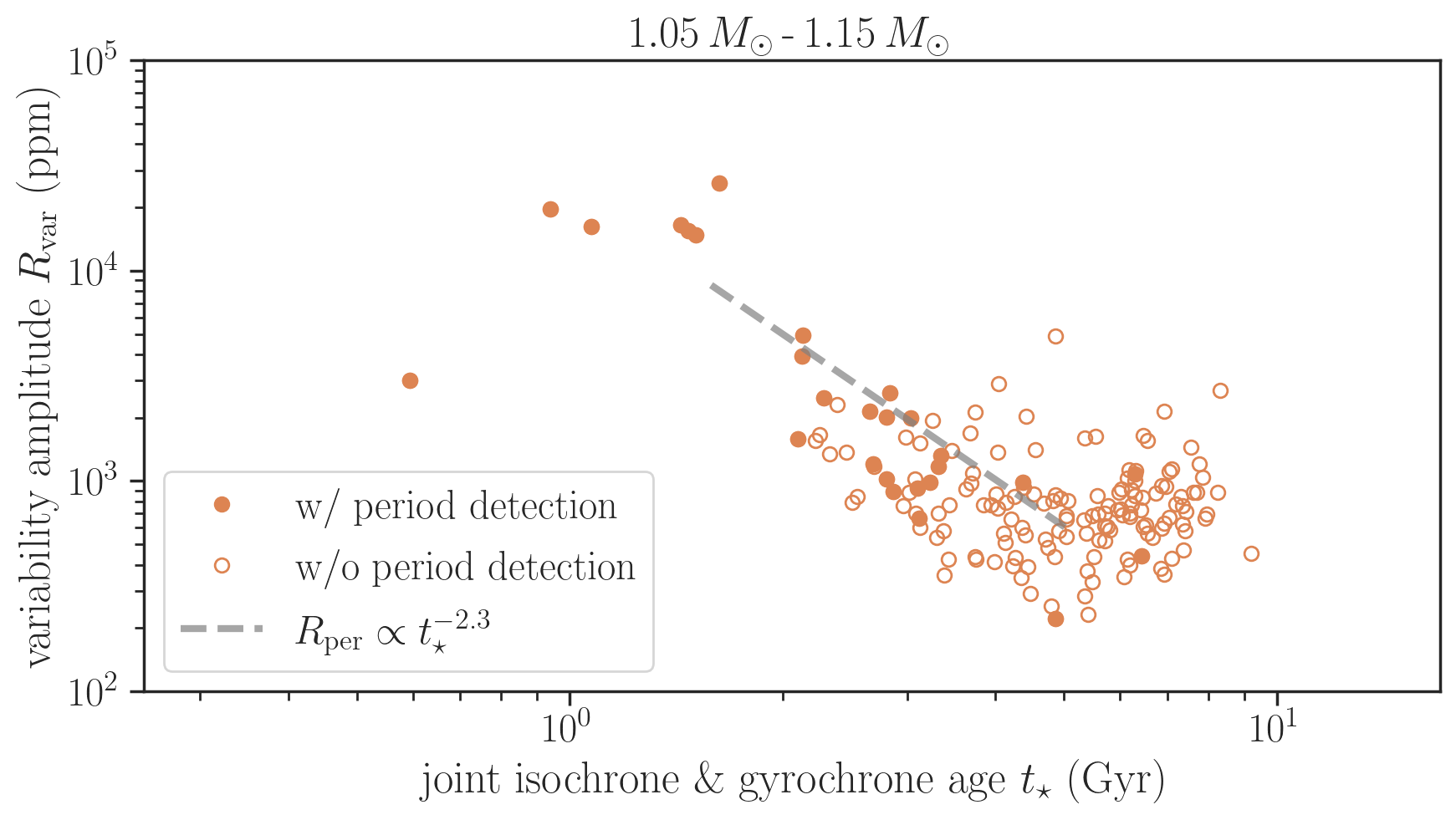}
    \caption{
    Amplitudes of photometric modulation versus ages for stars with different mass ranges shown in the title of each plot. Stars with robust period detection are shown with filled circles, for which ages and masses are from joint isochrone \& gyrochrone fitting (Section~\ref{ssec:cks_joint}); otherwise the parameters are from isochrone-only fitting (Section~\ref{ssec:cks_iso}). The ages shown here are medians of the marginal posterior PDFs.
    Gray dashed lines show the scaling $\amp \propto \age^{-2.3}$, which for $\prot \propto \age^{1/2}$ corresponds to $\amp \propto \prot^{-4.6}$ derived in \masuda\ using a larger sample of {\it Kepler} stars. Thus this is {\it not} a fit to the data shown here; see the text for details.}
    \label{fig:rvar_vs_age}
\end{figure*}

\begin{figure*}
    \epsscale{1.12}
    \plottwo{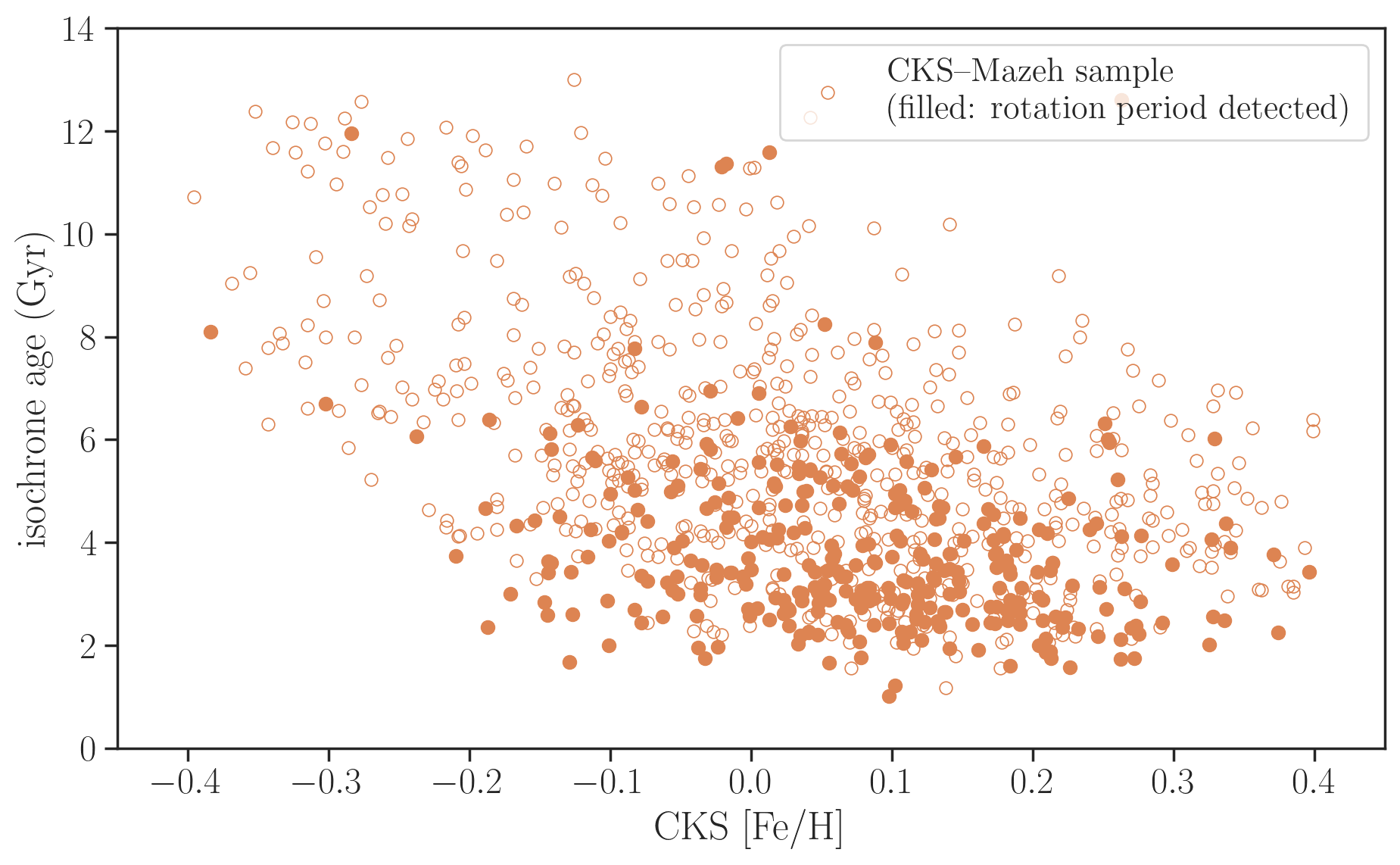}{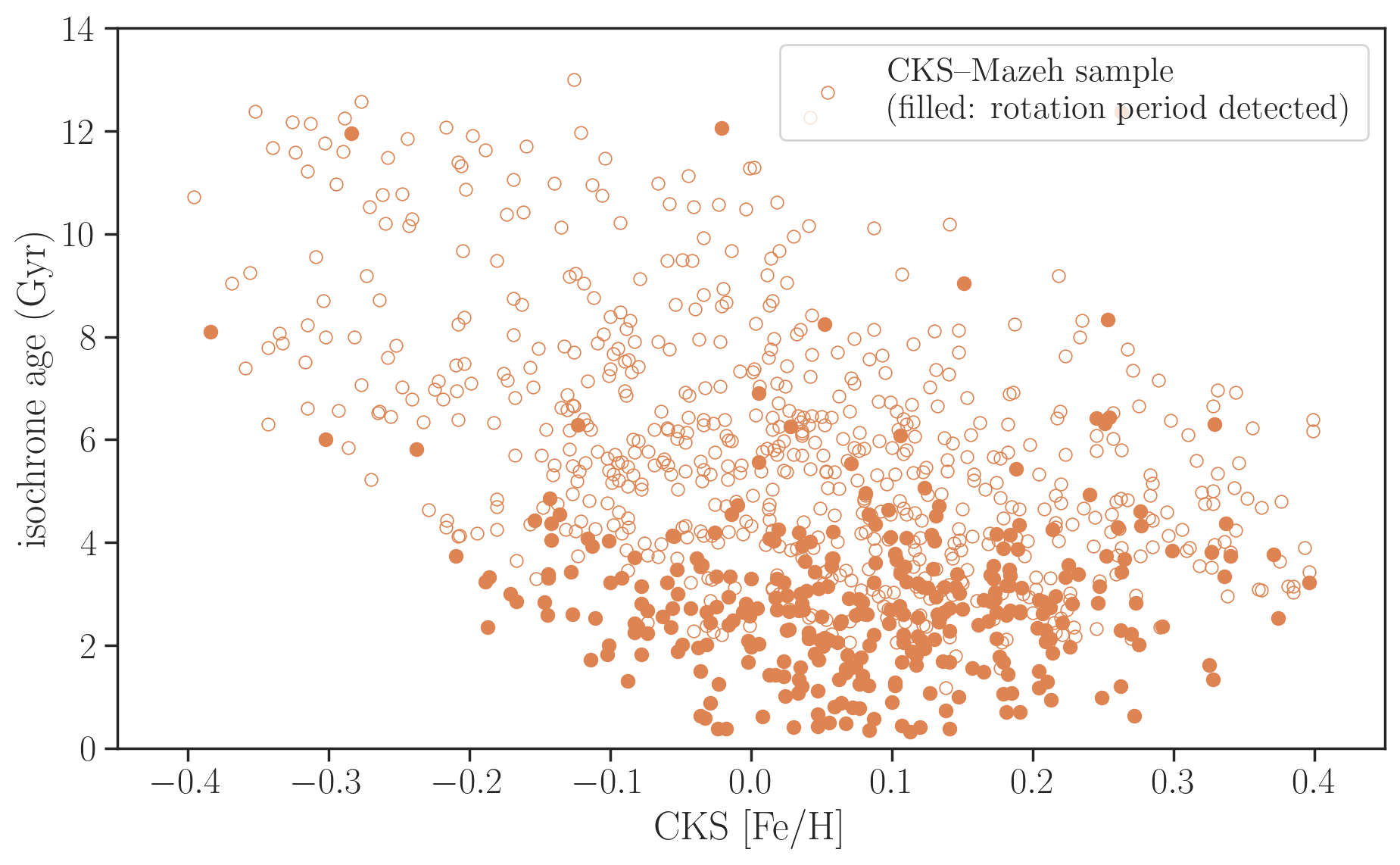}
    \caption{Ages and metallicities of the CKS stars. The metallicity comes from the CKS, and ages are medians of the marginal posteriors from isohcrone-only fitting {\it (Left)} and from joint isochrone \& gyrochrone fitting {\it (Right)}; see Section~\ref{sec:cks}. Filled circles show stars for which rotational modulation has been detected by \citet{2015ApJ...801....3M}.
    }
    \label{fig:age_vs_feh}
\end{figure*}

The simplest interpretation of this trend is that rotational modulation has been missed for older stars due to their weaker photometric variabilities. 
Although we will perform more careful analyses taking into account age and mass uncertainties in the following sections,
this is already supported by Figure~\ref{fig:rvar_vs_age}. Here we show how the photometric modulation amplitude $R_\mathrm{var}$ as reported by \citet{2015ApJ...801....3M}, which is defined as the median of the differences between the 95th and 5th percentiles of normalized flux in each rotation period cycle, changes as a function of the median posterior ages for stars with different masses. In \citet{2015ApJ...801....3M}, $R_\mathrm{var}$ is assigned also for stars without robust period detection (open circles), which is based on rotation periods that have not passed their criteria for robust detection (See Section~\ref{ssec:cks_joint}); the ages of these stars are thus estimated based on isochrone fitting alone (Section~\ref{ssec:cks_iso}). Figure~\ref{fig:rvar_vs_age} shows that robust detection is limited to stars with larger $R_\mathrm{var}$ that are younger.
The $R_\mathrm{var}$--$\age$ relation roughly follows $R_\mathrm{var} \propto \age^{-2.3}$, which corresponds to $\amp \propto \prot^{-4.6}$ derived by \masuda\ using the brightest solar-mass \kepler\ stars from \citet{2014ApJS..211...24M},\footnote{The modulation amplitude is denoted as $R_\mathrm{per}$ (rather than $R_\mathrm{var}$) in \citet{2014ApJS..211...24M}.} 
assuming the Skumanich relation $\prot \sim \age^{1/2}$.
The good correlation between $R_\mathrm{var}$ and $\age$ we see in Figure~\ref{fig:rvar_vs_age} also supports that we have correctly sorted ages of these stars despite that ages of stars shown with open circles are based on isochrone fitting alone.

Figure~\ref{fig:rvar_vs_age} shows that the ``modulation amplitudes" of stars with $\gtrsim 0.9\,M_\odot$ appear to plateau at older ages. Here we note that the detection of rotational modulation has not been considered to be significant in \citet{2015ApJ...801....3M} for these stars, and so it is not clear whether this variability is indeed associated with rotational modulation. In fact, the assigned amplitudes are close to the noise level of \kepler\ for these stars; so even if the detected periods are actual rotational periods, the modulation amplitudes may have been overestimated. For these reasons, it is not clear how the modulation amplitudes evolve at these older ages.

Could the detectable modulation be associated with metal-rich stars rather than young stars?
In Figure \ref{fig:age_vs_feh}, we show ages and metallicities for stars with and without robust detection of rotation periods. The ages are medians of the marginal posteriors as in the previous figures, but now $\feh$ is the value from the {\tt SpecMatch} pipeline, although the posterior constraints from the isochrone fitting are similar to those measured values. The plots show that the detectability of rotational modulation depends primarily on age rather than metallicity: at a given age, rotational modulation has been detected both for metal-rich and metal-poor stars, while at a given metallicity detection is limited to young stars.

\section{Inferring the Age--Mass Distribution}\label{sec:hbayes}

\subsection{Motivation}\label{ssec:hbayes_movitation}

The point estimates in Figure \ref{fig:age_vs_mass} based on isochrone (\& gyrochrone) analyses in Section~\ref{sec:cks} suggest that rotational modulation has been detected if and only if a star is young. 
On the other hand, the plots also show obvious issues associated with point estimates based on broad and skewed posterior PDFs. We saw in Figure \ref{fig:bias_maps} that youngest stars tended to be assigned older ages and vice versa especially for stars less massive than the Sun when the median of the posterior PDF is adopted, and the lack of points in the bottom and top-left parts of the left diagram is indeed consistent with this bias: in other words, the posterior age PDFs for many stars extend to these empty regions, while their medians do not (cf.~left panels in Figure~\ref{fig:examples}).
As we discussed in Section~\ref{sec:isochrone}, this implies that the constraints presented this way are sensitive to the adopted priors for age (e.g., uniform or log-uniform) and what statistics (e.g., median, mean, mode) was used to summarize the information from the posterior.

In the right panel of Figure \ref{fig:age_vs_mass} based on the joint isochrone \& gyrochrone fitting, those stars with rotation periods (i.e., filled circles) are assigned even younger ages than in the left panel, and the tendency of the young stars to show rotational modulation appears stronger. The ``shift" in the location of those stars shown with filled circles typically happened as follows: in the left panel, some stars have posterior PDFs similar to the two left cases in Figure~\ref{fig:examples}, and the gyrochrone information caused the PDF to shrink toward younger ages, decreasing both the median values and the widths of the PDFs. Thus the stars with and without detected rotation periods now have very different age uncertainties. This difference also needs to be accounted for carefully:
it might be the case, for example, that some of the stars without detected periods are in fact similarly young to those with detected periods, but are just appearing to be older due to large age uncertainties. Then we would overestimate the fraction of filled circles at the youngest ages.

To address these issues more carefully, we need to go beyond the point estimates using medians (or whatever summary statistics) of the posterior PDFs and to deal with the whole information provided by the likelihood function.
We present a framework in Section \ref{ssec:hbayes_method}, test it with simulated data sets in Section \ref{ssec:hbayes_tests}, and apply it to the CKS stars in Section \ref{sec:results}.

\subsection{Framework}\label{ssec:hbayes_method}

We formulate the problem as a simultaneous inference of
(i) the occurrence rate of stars in the mass--age plane (i.e., joint mass--age distribution) $p_{\rm occ}$ marginalized over other parameters including metallicity,\footnote{Here we are implicitly assuming that the joint distribution for $\mass$, $\age$, $\feh$, and distance is separable into that for ($\mass, \age$) and that for the other parameters. See Section~\ref{ssec:caveats} for the caveats associated with this assumption.
}
and (ii) the probability for a given star with mass $\mass$ and age $\age$ to exhibit rotational modulation detectable in \kepler\ light curves, $f(\mass, \age)$. We model the former in a ``non-parametric" way as a histogram in the mass--age plane:
\begin{equation}
\label{eq:pocc}
    p_\mathrm{occ}(x|\alpha) = \sum_{k=1}^M \exp(\alpha_k)\,\Pi_k(x),
\end{equation}
where $x=(\mass, \age)$, $\Pi_k$ is a step function ($1$ if $x$ is in the $k$th bin and 0 otherwise), 
and $\alpha$ is a set of parameters that controls the bin heights
\citep[see, e.g.,][for another example]{2014ApJ...795...64F}.
We assume that $\alpha$ is normalized so that $\sum_k \exp(\alpha_k)\,\Delta_k=1$ with $\Delta_k$ being the area of the $k$th bin; therefore $p_{\rm occ}$ is the probability density function and $\alpha_k$ denotes the log probability density in the $k$th bin.
Correspondingly, we assume that $f(\mass,\age)$ is constant in each bin and estimate the value of $0\leq f_k \leq 1$ in each bin.
In this paper, we set up 12 bins for masses spanning 0.7--1.3$\,M_\odot$ at $0.05\,M_\odot$ intervals and 14 bins for ages spanning 0--14$\,\mathrm{Gyr}$ at $1\,\mathrm{Gyr}$ intervals, which results in the total bin number of $M=168$.

Stated this way, our goal is to infer $\alpha=(\alpha_1, \alpha_2, \dots, \alpha_M)$ and $f=(f_1, f_2, \dots, f_M)$ using the data for $N$ stars. The data are (i) the observables $D_j$ used for isochrone fitting (and rotation period for joint fitting) defined in Section~\ref{ssec:isochrone_method}, and (ii) the result of rotation period search $R_j$ by \citet{2015ApJ...801....3M}, where $j$ is a label for individual stars in the sample. Here we assume that $R_j$ simply implies whether the period is detected (1) or not (0) for a given star.
We wish to sample from
\begin{equation}
p(\alpha, f, \epsilon|\bm{D}, \bm{R}) 
\propto p(\bm{D},\bm{R}|\alpha, f)\,\varpi(\alpha, f, \epsilon),
\end{equation}
where $p(\bm{D}, \bm{R}|\alpha, f)$ is the likelihood function, i.e., the probability to obtain the whole data $\bm{D}=\{D_j\}_{j=1}^N$ and $\bm{R}=\{R_j\}_{j=1}^N$ for given $\alpha$ and $f$, and 
$\varpi(\alpha, f, \epsilon)$ is the prior PDF for $\alpha$, $f$, and the parameter $\epsilon$ that encodes the expected smoothness of $p_{\rm occ}$ (see below).

The likelihood is separable as
\begin{equation}
    p(\bm{D},\bm{R}|\alpha, f) = \prod_{j=1}^N p(D_j, R_j|\alpha, f),
\end{equation}
and we need to compute 
\begin{align}
\notag
    &p(D_j, R_j|\alpha, f) 
    = \int p(D_j, R_j, x_j|\alpha, f) \,\mathrm{d}x_j\\
 \label{eq:jlikelihood}
    &= \int p(D_j, R_j|x_j, \alpha, f)\,p(x_j|\alpha, f)\,\mathrm{d}x_j
\end{align}
for each star.
In $p(x_j|\alpha, f)$ above, the PDF for $x_j$ does not depend on $f$ when conditioned on $\alpha$, and so this is equivalent to $p_{\rm occ}(x_j|\alpha)$ defined in Equation~\ref{eq:pocc}. Because the data $D_j$ and $R_j$ are independent, we also have
\begin{align}
\notag
    &p(D_j, R_j|x_j, \alpha, f) = p(D_j|x_j, \alpha, f)\,p(R_j|x_j, \alpha, f)\\
    &= p(D_j|x_j)\,p(R_j|x_j,f),
\end{align}
where by definition of $R$
\begin{equation}
\label{eq:Rlikelihood}
    p(R_j|x_j, f) = \sum_{k=1}^M f_k^{R_j}\,(1-f_k)^{1-R_j} \Pi_k(x_j).
\end{equation}
The above $p(R_j|x_j,f)$ reduces to $f_k$ for $R_j=1$ and $1-f_k$ for $R_j=0$, where $k$ is the index of the bin that $x_j$ falls in.
Then the likelihood for the $j$th star (Equation~\ref{eq:jlikelihood}) is
\begin{align}
\notag
    &p(D_j, R_j|\alpha, f) = \sum_{k=1}^M f_k^{R_j}(1-f_k)^{1-{R_j}}\exp(\alpha_k)\,L_{jk},\\
    & \quad L_{jk} \equiv \int p(D_j|x_j)\,\Pi_k(x_j)\,\mathrm{d}x_j
\end{align}
where we have used $\Pi_k(x)\Pi_{k'}(x)=\delta_{kk'}\Pi_k(x)$ with $\delta_{kk'}$ being the Kronecker delta.
We follow \citet{2010ApJ...725.2166H} to evaluate $L_{jk}$ using the samples from the posterior $\pi(x_j|D_j)$ conditioned on a certain uninformative prior $\pi(x)$, which we already obtained in Section \ref{sec:cks} for individual stars. Since
\begin{equation}
    p(D_j|x_j) = {{\pi (x_j|D_j) \pi(D_j)} \over \pi(x_j)},
\end{equation}
$L_{jk}$ may be evaluated as
\begin{align}
\notag
    L_{jk} &= \pi(D_j) \int {\Pi_k(x_j) \over \pi(x_j)}\,\pi(x_j|D_j)\,\mathrm{d}x_j\\
\label{eq:ljk}
    &\approx \pi(D_j){1\over K}\sum_{n=1}^K {\Pi_k(x_j^{(n)}) \over \pi(x_j^{(n)})},
    \quad x_j^{(n)} \sim \pi(x_j|D_j)
\end{align}
with $K$ being the number of posterior samples used.
Here the constant $\pi(D_j)$ is irrelevant to the inference. In Section~\ref{sec:cks} we adopted the prior $\pi(x)=\mathrm{const}.$, so $L_{jk}$ is proportional to the number of posterior samples $x_j$ that falls in $k$th bin. 
Although here we use the posterior samples drawn from $\pi(x_j|D_j)$, 
the integral in Equation~\ref{eq:ljk} does not explicitly depend on the choice of the prior $\pi(x_j)$, because it only involves $\pi(x|D)/\pi(x)\propto p(D|x)$ (i.e., likelihood function) that is a function of the data alone.

We choose the prior $\varpi(\alpha, f, \epsilon)=\varpi(\alpha, f|\epsilon)\,\varpi(\epsilon)$ as follows:
\begin{align}
\notag
    \varpi(\alpha, f|\epsilon) 
    \propto &\ \delta\left(\sum_{k=1}^M \exp(\alpha_k)\Delta_k - 1\right) \\
\notag
    &\ \times {1\over \epsilon^{(N-1)/2}}\exp\left[-{\epsilon \over 2}\sum_{k,k' \mathrm{neighbors}}(\alpha_k-\alpha_{k'})^2 \right]\\
    \label{eq:hprior}
    &\ \times \prod_{k=1}^M\left[\,\mathcal{U}(f_k; 0, 1) \cdot \mathcal{U}(\alpha_k; -10, \alpha_\mathrm{max})\,\right],
\end{align}
where the sum in the second row is computed for all neighboring bins, and $\mathcal{U}(x; a,b)$ denotes the uniform distribution for $x$ between $a$ and $b$.
Here the first delta function ensures the normalization of $p_\mathrm{occ}(x|\alpha)$, 
and the third term represents uniform priors for $f$ and $\alpha$, where $\alpha_{\rm max}$ is chosen so that the normalized $p_{\rm occ}(x|\alpha)$ is positive.
The second term encodes the expected smoothness of $p_{\rm occ}$ by penalizing solutions with large differences of $\alpha$ in neighboring bins, unless significantly favored by the likelihood $p(\bm{D}, \bm{R}|\alpha, f)$.
The parameter $\epsilon$ controls the balance between the two, and was also inferred from the data by choosing $\varpi(\epsilon)$ to be a log-uniform distribution between $\exp(-5)$ and $\exp(5)$.
Note that here we impose the smoothness for $p_{\rm occ}$ alone, and do not assume how the values of $f_k$ in different bins might be related a priori.
Given the value of $p_{\rm occ}(M_\star, t_\star)$, $f(M_\star, t_\star)$ is inferred purely from the binomial statistics (see Equation~\ref{eq:Rlikelihood}), but the above framework takes into account the uncertainties in $p_{\rm occ}$ that is inferred from the likelihood function for the masses and ages of individual stars.

\subsection{Tests using Simulated Data Sets}\label{ssec:hbayes_tests}

We test the framework in Section \ref{ssec:hbayes_method} using the results of injection-and-recovery tests in Section \ref{sec:isochrone}.
To do so, we pick up 850 stars (close to the CKS sample size) from $5,000$ simulated stars so that they follow several different mass--age distributions as described below, perform the hierarchical inference using the posterior samples obtained by fitting those simulated measurements, and compare the inferred distribution with the ground truths. Here we ignore the data $R_j$ and infer only $\alpha$ and $\epsilon$: we test whether $p_{\rm occ}$ can be inferred correctly, because once this is done the inference of $f$ relies simply on the binomial distribution (Equation~\ref{eq:Rlikelihood}).

In choosing the subsamples, we use only the stars that satisfy
\begin{equation}
\label{eq:agecut}
    t_\star(\mathrm{Gyr}) < -30 (M_\star/M_\odot-1.25) + 5,
\end{equation} 
above which few simulated stars exist due to our EEP cut at 600 (see Table~\ref{tab:simulated}).
We fixed $\alpha_k=-10$ (i.e., essentially zero probability density) for the bins falling in this region (i.e., we incorporate the prior knowledge that the stars do not exist in this region). We also drop stars falling outside the assumed mass (0.7--$1.3\,M_\odot$) and age (0--14$\,\mathrm{Gyr}$) ranges. By this we are implicitly assuming that the target stars are known a priori to fall in a certain mass range. Practically this is not a problem, because the masses can be estimated reasonably well via isochrone fitting, as we saw in Section \ref{sec:isochrone}.

\begin{figure*}
\epsscale{1.15}
\plottwo{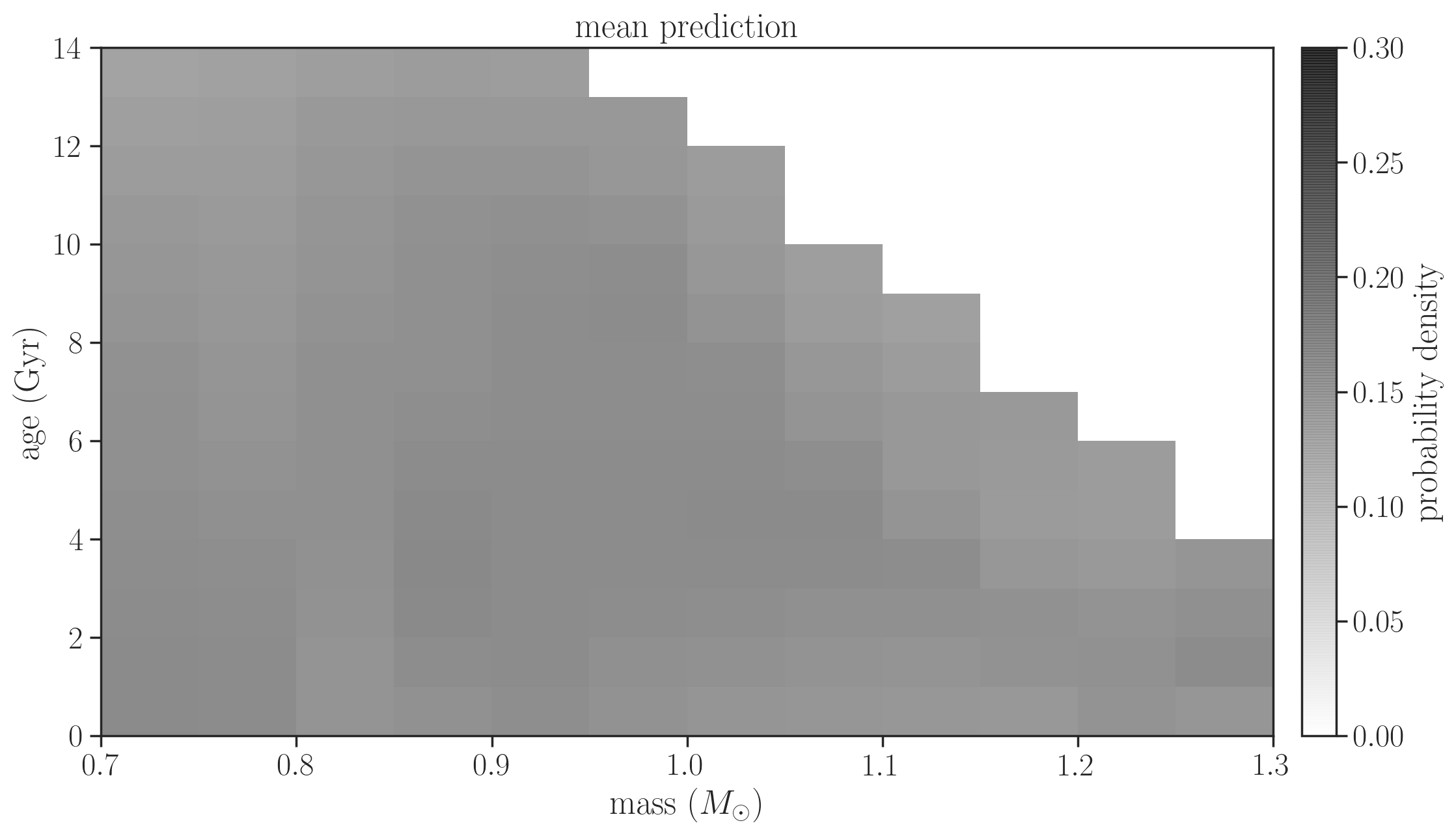}{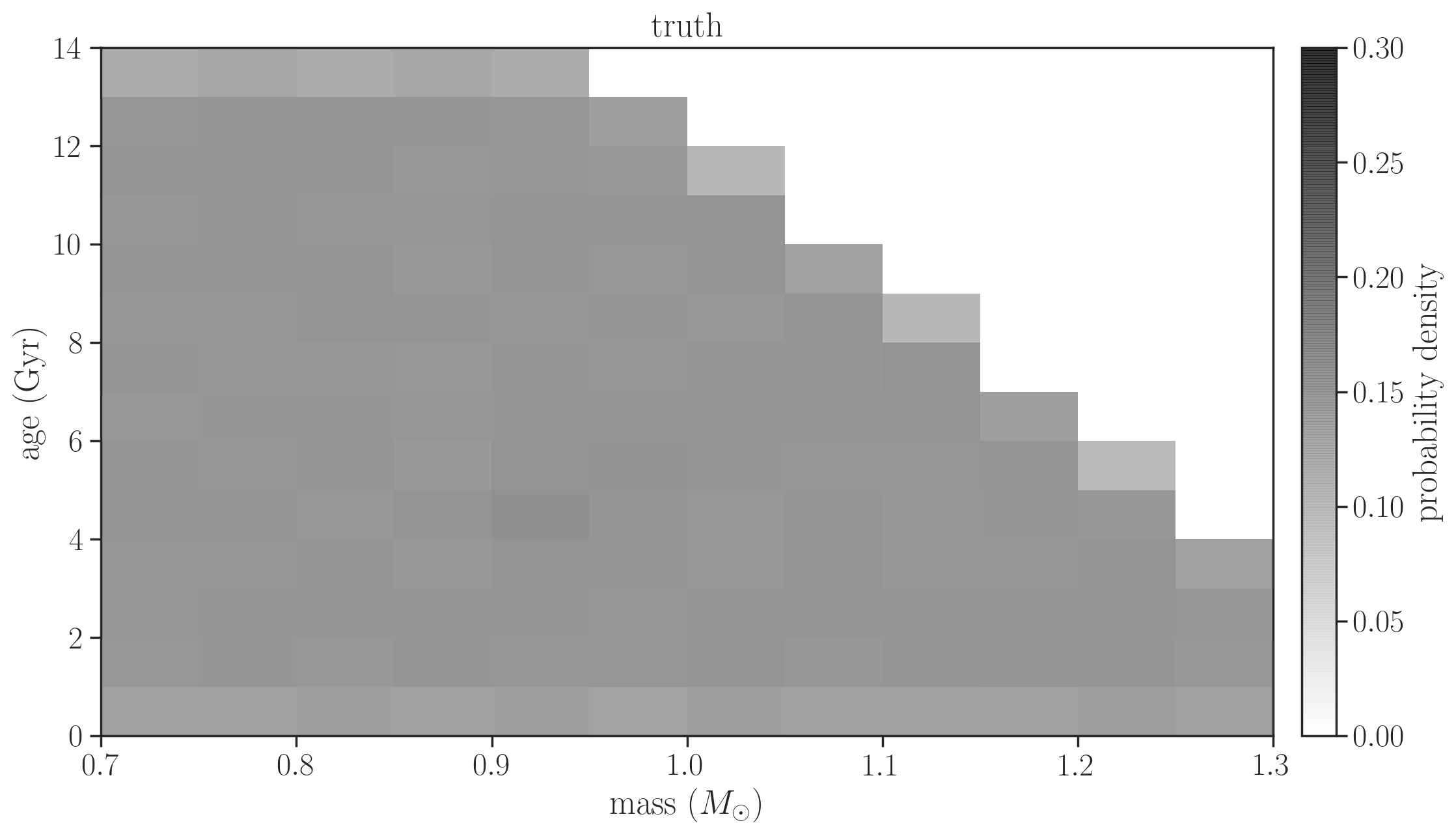}
\plottwo{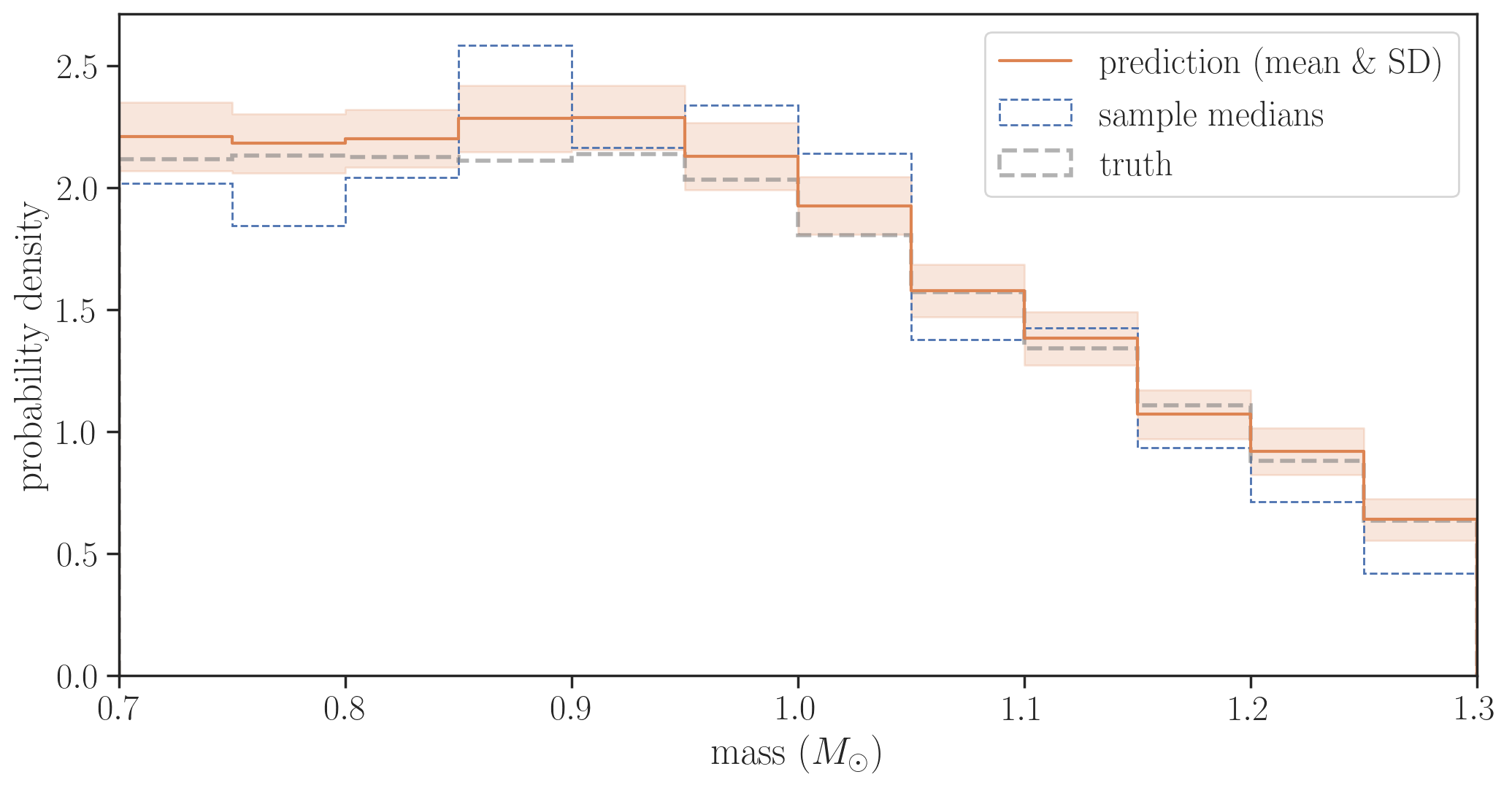}{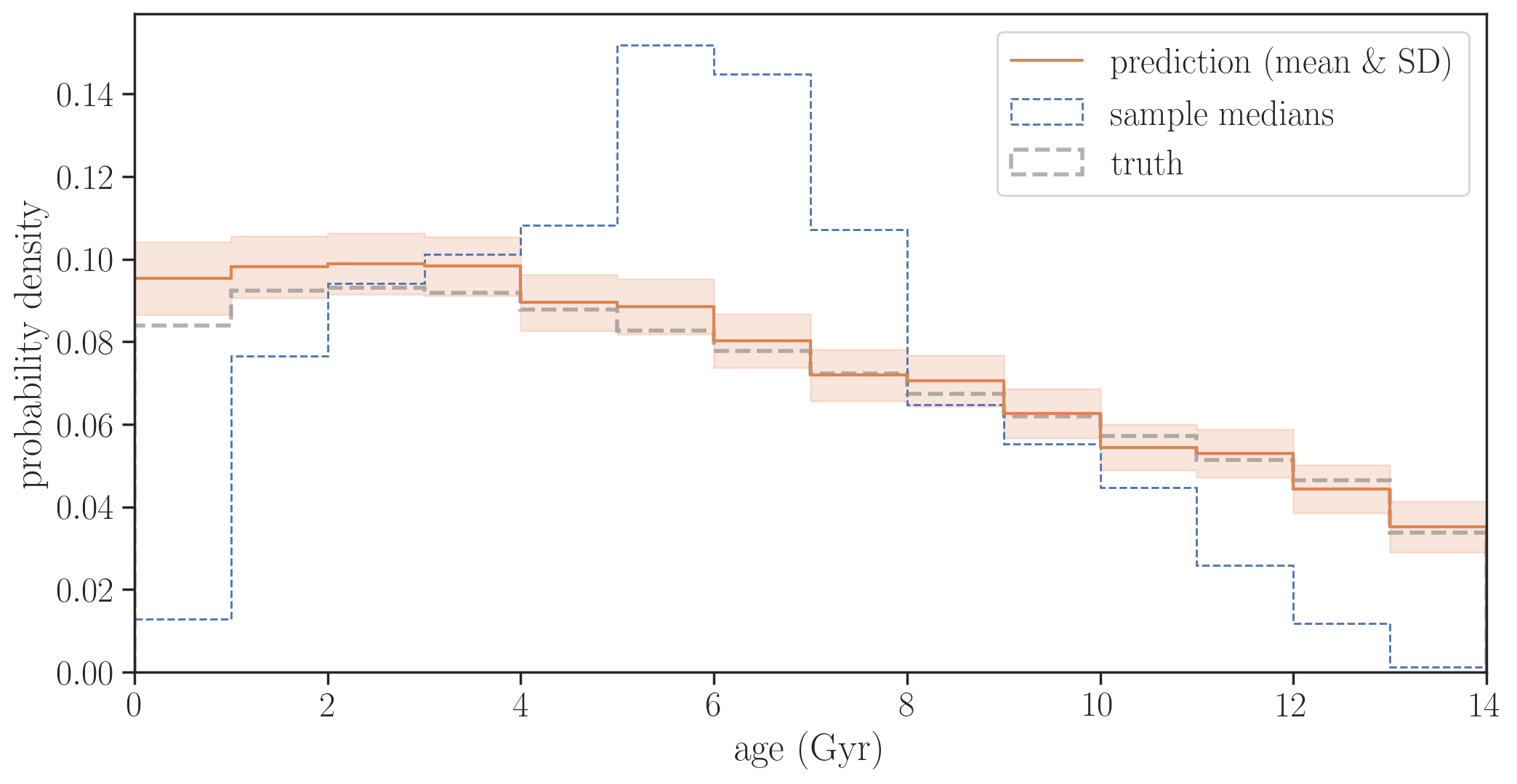}
\caption{Test results for the uniform mass--age distribution in Section \ref{sssec:uniform}.
{\it (Top-left)} The mean of the posterior PDF for $p_{\rm occ}$ as a function of mass and age. {\it (Top-right)} The true distribution that was used to simulate the sample.
{\it (Bottom-left)} Marginal distribution for stellar mass. The orange line and shading show the mean and standard deviation of the prediction. The gray dashed histogram shows the ground truth (i.e., top right panel). The blue dashed histogram shows the distribution of the medians of marginal mass posterior in the sample.
{\it (Bottom-right)} Same as the bottom left, but for age.
}
\label{fig:sim-uni}
\epsscale{1.15}
\plottwo{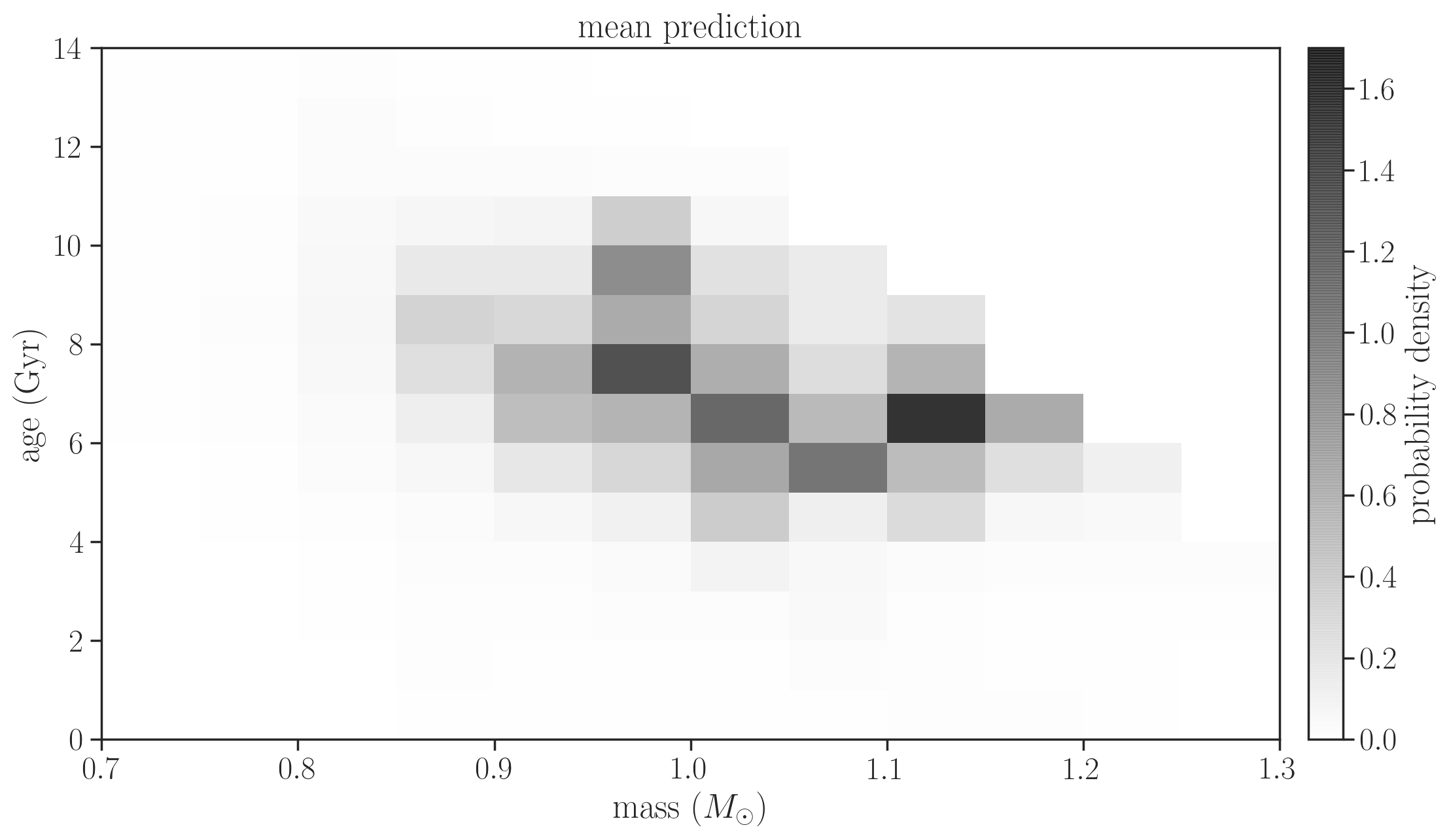}{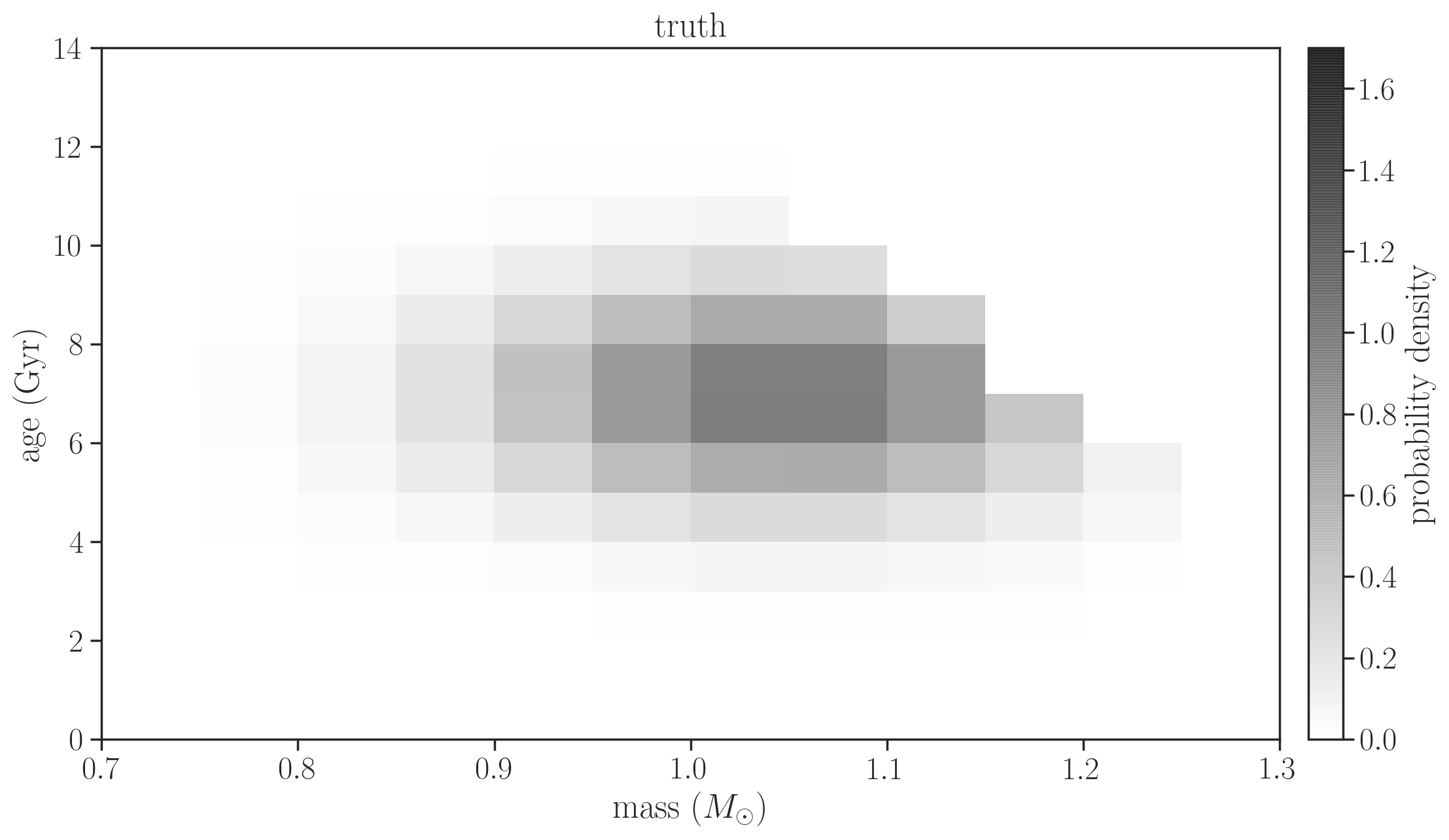}
\plottwo{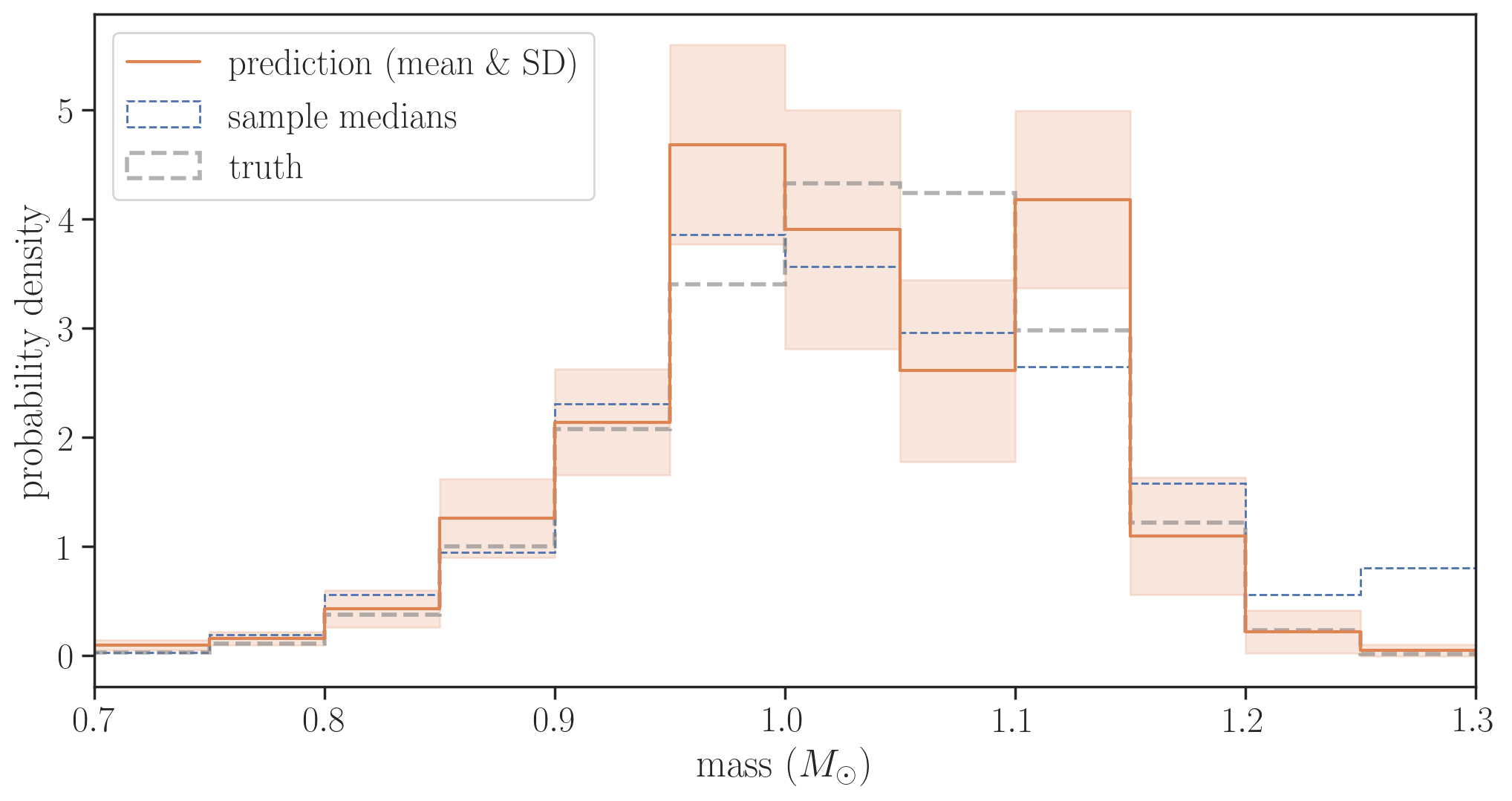}{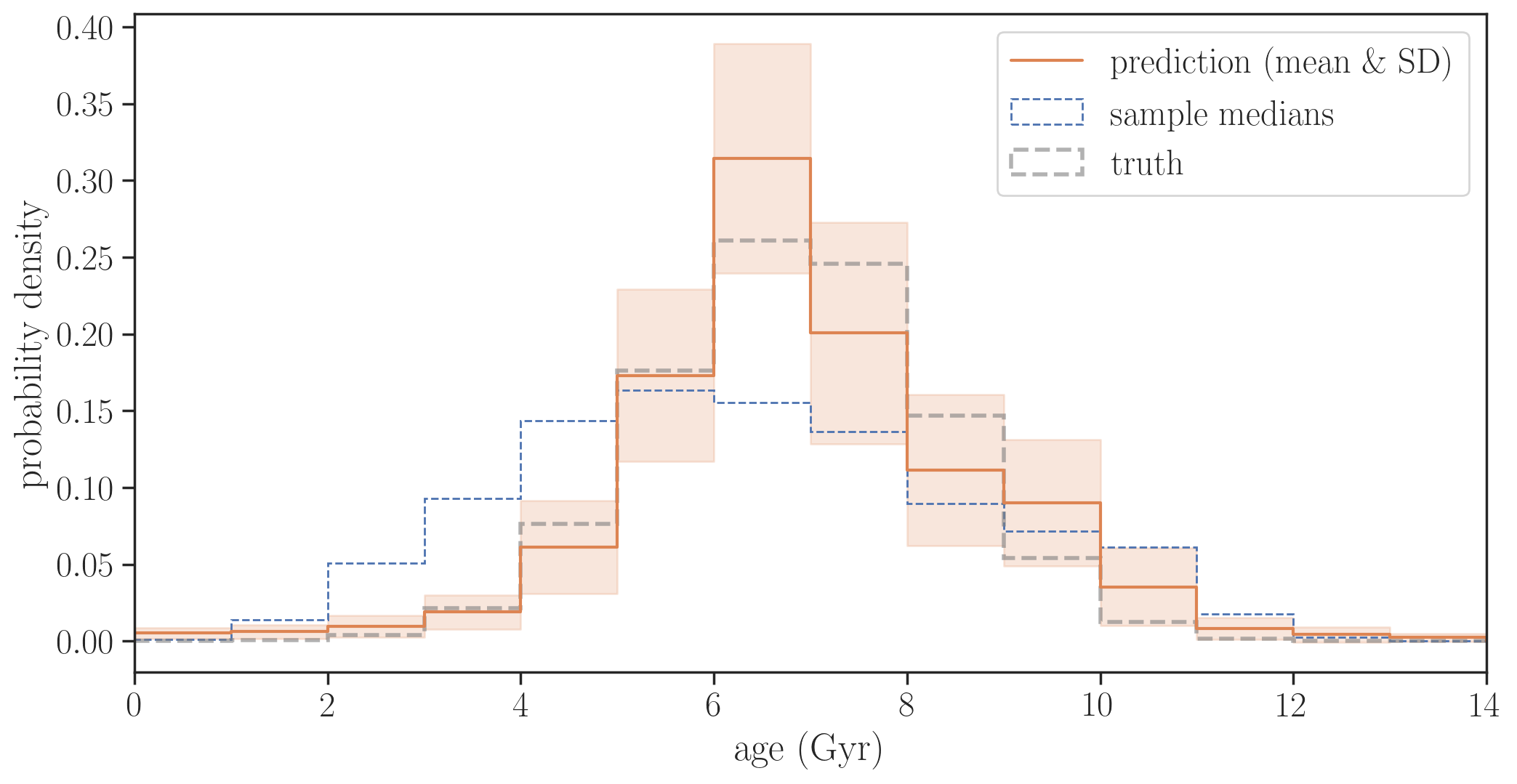}
\caption{Test results for the Gaussian distribution in Section \ref{sssec:gauss}. See the caption of Figure~\ref{fig:sim-uni}.}
\label{fig:sim-gauss}
\end{figure*}

\subsubsection{Nearly Uniform Mass--age Distribution} \label{sssec:uniform}

We first select a subsample drawn from a uniform distribution in the mass--age plane that satisfies Equation~\ref{eq:agecut}. Because of this cut, the marginal distributions for mass and age are not uniform. 

The recovered mass--age distributions (posterior mean of $\exp(\alpha_k)$ in each bin) are compared with the ground truths in Figure \ref{fig:sim-uni}. 
The top left panel shows the mean $p_{\rm pred}$ of the posterior distribution for $p_{\rm occ}$ in each bin, which agrees well with the true probability density $p_{\rm true}$ in the top right panel.
We find $(p_{\rm pred}-p_{\rm true})/\sigma_{\rm pred}=0.3\pm0.3$ as the mean and standard deviation of the values in all the bins, where $\sigma_{\rm pred}$ is the standard deviation of the $p_{\rm occ}$ posterior in each bin.
Also shown are the marginal distributions for mass (bottom left) and age (bottom right). Here the orange line and shading show the mean and standard deviation of the recovered distribution, the gray dashed line shows the ground truth, and the blue thin line shows the histogram of medians of the marginal posteriors (i.e., point estimates as shown in Figure~\ref{fig:age_vs_mass}) for comparison. Both our inference and posterior medians work well for the mass distribution, because this parameter is well-constrained from the isochrone analysis (cf. Section \ref{sec:isochrone}). For ages, on the other hand, the hierarchical inference performs better than simply making a histogram of the medians of the marginal posteriors; the latter are clustered around the middle of the prior range, due to the bias shown in Figure \ref{fig:bias_maps}.

\subsubsection{Nearly Gaussian Distribution}\label{sssec:gauss}

Next we test the case where the masses and ages are clustered around certain values, although we might not expect such a sharp concentration in the real sample. 
We draw stars from independent Gaussians for mass and age: $\mass/M_\odot \sim \mathcal{N}(1.05, 0.1)$ and $\age/\mathrm{Gyr} \sim \mathcal{N}(7, 1.5)$, again dropping stars older than given by Equation \ref{eq:agecut}. 
The results in Figure \ref{fig:sim-gauss} show that the hierarchical inference works reasonably well in this case as well, with $(p_{\rm pred}-p_{\rm true})/\sigma_{\rm pred}=0.1\pm0.7$. 
We note that the simulated mass and age are around the region where the medians of the posteriors work best (Figure \ref{fig:bias_maps}). Nevertheless, the histogram of the median ages (blue dashed) is significantly flatter than the truth (gray dashed) due to wide and skewed marginal posteriors; they are ``blurred" due to large uncertainties. The hierarchical inference (orange solid and shaded region) better reproduces the actual sharper peak, because it correctly takes into account this blurring and provide a ``deconvolved" distribution.

\begin{figure*}
\epsscale{1.15}
\plottwo{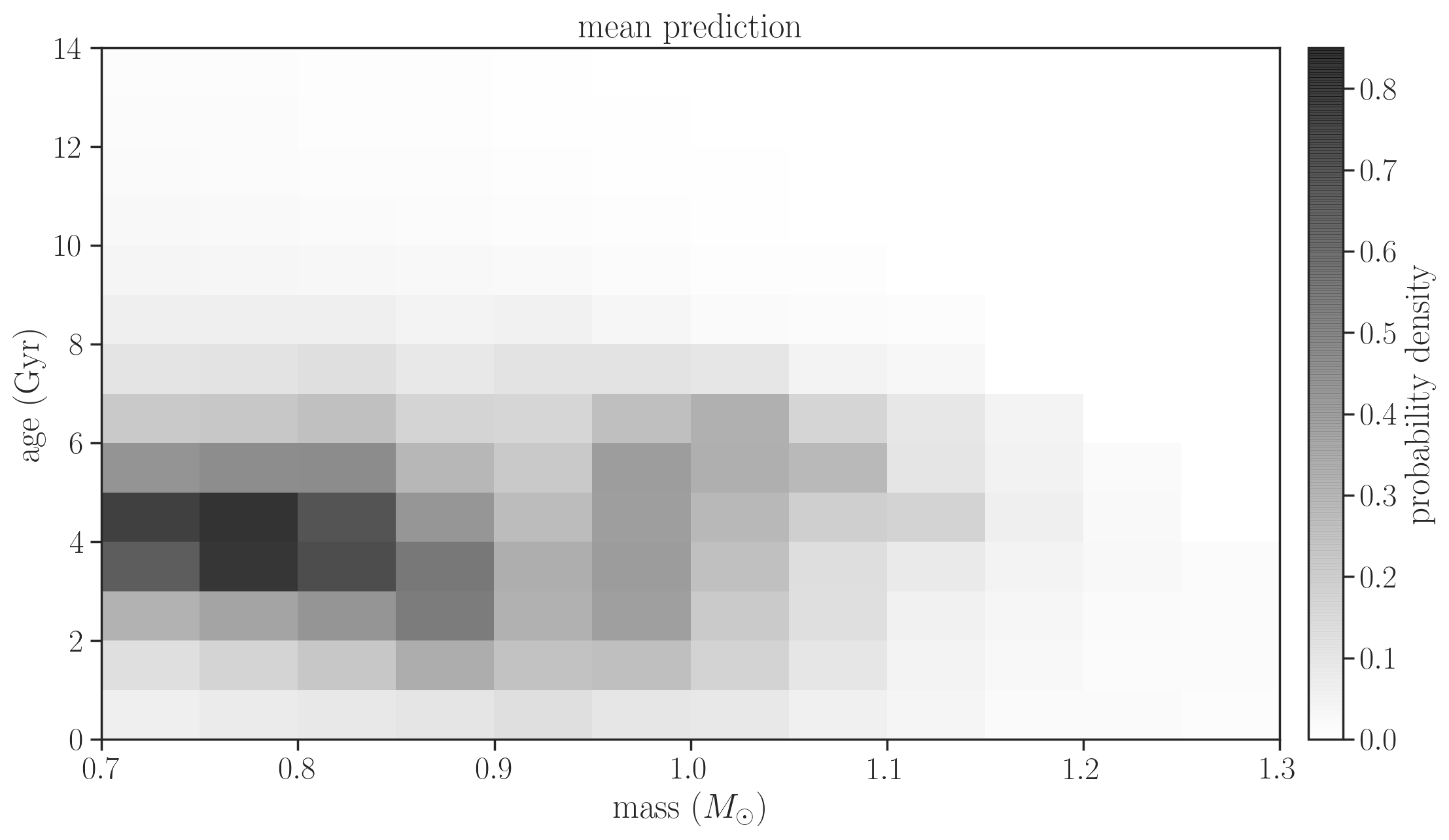}{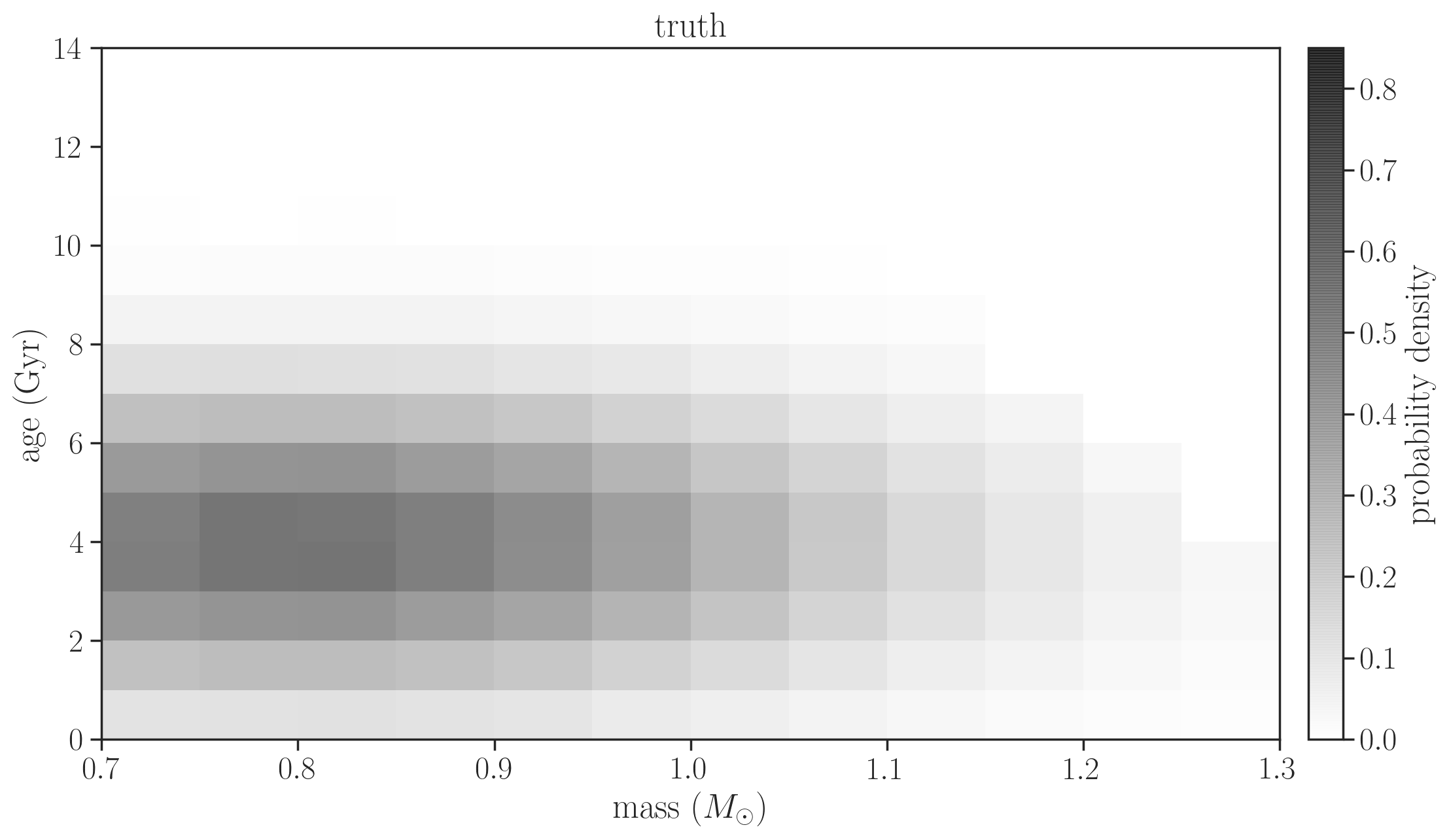}
\plottwo{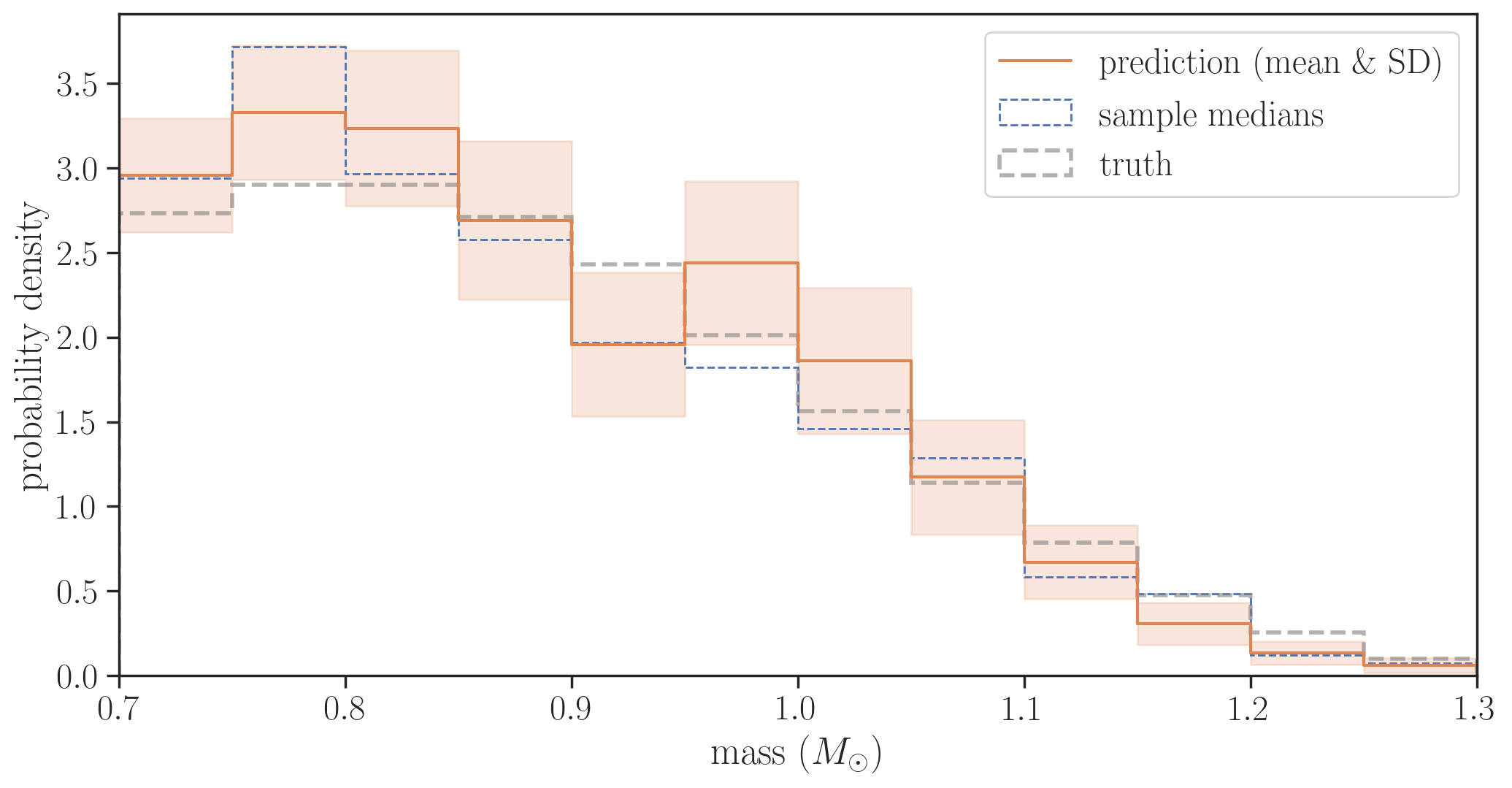}{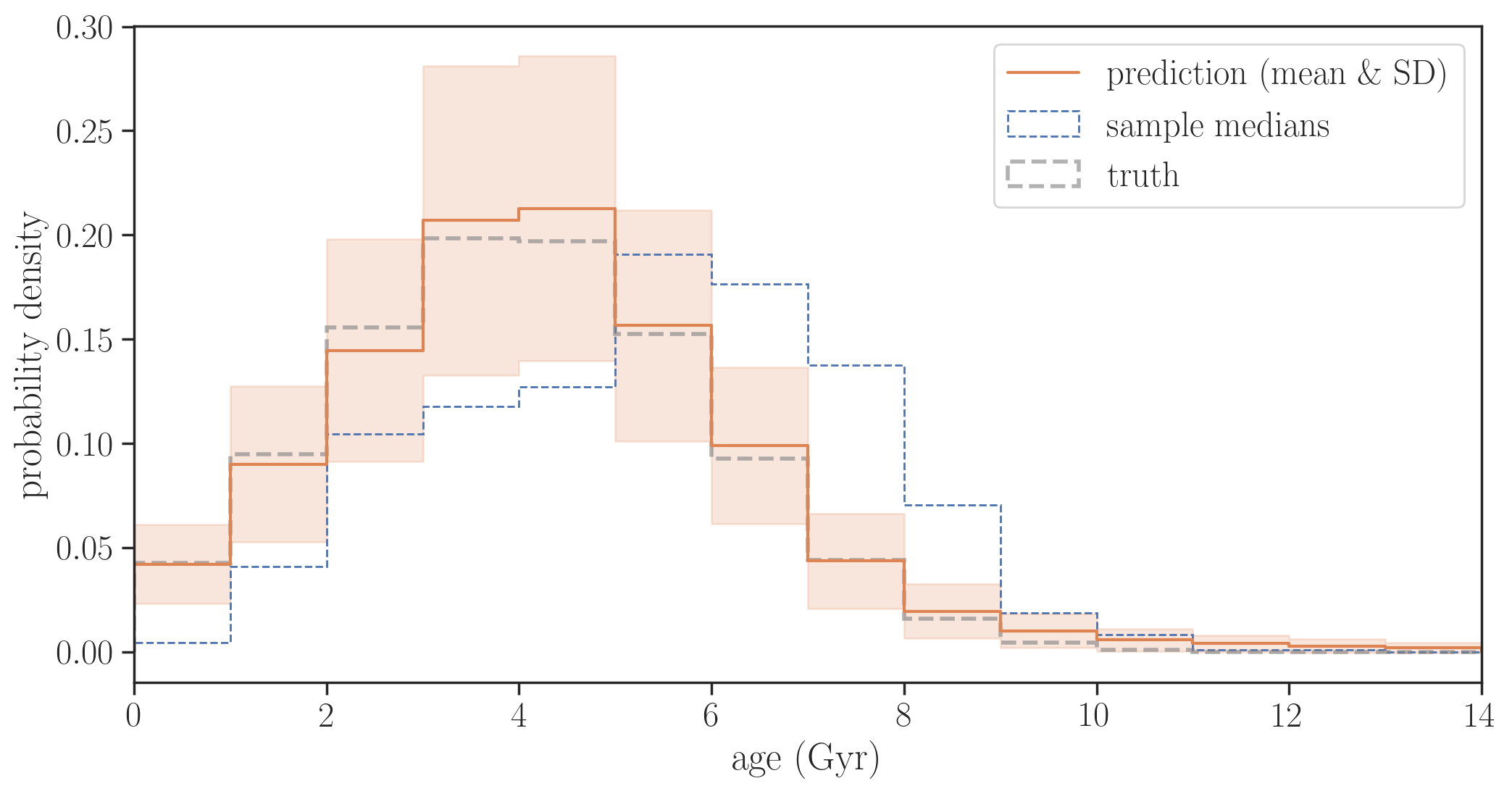}
\caption{Test results for the young low-mass stars (Section \ref{sssec:lowmass}). See the caption of Figure~\ref{fig:sim-uni}.}
\label{fig:sim-lowleft}
\epsscale{1.15}
\plottwo{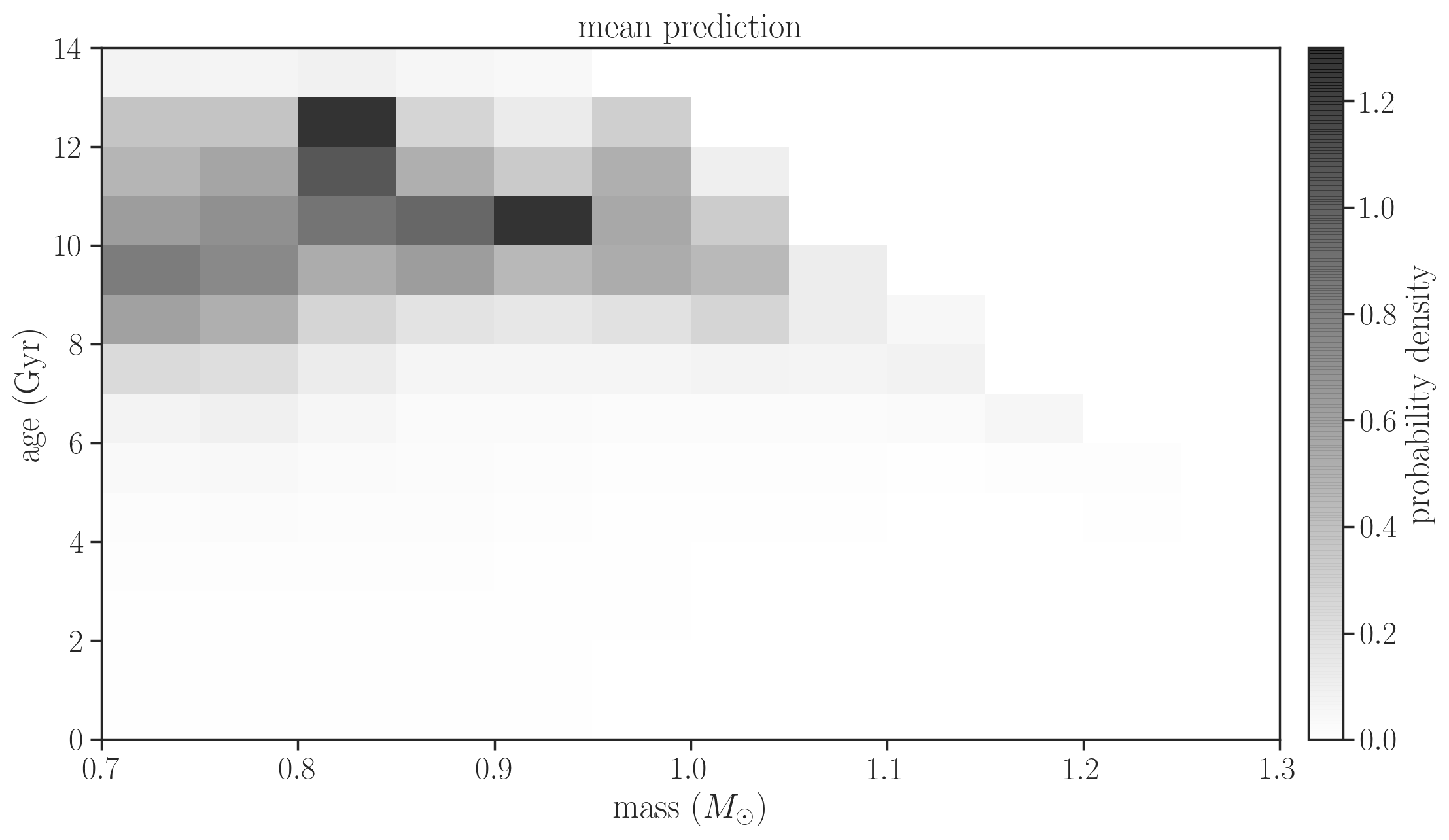}{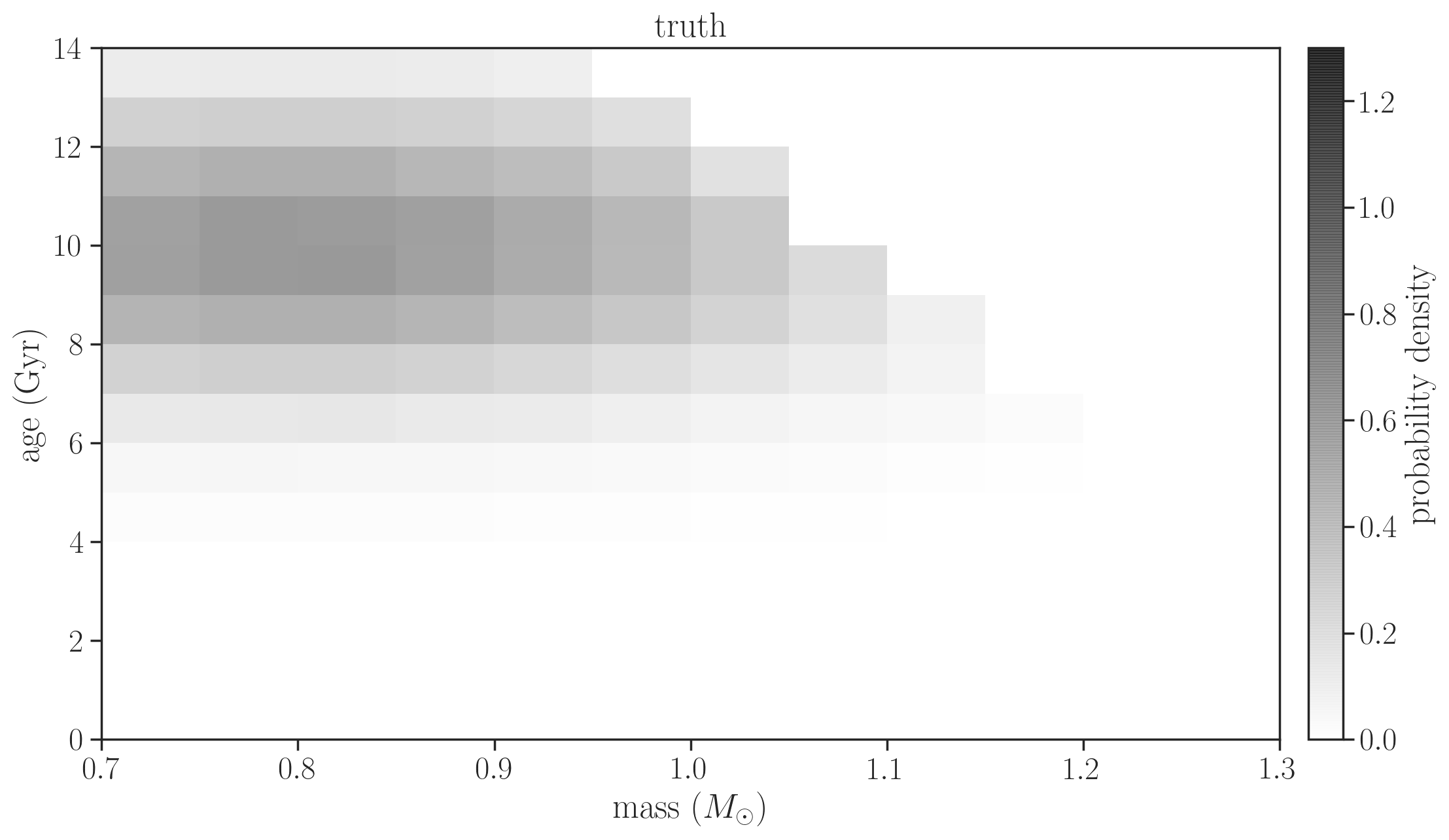}
\plottwo{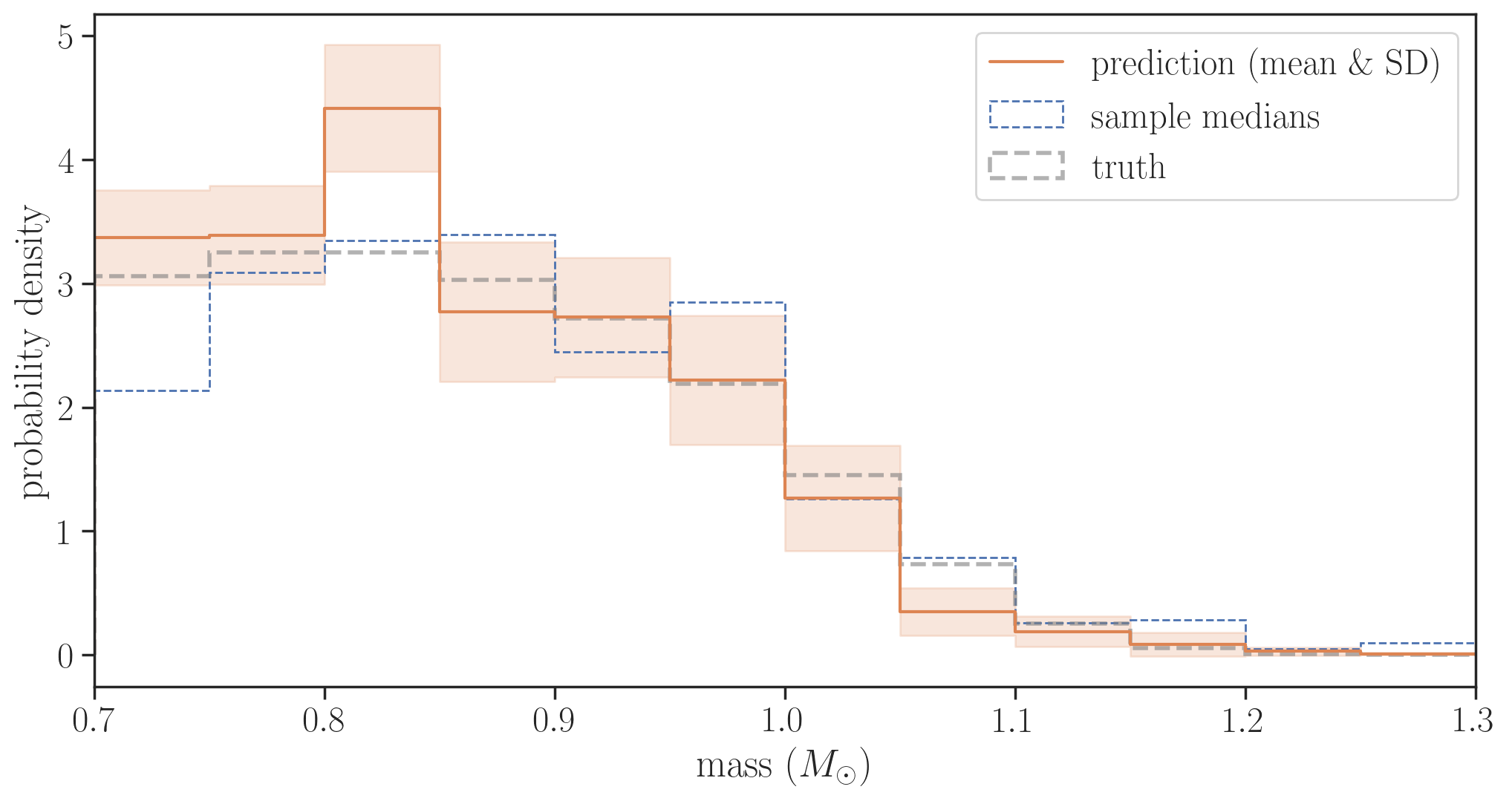}{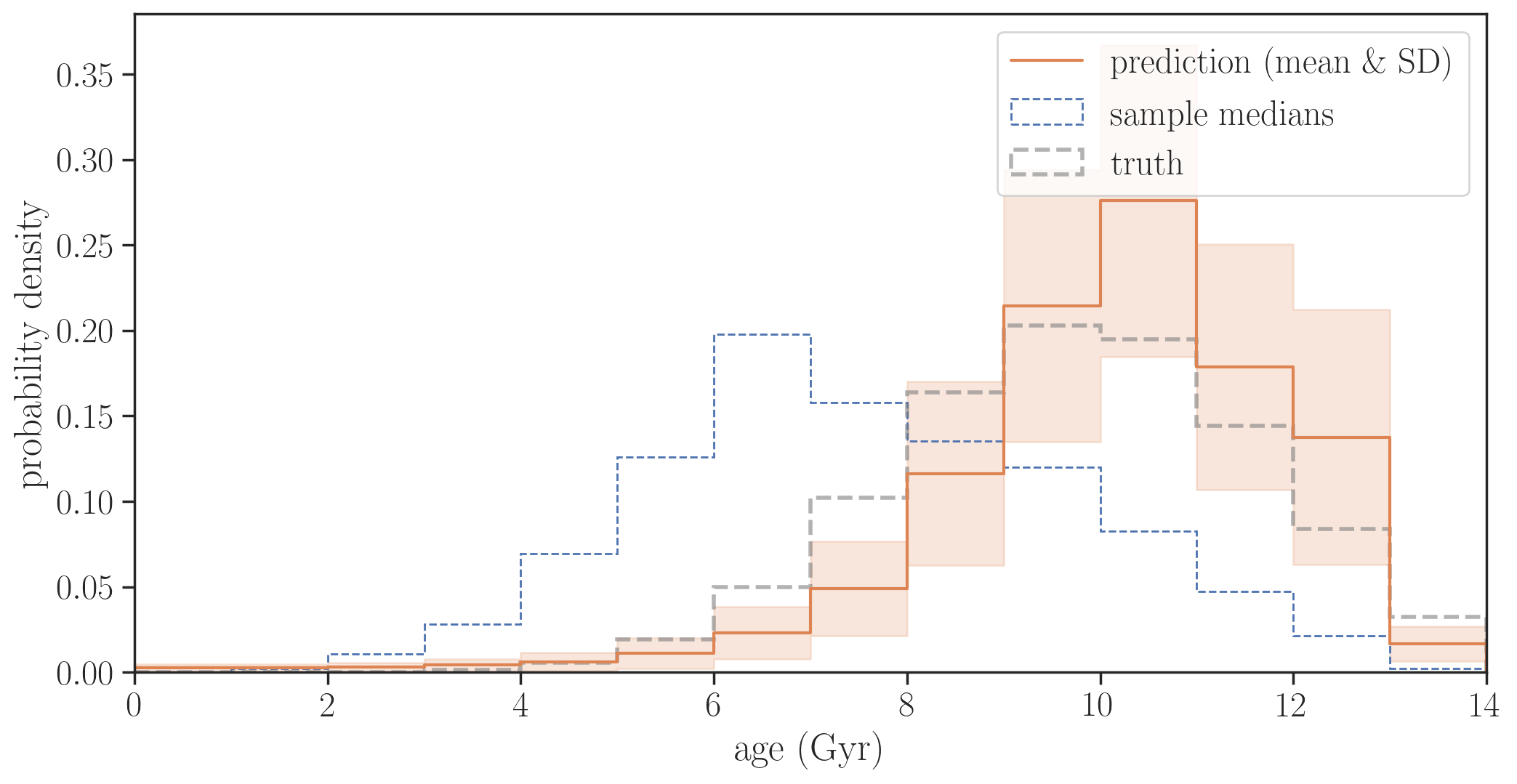}
\caption{Test results for the old low-mass stars (Section \ref{sssec:lowmass}). See the caption of Figure~\ref{fig:sim-uni}.}
\label{fig:sim-uppleft}
\end{figure*}

\subsubsection{Young and Old Low-mass Stars}\label{sssec:lowmass}

As shown in Section \ref{sec:isochrone} and Figure \ref{fig:bias_maps}, age estimation with isochrone fitting is most challenging for youngest and oldest low-mass stars. 
Here we test these cases by drawing masses from $\mass/M_\odot\sim\mathcal{N}(0.8, 0.2)$, and ages from  $\age/\mathrm{Gyr}\sim\mathcal{N}(4, 2)$ or $\age/\mathrm{Gyr}\sim\mathcal{N}(10, 2)$. 
The results are shown in Figures \ref{fig:sim-lowleft} and \ref{fig:sim-uppleft}, respectively.
We find $(p_{\rm pred}-p_{\rm true})/\sigma_{\rm pred}=0.0\pm0.5$ for the younger stars (Figure \ref{fig:sim-lowleft}), and $(p_{\rm pred}-p_{\rm true})/\sigma_{\rm pred}=-0.1\pm0.8$ for the older stars (Figure \ref{fig:sim-uppleft}).

These cases most clearly illustrate the advantage of the inference based on the entire likelihood functions. The age histograms based on medians of the posteriors (blue dashed lines) in Figures \ref{fig:sim-lowleft} and \ref{fig:sim-uppleft} are both far from the truths and even similar to each other despite the very different underlying age distributions. This is because ages are not well constrained and the posteriors tend to be flat (as is the prior), resulting in their medians closer to the middle of the allowed range (see Section \ref{sec:isochrone} and Figure \ref{fig:bias_maps}).
The hierarchical framework exploits the subtle but meaningful information in the entire likelihood function and correctly detects the difference in the true age distributions.

\subsubsection{Summary and Caveats}\label{ssec:caveats}

We have shown that the framework works reasonably well for smooth distributions as assumed above. The result captures the underlying distribution better than the point estimates, and the width of the posterior distributions provide a useful measure for the uncertainties in our inference in that the deviations from the truths are roughly one standard deviation of the prediction. 

We do not claim, though, that this framework works for any distribution. If the true distribution has sharp discontinuities, for example, it is difficult for the model to capture such features because the prior in Equation~\ref{eq:hprior} assumes that $p_{\rm occ}$ is smooth: the prediction will thus be more blurred compared to the truth. We do not believe this is the case in the real sample analyzed in Section~\ref{sec:results}, but our framework cannot prove this. 
This is a limitation of this work. 

Another important caveat is that we have implicitly assumed that the prior is separable: we have assumed that the mass--age distribution is independent from metallicity.
This is not true in general, and the masses and ages of stars are likely correlated with their metallicities and distances.
We do not believe this assumption significantly affects our conclusion: 
we repeated the same analyses splitting the samples into metal-rich and metal-poor stars and found the same trends, although the constraints in each population became weaker due to smaller sample sizes.
This limitation is not inherent in the formulation; 
one could in principle infer the joint mass--age--metallicity distribution, or more practically infer the mass--age distribution for 
separate subsets of stars with almost the same metallicity if the sample is large enough.

\section{Fraction of the CKS Stars with Detected Rotational Modulation as a Function of Mass and Age}\label{sec:results}

\begin{figure*}
\epsscale{1.15}
\plottwo{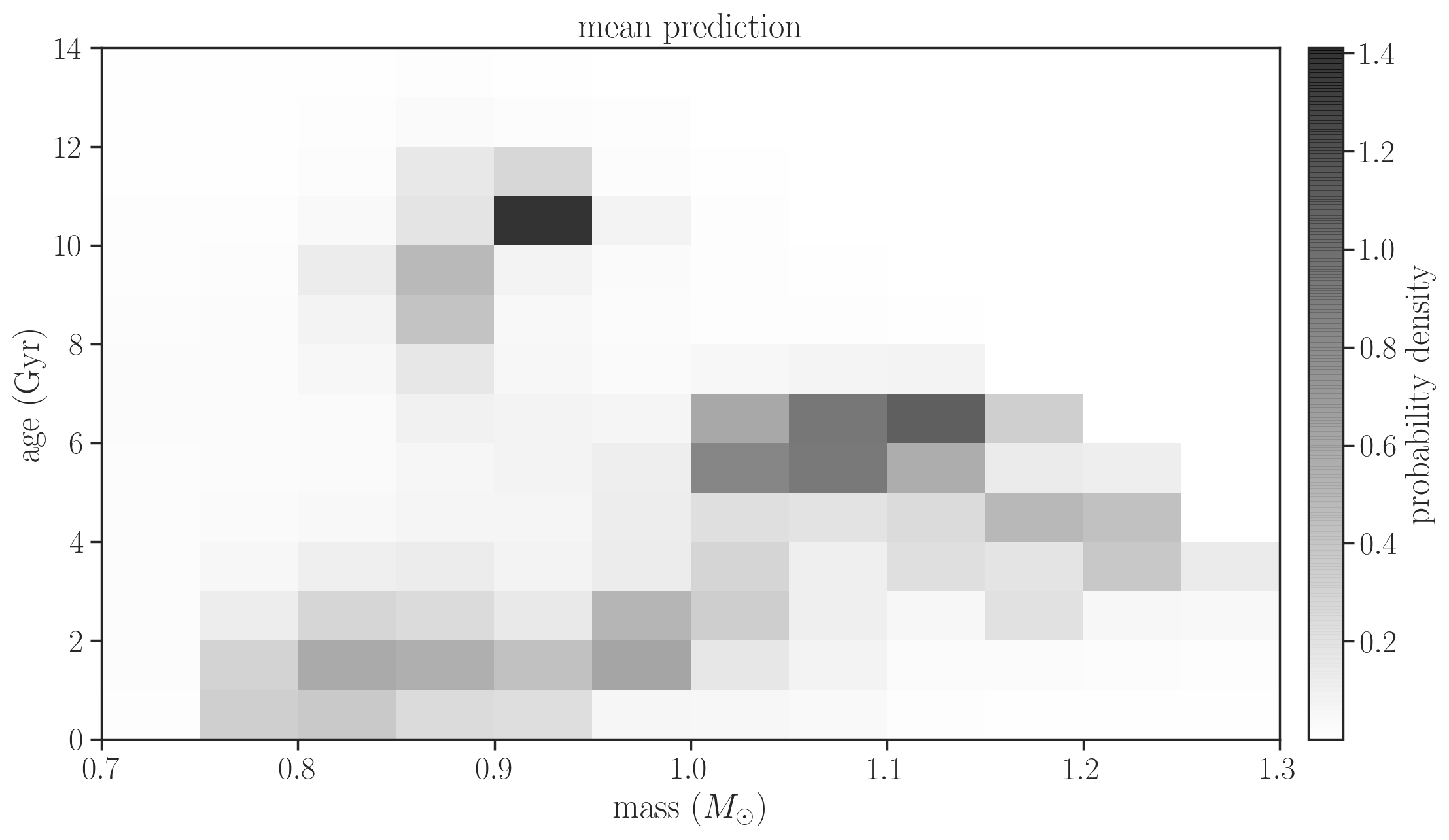}{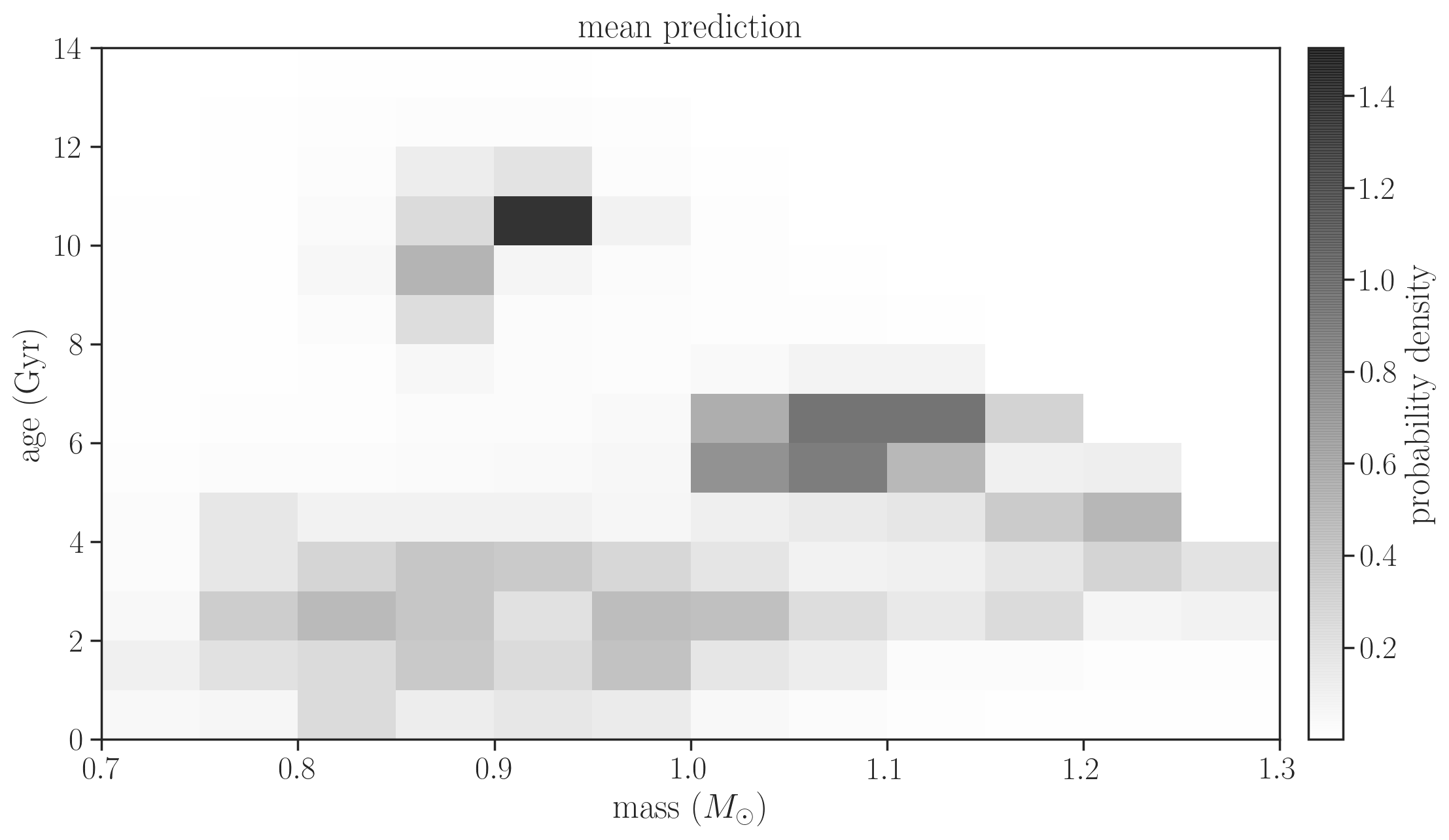}
\plottwo{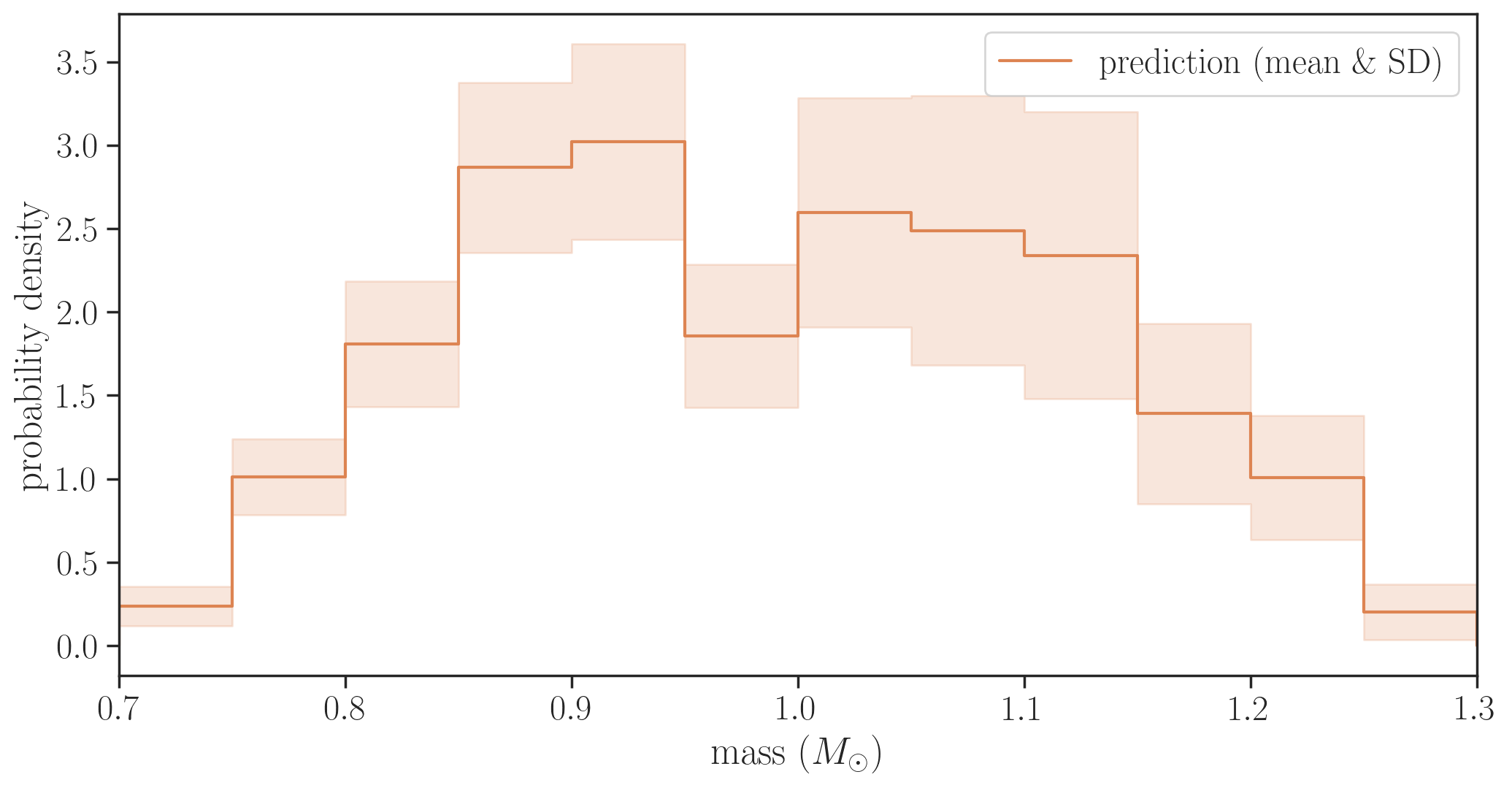}{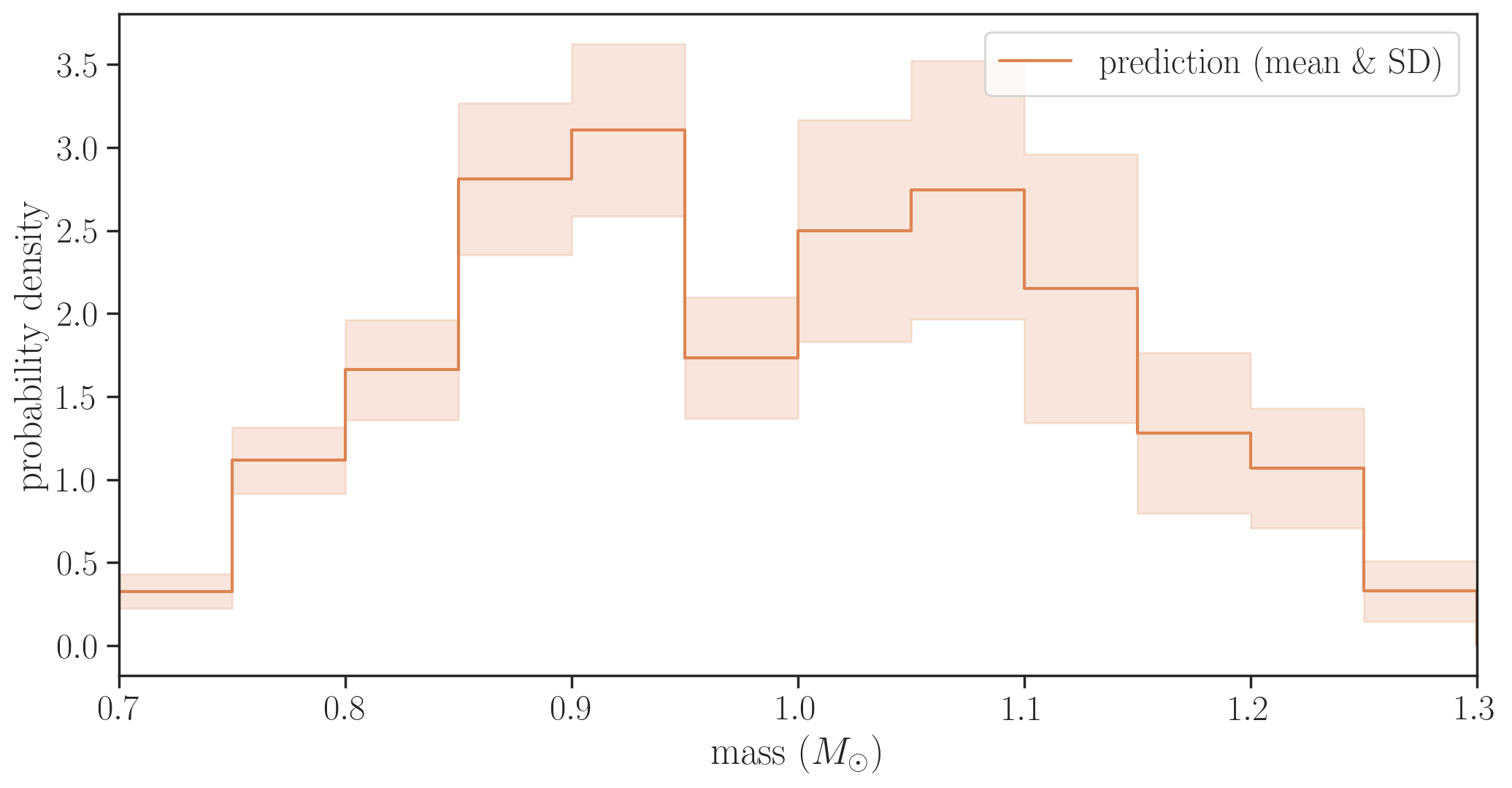}
\plottwo{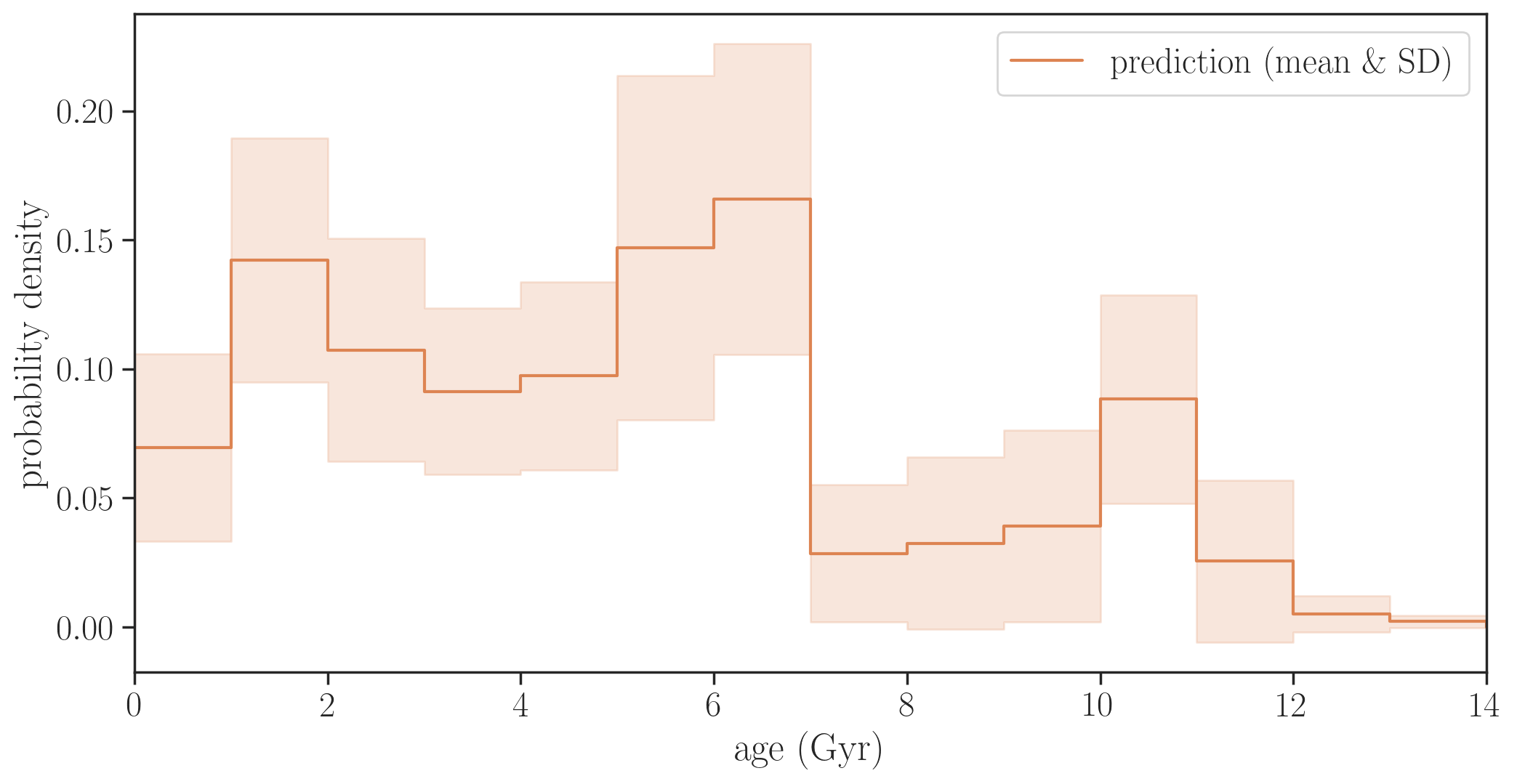}{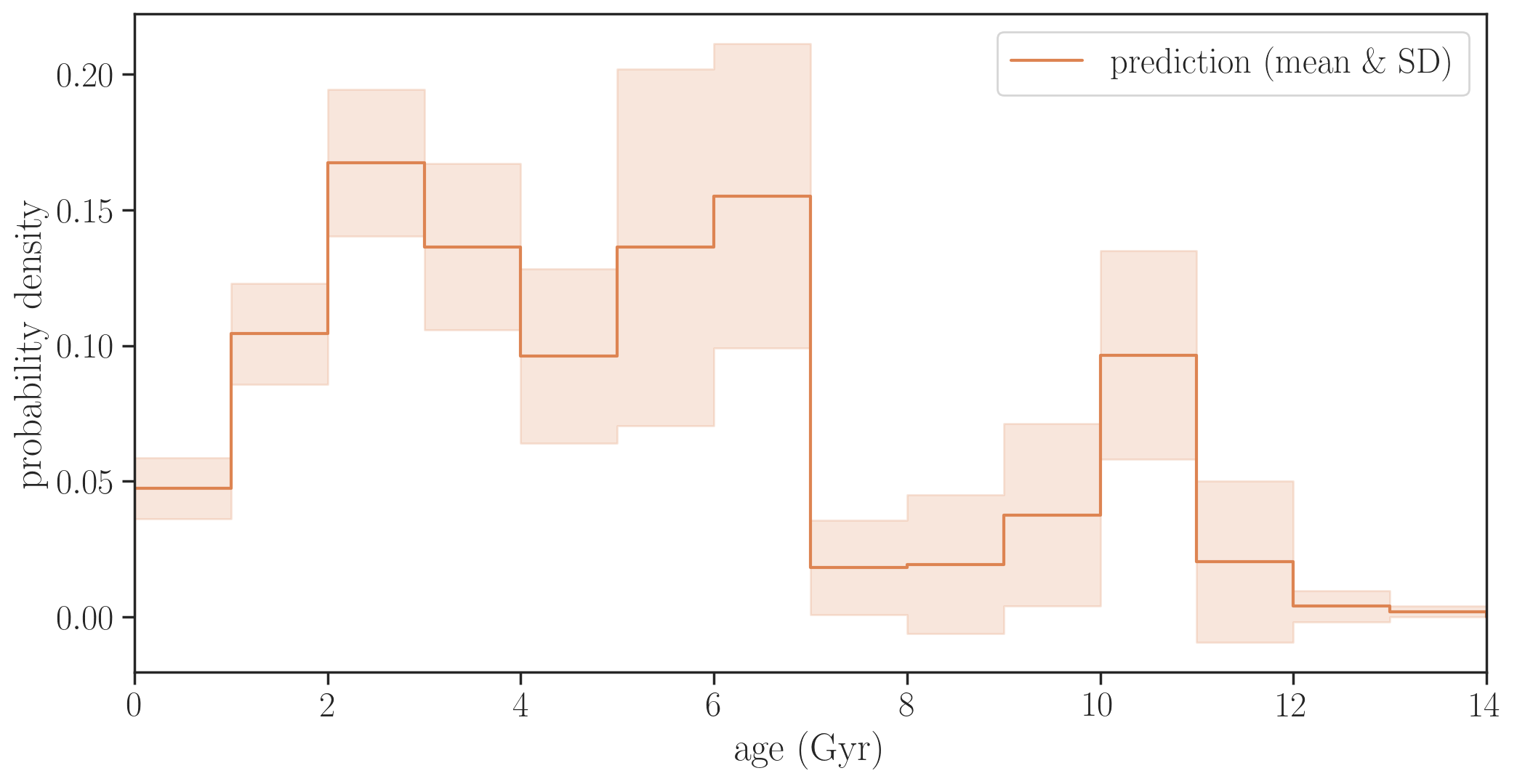}
\caption{Age--mass distribution for isochrone-only sample defined in Section~\ref{ssec:cks_iso} {\it (left column)} and joint isochrone \& gyrochrone sample defined in Section~\ref{ssec:cks_joint} {\it (right column)}. From top to bottom, the two-dimensional age--mass distribution (mean of the posterior), marginalized mass distribution (mean and standard deviation), and marginalized age distribution (mean and standard deviation) are shown.
}
\label{fig:cks_pocc}
\end{figure*}

\begin{figure*}
\epsscale{1.15}
\plotone{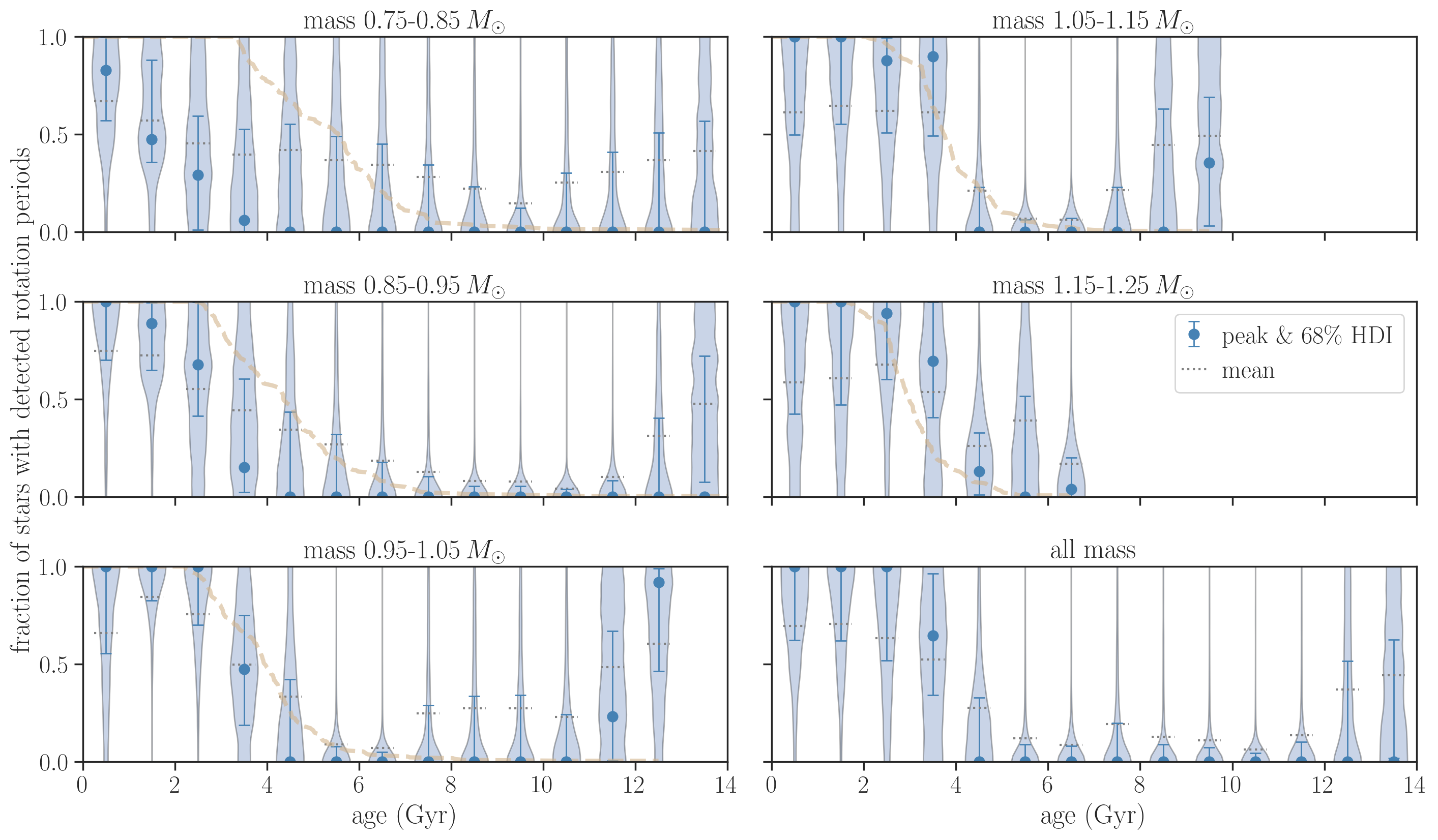}
\caption{
Fraction $f$ of stars with detected rotational modulation as a function of mass and age, for the isochrone-only sample in Section~\ref{ssec:cks_iso}. The filled vertical shaded region (violins) shows the posterior PDF for $f$ in each age--mass bin, where the results from two mass bins are combined here. The blue circles and vertical error bars show the peaks and 68\% highest density intervals of these PDFs, and the short horizontal gray dashed lines show their means. The tan dasehd line shows the $f$--$\age$ relation predicted by the simple detection model; see text in Section~\ref{sec:results}.
}
\label{fig:cks_frot_iso}
\epsscale{1.15}
\plotone{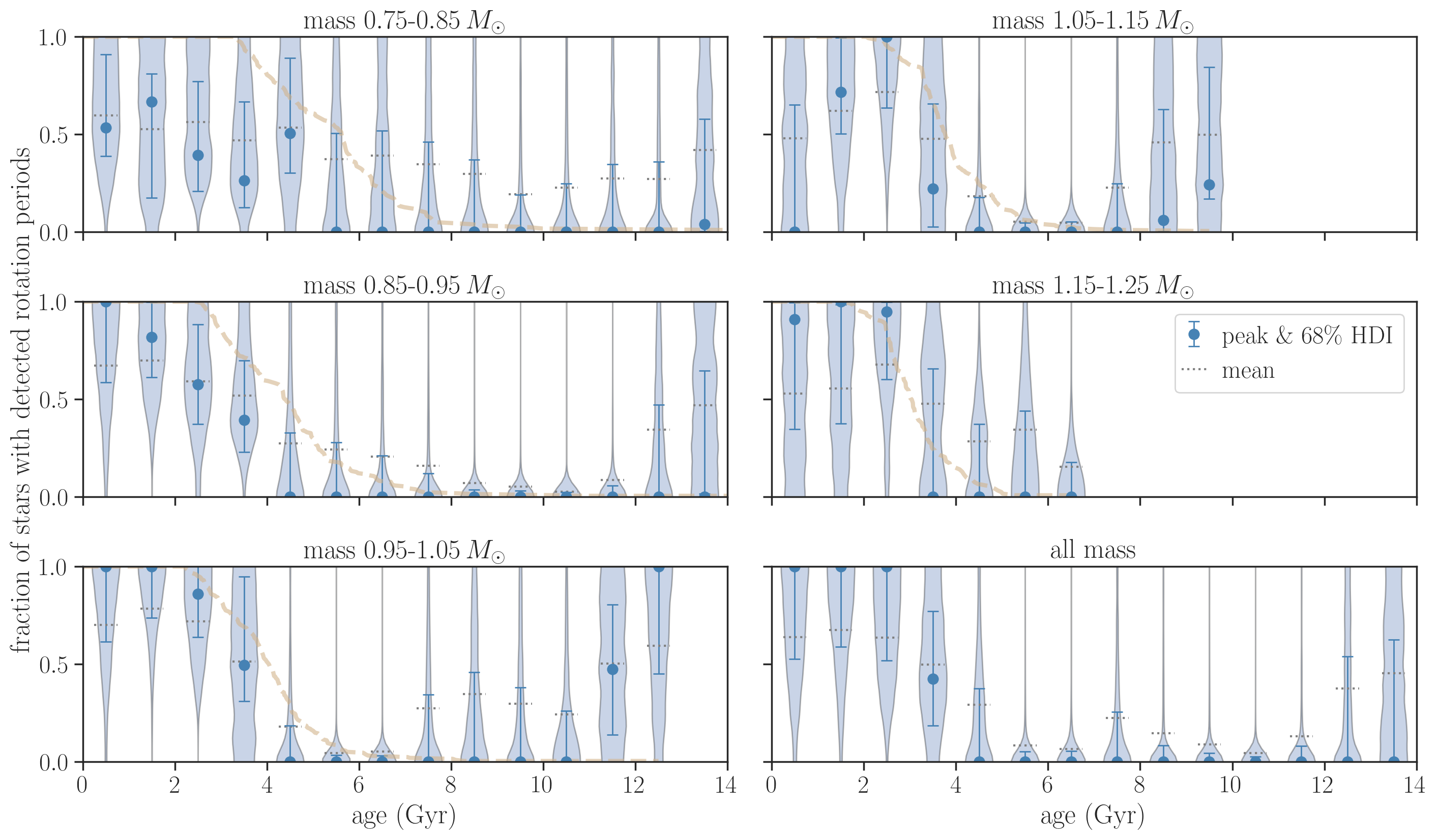}
\caption{Same as Figure~\ref{fig:cks_frot_iso}, but for the joint isochrone \& gyrochrone sample defined in Section~\ref{ssec:cks_joint}.
}
\label{fig:cks_frot_joint}
\end{figure*}

We apply the method described in Section \ref{sec:hbayes} to two sets of posterior samples for the CKS stars derived in Section \ref{sec:cks}, with and without gyrochronology information. Here we use the same mass/age bins and age cut as adopted in Section \ref{sec:hbayes}, but now we use the results of rotational modulation search $R$ and infer the fraction of stars with detected modulation $f$ simultaneously with $\alpha$. 

Figure \ref{fig:cks_pocc} shows the age--mass distribution inferred 
from the isochrone-only posterior samples (left column) and from the samples incorporating gyrochrone information when available (right column).
In Figure \ref{fig:age_vs_mass}, we saw that gyrochronology generally provides more precise constraints for ages of young stars, thereby making the distribution of median ages very different in the lower-left part of the diagram. Nevertheless, Figure \ref{fig:cks_pocc} shows that the samples with and without gyrochrone information imply similar age--mass distributions (top panel), including their marginalized distributions (middle and bottom panels).
This agreement suggests that our framework is working properly:
although the isohcrone-only fit provides weaker constraints for individual stars than the joint fit, the former results contain similar information about the population-level distribution that has been successfully retrieved by our hierarchical inference.

Figures~\ref{fig:cks_frot_iso} and \ref{fig:cks_frot_joint} show the fraction of stars with robust detection of rotation periods ($f_k$) that was inferred simultaneously with the age--mass distribution in Figure~\ref{fig:cks_pocc}, with and without gyrochrone information, respectively. Here we show the posterior probability distribution for $f$ as the vertical filled ``violins" as a function of age for different mass ranges shown in the title of each subplot, combining the information from two $0.05\,M_\odot$ mass bins for clarity. The widths of the violins correspond to the values of the PDFs. The filled circles and vertical error bars show the peak and 68\% highest density interval of each distribution, although we note that the ``peak" is not very well defined from a finite number of samples drawn from the distribution and that these representations are not very informative when $f$ is not well constrained by the data.\footnote{Here we note that the filled circles and error bars cannot be interpreted in the same way as the intervals defined for the Gaussian distribution, since the posterior PDFs for $f$ are far from the Gaussian. For example, $f=1$ is above the circle by two units of the error bar for the $2.5\,\mathrm{Gyr}$ old stars with $0.8\,M_\odot$ in the top left panel of Figure~\ref{fig:cks_frot_iso}, but $f\sim 1$ is in fact almost equally plausible with smaller values as the PDF is almost flat over $[0,1]$. In this case, the value of $f$ is simply not well constrained. This is exactly the reason why we show the entire PDFs in Figures~\ref{fig:cks_frot_iso} and \ref{fig:cks_frot_joint}, rather than their simple summary statistics. The latter alone can be misleading in this case.}
The horizontal short gray dashed lines show the means of the distribution that are better defined.
Again, both results are consistent with each other, and show that $f$ is consistent with unity at younger ages and drops rapidly to zero by $\age \sim 5\,\mathrm{Gyr}$.
The trend is most clearly seen in the nearly solar-mass stars, presumably because the isochrone constrains their ages relatively well (Figure~\ref{fig:bias_maps}), 
and because the sample covers a sufficiently wide range of ages.
The transition from $f\sim 1$ to $f\sim 0$ is not clearly seen for the lowest-mass stars in the sample, although the result does suggest that $f$ decreases with age.
We also see a hint that $f$ might increase again once the star evolves off the main sequence, but the constraint on $f$ is weak and the significance is modest due to a small number of evolved stars with detected rotation periods in the sample (see Figure~\ref{fig:age_vs_mass}). This trend, if real, could be due to the decrease of the Rossby number associated with the thickening of the convective envelopes \citep[cf.][]{2020NatAs...4..658L}. The \citet{2021ApJS..255...17S} catalog that provides more rotation periods for evolved stars may be useful to shed further light on this possibility.

Overall, the results in Figures~\ref{fig:cks_frot_iso} and \ref{fig:cks_frot_joint} support our tentative conclusion in Section~\ref{sec:cks}, as well as the prediction by \masuda, that the rotational modulation is detected if and only if a star is young and the photometric modulation amplitude is large. Unlike in Section~\ref{sec:cks}, though,
here we performed the analysis that fully incorporates the mass and age uncertainties associated with isochrone fitting.
The tan thick dashed lines in Figures~\ref{fig:cks_frot_iso} and \ref{fig:cks_frot_joint} show the expected fraction of stars with rotation periods based on a simple detection model presented in \masuda: 
here we (i) compute expected rotational modulation amplitude as a function of $\prot$ and stellar mass using the $\amp$--$\prot$ relation derived from the brightest \kepler\ stars, (ii) compute the fraction of stars $f$ in the sample for which the modulation would be detectable as a function of $\prot$, assuming that the detection threshold is three times the long-cadence photometric precision of \kepler, 
and (iii) convert the above $f(\prot)$ into $f(\age)$ using the gyrochrone calibrated to the Praesepe cluster and the Sun by \citet{2019AJ....158..173A}.
Although this is a simplified detection model, 
the prediction is consistent with the inferred fraction, thus further supporting this view. We note that the detection model in \masuda\ is based on main-sequence stars and by construction does not take into account the possible enhancement of activity in evolved stars as mentioned in the previous paragraph.

Here the model of $f$ does not take into account the weakened magnetic  braking when converting $\prot$ to $\age$, but 
the prediction is in any case insensitive to the late-time spin evolution for stars with $\gtrsim 1\,M_\odot$ because $f$ is already nearly zero when $\ro \sim \ro_\odot$. This implies that the available photometric data provide little information on how the modulation amplitude of solar-mass stars evolves at $\ro \gtrsim \ro_\odot$.
On the other hand, we find that the weakened magnetic braking may affect the late-time evolution of $f$ for stars with $\lesssim 0.9\,M_\odot$ depending on how the modulation amplitude evolves at $\ro \gtrsim \ro_\odot$, although the presence of the weakened magnetic braking is less well established for those sub-solar mass stars.
Thus the value of $f$ could be useful to study how their rotation periods and modulation amplitudes evolve at older ages.

At least for nearly solar-mass stars, $f$ is nearly unity when the star is young. This implies that, as long as the star is young and active, the detectability of rotational modulation in the \kepler\ data is insensitive to long-/short-term changes in the spot-modulation amplitudes due to activity cycles and/or spot evolution, and to the dependecne of spot modulation amplitude on surface metallicity (see also Section~\ref{ssec:cks_result}).

Recently, \citet{2022arXiv220308920D} reported the presence of a pile-up around the upper edge of the $\prot$--$\teff$ distribution combining the \citet{2014ApJS..211...24M} sample and $\teff$ from spectroscopic surveys, and argued that this pile-up --- as predicted by \citet{2019ApJ...872..128V} --- provides further evidence for the weakened magnetic braking. Our conclusion that the upper edge is due to detection bias may appear to be in conflict with the reported pile-up, but it is not. 
The pile-up they detected is located at shorter $\prot$ than the detection edge, and such a pile-up results as long as the $\prot$ distribution increases toward longer periods across the detection edge, as shown in \masuda: this is what we generally expect from the Skumanich law $\prot \sim \age^{1/2}$ and a roughly flat age distribution, as we have inferred. Thus the pile-up itself is consistent with the simple detection edge, although the detailed shape of the pile-up may also provide information on the spin distribution in this region (see Section 4 of \masuda\ for a more detailed discussion).
\citet{2022arXiv220308920D} also pointed out that stars hotter than the Sun (with $\teff \sim 5800$--$6400\,\mathrm{K}$) around the pile-up may have a somewhat broad age distribution spanning $2$--$6\,\mathrm{Gyr}$.
This might appear to favor the weakened magnetic braking origin for the upper edge, but this is not necessarily the case. Stars with different apparent magnitudes have different detection thresholds, and so a certain amount of age scatter is naturally expected for stars around the detection edge.
Considering that the age uncertainty of at least $\sim 1\,\mathrm{Gyr}$ contributes to this scatter, as they also note, the age range they report is indeed compatible with our finding that $f$ of $\approx 1.1\,M_\odot$ stars (corresponding to the $\teff$ range above) drops from $\sim 1$ to $\sim 0$ at ages spanning $3$--$5\,\mathrm{Gyr}$ (top-right panels in Figures~\ref{fig:cks_frot_iso} and \ref{fig:cks_frot_joint}). 

Below we comment on potential subtleties in interpreting these results and argue that they do not affect our main conclusion.

\subsection{Systematics in Stellar Models}

We adopted the MIST models in our analysis and did not examine the dependence of the results on stellar models. \citet{2022ApJ...927...31T} showed that such model-dependent offsets are typically $\sim 5\%$ in mass and $\sim 20\%$ in age for main-sequence and sub-giant stars. Our conclusions are based on the arguments less precise than these potential systematic effects. 

We also reiterate that 
we find a good agreement between our isochrone ages and those from  asteroseismology for older stars with $\gtrsim 0.9\,M_\odot$ (Section~\ref{ssec:isochrone_seismic}).
For younger stars, we found consistent age distributions with and without gyrochronal constraints, thus statistically validating the isochronal age scale.
A good correlation found between photometric amplitude and age presented in Section~\ref{ssec:cks_result} also supports the accuracy of relative ages.
We note that most of our main conclusions, except for quantitative comparison with the detection model, rely only on relative ages.

\subsection{Systematics in Effective Temperatures}

For young T Tauri stars, there is evidence that $\teff$ from optical spectra could be biased by a few 100~K due to large spots on the surface \citep[e.g.,][]{2022ApJ...925...21F}. 
The youngest stars in our sample are a few 100$\,\mathrm{Myr}$ old, at which ages typical photometric modulation amplitudes are smaller roughly by an order of magnitude \citep[e.g.,][]{2020ApJ...893...67M}.
Therefore we argue that the systematics in $\teff$ is likely smaller than our assigned error of $\sim100\,\mathrm{K}$ even for the youngest stars in the sample.

\section{Summary and Conclusion}\label{sec:discussion}

We performed a probabilistic inference for the masses and ages of FKG stars in the CKS sample
by fitting stellar models to their atmospheric parameters from high-resolution spectra, {\it Gaia} EDR3 parallax, and $K_s$ magnitudes from 2MASS (Section~\ref{sec:cks}). 
We presented internal and external tests to validate the procedure, and discussed caveats in interpreting such probabilistic constraints as obtained in a Bayesian manner (Section~\ref{sec:isochrone}).
We then presented a framework to infer the occurrence rate of a certain property of stars as a function of their masses and ages, 
leveraging imprecise but statistically well-defined constraints on those parameters as are typically available from isochrone fitting (Section~\ref{sec:hbayes}).
We applied the framework to derive the fraction $f$ of stars exhibiting detectable rotational modulation in the \kepler\ data, focusing on the subset of the CKS stars for which rotational modulation has been searched homogeneously by \citet{2015ApJ...801....3M}.
For nearly solar-mass stars, we found that $f$ is near unity at $\age \lesssim 3\,\mathrm{Gyr}$ and drops rapidly to zero by $\age \sim 5\,\mathrm{Gyr}$, although the trend is less clear for lower-mass stars (Section~\ref{sec:results}).
We also showed that photometric modulation amplitudes of the sample stars older than $\sim 2\,\mathrm{Gyr}$ decrease monotonically with age, and that the age cuts found above correspond to the amplitude cuts (Section~\ref{ssec:cks_result}). 

These findings are consistent with a view that the detection is simply limited by photometric precision to younger stars that exhibit rotational modulation with larger amplitudes, as proposed by \masuda.
This argues against the hypothesis that the longest detected rotation periods are 
determined by the weakened magnetic braking, in which case the rotational modulation should have been detected for stars with a wide range of ages at a given mass.
Rather, the photometric sample provides limited information on the rotational evolution of solar-mass stars in the latter halves of their lives, for which weakened magnetic braking has been considered to be important.
Although our analysis 
is conditioned on the periodicity search by
\citet{2015ApJ...801....3M} focusing on $\approx 3,000$ \kepler\ stars with transiting planet candidates, we find evidence that the detection function is similar to that in the sample of \citet{2014ApJS..211...24M} who searched rotation modulation with the same method but for generic ($\gtrsim 100,000$) \kepler\ stars (see Section~\ref{ssec:mazeh-mcquillan} below). 
Our analyses also suggest that the detectability of rotational modulation is insensitive to the parameters other than the age, such as metallicity and activity cycles, at least in the \kepler\ photometric data for stars younger than the Sun. 

The results in this paper, as well as those in \masuda, consistently indicate that the distribution of photometrically determined rotation periods of solar-mass stars older than a few Gyr is significantly affected by the detection bias, even in the Kepler prime mission sample.
The rapid decrease of rotational modulation amplitude with increasing Rossby number or rotation period \masudap\ suggests that it is difficult to obtain a nearly unbiased sample of photometric rotation periods for those stars in the near future. 
Thus it is crucial to explicitly model the detection bias in any attempt to quantitatively interpret the available photometric rotation period distribution of older Sun-like stars, such as the derivation of the critical Rossby number for the onset of weakened magnetic braking \citep[e.g.,][]{2019ApJ...872..128V,2022arXiv220308920D}. Other probes of rotation that are applicable to older stars and are subject to different detection biases, such as asteroseismology \citep{2021NatAs...5..707H} and $v\sin i$ \citep{2022MNRAS.510.5623M}, will also remain important for the study of their spin evolution. 
For Sun-like stars younger than a few Gyr, on the other hand, our results suggest that the Kepler sample provides a relatively unbiased view on their rotation period distribution, as long as an appropriate magnitude cut is applied to the sample.

\subsection{How General Is the Result?}\label{ssec:mazeh-mcquillan}

In this paper, we focused on the detectability of rotational modulation in the \citet{2015ApJ...801....3M} sample, who searched rotational modulation for a subset of \kepler\ stars with transiting planet candidates. However, the method used is the same as the one adopted by \citet{2014ApJS..211...24M} who searched rotational modulation in generic \kepler\ stars and determined rotation periods for $\approx 34,000$ of them. Therefore the conclusions in this paper likely apply to the \citet{2014ApJS..211...24M} sample as well.

Of course, the samples of stars for which rotational modulation has been searched in the two works (i.e., stars with and without transiting planets) may well have different properties, including the distributions of rotation periods and modulation amplitudes \citep[see, e.g.,][]{2015ApJ...801....3M}. 
Nevertheless, what we focus on here is the detectability of the modulation from a star with the given mass and age, and so the intrinsic differences in the stellar populations do not matter as long as the period detections in the two samples are subject to the same detectability thresholds as a function of signal to noise.

\begin{figure}
\epsscale{1.15}
\plotone{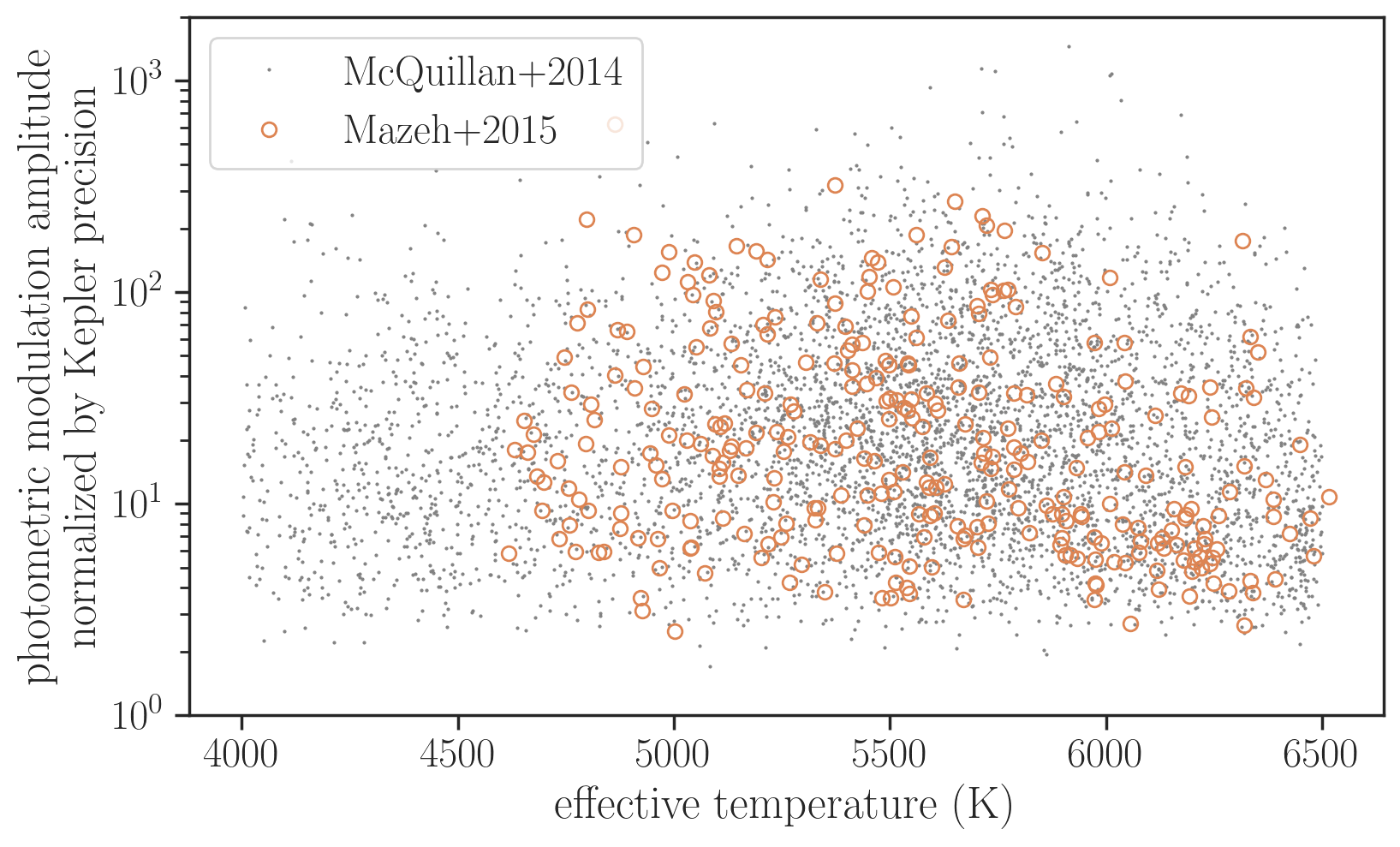}
\plotone{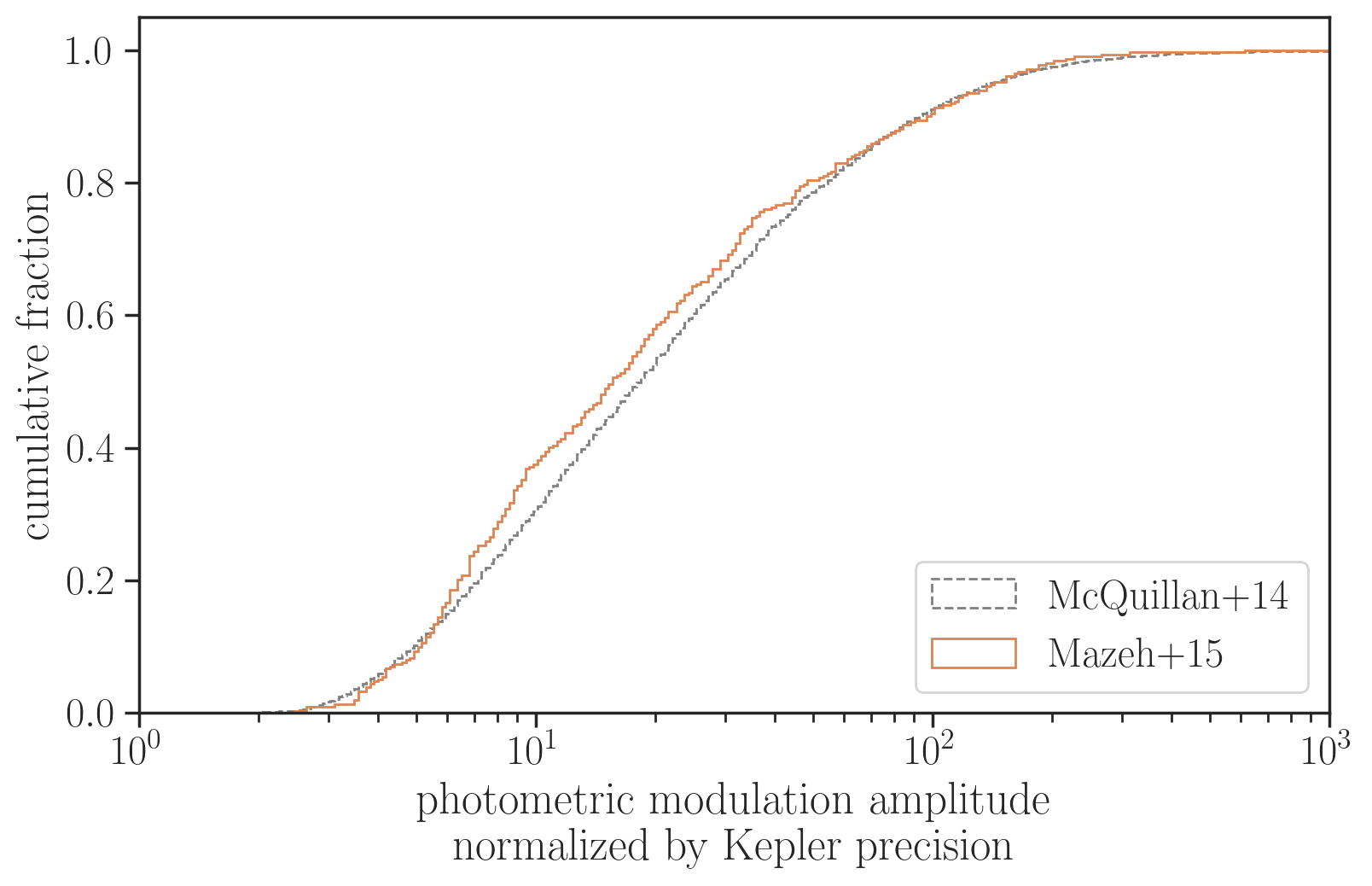}
\caption{
{\it (Top)} Spot-modulation amplitudes normalized by {\it Kepler} photometric precision and $\teff$ of stars for which rotational modulation has been detected in the searches by \cite{2015ApJ...801....3M} (open orange circles)
and by \cite{2014ApJS..211...24M} (gray dots); see text for details of the samples.
{\it (Bottom)} Normalized cumulative distributions for the normalized modulation amplitudes in the two samples shown in the top panel.
}
\label{fig:rnorm}
\end{figure}

To check on the possible difference in the detection functions, in the top panel of Figure~\ref{fig:rnorm} we check the distributions of photometric modulation amplitudes normalized by the long-cadence photometric precision of \kepler\ as a function of $\teff$, for the stars with detected rotation periods in the samples of \citet{2015ApJ...801....3M} and \citet{2014ApJS..211...24M}. 
For the former, we adopt the CKS--Mazeh sample defined in this study and  $\teff$ is from the CKS; for the latter, we show the subset of the \citet{2014ApJS..211...24M} sample with LAMOST $\teff$, as defined in \masuda. 
We see that both samples are truncated around the normalized amplitude of $\sim 3$ without strong dependence on $\teff$, suggesting that both samples are subject to similar detectability thresholds \citep[see also][]{2022ApJ...933..195M}.
In the bottom panel, we show the normalized cumulative distribution functions for the noise-normalized modulation amplitudes. The plot shows that the distributions are not exactly the same in the two samples, but the normalized amplitudes at lower percentiles (e.g., lower than the 20th) agree within 20\%. 
Thus we conclude that the detection thresholds in the two samples are not significantly different as to affect $f(\mass, \age)$ inferred in this work: the $20\%$ change in the amplitude corresponds to only $10\%$ difference in age, given $\amp \sim \age^{-2}$ (see Section~\ref{ssec:cks_result}).

\vspace{0.5cm}
The data and the code underlying this article are available through GitHub.\footnote{\url{https://github.com/kemasuda/acheron}}
Posterior samples from isochrone fitting for the sample stars are available from the author upon request.

\begin{acknowledgements}

The author thanks Shinsuke Takasao for illuminating conversations on this subject. The author also thanks the anonymous reviewer for careful reading of the manuscript and for a helpful report. This work was supported by JSPS KAKENHI Grant Number 21K13980.
This work has made use of data from the European Space Agency (ESA) mission {\it Gaia} (\url{https://www.cosmos.esa.int/gaia}), processed by the {\it Gaia} Data Processing and Analysis Consortium (DPAC, \url{https://www.cosmos.esa.int/web/gaia/dpac/consortium}). Funding for the DPAC has been provided by national institutions, in particular the institutions participating in the {\it Gaia} Multilateral Agreement.

\end{acknowledgements}

\software{ 
corner \citep{corner}, JAX \citep{jax2018github}, NumPyro \citep{bingham2018pyro, phan2019composable}
}



\bibliography{age}

\begin{thebibliography}{}
\expandafter\ifx\csname natexlab\endcsname\relax\def\natexlab#1{#1}\fi
\providecommand{\url}[1]{\href{#1}{#1}}
\providecommand{\dodoi}[1]{doi:~\href{http://doi.org/#1}{\nolinkurl{#1}}}
\providecommand{\doeprint}[1]{\href{http://ascl.net/#1}{\nolinkurl{http://ascl.net/#1}}}
\providecommand{\doarXiv}[1]{\href{https://arxiv.org/abs/#1}{\nolinkurl{https://arxiv.org/abs/#1}}}

\bibitem[{{Amard} {et~al.}(2020){Amard}, {Roquette}, \&
  {Matt}}]{2020MNRAS.499.3481A}
{Amard}, L., {Roquette}, J., \& {Matt}, S.~P. 2020, \mnras, 499, 3481,
  \dodoi{10.1093/mnras/staa3038}

\bibitem[{{Angus} {et~al.}(2015){Angus}, {Aigrain}, {Foreman-Mackey}, \&
  {McQuillan}}]{2015MNRAS.450.1787A}
{Angus}, R., {Aigrain}, S., {Foreman-Mackey}, D., \& {McQuillan}, A. 2015,
  \mnras, 450, 1787, \dodoi{10.1093/mnras/stv423}

\bibitem[{{Angus} {et~al.}(2019){Angus}, {Morton}, {Foreman-Mackey}, {van
  Saders}, {Curtis}, {Kane}, {Bedell}, {Kiman}, {Hogg}, \&
  {Brewer}}]{2019AJ....158..173A}
{Angus}, R., {Morton}, T.~D., {Foreman-Mackey}, D., {et~al.} 2019, \aj, 158,
  173, \dodoi{10.3847/1538-3881/ab3c53}

\bibitem[{{Astraatmadja} \& {Bailer-Jones}(2016)}]{2016ApJ...832..137A}
{Astraatmadja}, T.~L., \& {Bailer-Jones}, C. A.~L. 2016, \apj, 832, 137,
  \dodoi{10.3847/0004-637X/832/2/137}

\bibitem[{{Bailer-Jones}(2015)}]{2015PASP..127..994B}
{Bailer-Jones}, C. A.~L. 2015, \pasp, 127, 994, \dodoi{10.1086/683116}

\bibitem[{{Betancourt}(2017)}]{2017arXiv170102434B}
{Betancourt}, M. 2017, arXiv e-prints, arXiv:1701.02434.
\newblock \doarXiv{1701.02434}

\bibitem[{Bingham {et~al.}(2018)Bingham, Chen, Jankowiak, Obermeyer, Pradhan,
  Karaletsos, Singh, Szerlip, Horsfall, \& Goodman}]{bingham2018pyro}
Bingham, E., Chen, J.~P., Jankowiak, M., {et~al.} 2018, arXiv preprint
  arXiv:1810.09538

\bibitem[{{Borucki} {et~al.}(2010){Borucki}, {Koch}, {Basri}, {Batalha},
  {Brown}, {Caldwell}, {Caldwell}, {Christensen-Dalsgaard}, {Cochran},
  {DeVore}, {Dunham}, {Dupree}, {Gautier}, {Geary}, {Gilliland}, {Gould},
  {Howell}, {Jenkins}, {Kondo}, {Latham}, {Marcy}, {Meibom}, {Kjeldsen},
  {Lissauer}, {Monet}, {Morrison}, {Sasselov}, {Tarter}, {Boss}, {Brownlee},
  {Owen}, {Buzasi}, {Charbonneau}, {Doyle}, {Fortney}, {Ford}, {Holman},
  {Seager}, {Steffen}, {Welsh}, {Rowe}, {Anderson}, {Buchhave}, {Ciardi},
  {Walkowicz}, {Sherry}, {Horch}, {Isaacson}, {Everett}, {Fischer}, {Torres},
  {Johnson}, {Endl}, {MacQueen}, {Bryson}, {Dotson}, {Haas}, {Kolodziejczak},
  {Van Cleve}, {Chandrasekaran}, {Twicken}, {Quintana}, {Clarke}, {Allen},
  {Li}, {Wu}, {Tenenbaum}, {Verner}, {Bruhweiler}, {Barnes}, \&
  {Prsa}}]{2010Sci...327..977B}
{Borucki}, W.~J., {Koch}, D., {Basri}, G., {et~al.} 2010, Science, 327, 977,
  \dodoi{10.1126/science.1185402}

\bibitem[{Bradbury {et~al.}(2018)Bradbury, Frostig, Hawkins, Johnson, Leary,
  Maclaurin, Necula, Paszke, Vander{P}las, Wanderman-{M}ilne, \&
  Zhang}]{jax2018github}
Bradbury, J., Frostig, R., Hawkins, P., {et~al.} 2018, {JAX}: composable
  transformations of {P}ython+{N}um{P}y programs, 0.2.5.
\newblock \url{http://github.com/google/jax}

\bibitem[{{Choi} {et~al.}(2016){Choi}, {Dotter}, {Conroy}, {Cantiello},
  {Paxton}, \& {Johnson}}]{2016ApJ...823..102C}
{Choi}, J., {Dotter}, A., {Conroy}, C., {et~al.} 2016, \apj, 823, 102,
  \dodoi{10.3847/0004-637X/823/2/102}

\bibitem[{{David} {et~al.}(2022){David}, {Angus}, {Curtis}, {van Saders},
  {Colman}, {Contardo}, {Lu}, \& {Zinn}}]{2022arXiv220308920D}
{David}, T.~J., {Angus}, R., {Curtis}, J.~L., {et~al.} 2022, arXiv e-prints,
  arXiv:2203.08920.
\newblock \doarXiv{2203.08920}

\bibitem[{{Dotter}(2016)}]{2016ApJS..222....8D}
{Dotter}, A. 2016, \apjs, 222, 8, \dodoi{10.3847/0067-0049/222/1/8}

\bibitem[{Duane {et~al.}(1987)Duane, Kennedy, Pendleton, \&
  Roweth}]{DUANE1987216}
Duane, S., Kennedy, A., Pendleton, B.~J., \& Roweth, D. 1987, Physics Letters
  B, 195, 216 , \dodoi{https://doi.org/10.1016/0370-2693(87)91197-X}

\bibitem[{{El-Badry} {et~al.}(2021){El-Badry}, {Rix}, \&
  {Heintz}}]{2021MNRAS.506.2269E}
{El-Badry}, K., {Rix}, H.-W., \& {Heintz}, T.~M. 2021, \mnras, 506, 2269,
  \dodoi{10.1093/mnras/stab323}

\bibitem[{{Flores} {et~al.}(2022){Flores}, {Connelley}, {Reipurth}, \&
  {Duch{\^e}ne}}]{2022ApJ...925...21F}
{Flores}, C., {Connelley}, M.~S., {Reipurth}, B., \& {Duch{\^e}ne}, G. 2022,
  \apj, 925, 21, \dodoi{10.3847/1538-4357/ac37bd}

\bibitem[{Foreman-Mackey(2016)}]{corner}
Foreman-Mackey, D. 2016, The Journal of Open Source Software, 24,
  \dodoi{10.21105/joss.00024}

\bibitem[{{Foreman-Mackey} {et~al.}(2014){Foreman-Mackey}, {Hogg}, \&
  {Morton}}]{2014ApJ...795...64F}
{Foreman-Mackey}, D., {Hogg}, D.~W., \& {Morton}, T.~D. 2014, \apj, 795, 64,
  \dodoi{10.1088/0004-637X/795/1/64}

\bibitem[{{Fulton} \& {Petigura}(2018)}]{2018AJ....156..264F}
{Fulton}, B.~J., \& {Petigura}, E.~A. 2018, \aj, 156, 264,
  \dodoi{10.3847/1538-3881/aae828}

\bibitem[{{Gaia Collaboration} {et~al.}(2021){Gaia Collaboration}, {Brown},
  {Vallenari}, {Prusti}, {de Bruijne}, {Babusiaux}, {Biermann}, {Creevey},
  {Evans}, {Eyer}, {Hutton}, {Jansen}, {Jordi}, {Klioner}, {Lammers},
  {Lindegren}, {Luri}, {Mignard}, {Panem}, {Pourbaix}, {Randich}, {Sartoretti},
  {Soubiran}, {Walton}, {Arenou}, {Bailer-Jones}, {Bastian}, {Cropper},
  {Drimmel}, {Katz}, {Lattanzi}, {van Leeuwen}, {Bakker}, {Cacciari},
  {Casta{\~n}eda}, {De Angeli}, {Ducourant}, {Fabricius}, {Fouesneau},
  {Fr{\'e}mat}, {Guerra}, {Guerrier}, {Guiraud}, {Jean-Antoine Piccolo},
  {Masana}, {Messineo}, {Mowlavi}, {Nicolas}, {Nienartowicz}, {Pailler},
  {Panuzzo}, {Riclet}, {Roux}, {Seabroke}, {Sordo}, {Tanga}, {Th{\'e}venin},
  {Gracia-Abril}, {Portell}, {Teyssier}, {Altmann}, {Andrae}, {Bellas-Velidis},
  {Benson}, {Berthier}, {Blomme}, {Brugaletta}, {Burgess}, {Busso}, {Carry},
  {Cellino}, {Cheek}, {Clementini}, {Damerdji}, {Davidson}, {Delchambre},
  {Dell'Oro}, {Fern{\'a}ndez-Hern{\'a}ndez}, {Galluccio}, {Garc{\'\i}a-Lario},
  {Garcia-Reinaldos}, {Gonz{\'a}lez-N{\'u}{\~n}ez}, {Gosset}, {Haigron},
  {Halbwachs}, {Hambly}, {Harrison}, {Hatzidimitriou}, {Heiter},
  {Hern{\'a}ndez}, {Hestroffer}, {Hodgkin}, {Holl}, {Jan{\ss}en}, {Jevardat de
  Fombelle}, {Jordan}, {Krone-Martins}, {Lanzafame}, {L{\"o}ffler}, {Lorca},
  {Manteiga}, {Marchal}, {Marrese}, {Moitinho}, {Mora}, {Muinonen}, {Osborne},
  {Pancino}, {Pauwels}, {Petit}, {Recio-Blanco}, {Richards}, {Riello},
  {Rimoldini}, {Robin}, {Roegiers}, {Rybizki}, {Sarro}, {Siopis}, {Smith},
  {Sozzetti}, {Ulla}, {Utrilla}, {van Leeuwen}, {van Reeven}, {Abbas}, {Abreu
  Aramburu}, {Accart}, {Aerts}, {Aguado}, {Ajaj}, {Altavilla}, {{\'A}lvarez},
  {{\'A}lvarez Cid-Fuentes}, {Alves}, {Anderson}, {Anglada Varela}, {Antoja},
  {Audard}, {Baines}, {Baker}, {Balaguer-N{\'u}{\~n}ez}, {Balbinot}, {Balog},
  {Barache}, {Barbato}, {Barros}, {Barstow}, {Bartolom{\'e}}, {Bassilana},
  {Bauchet}, {Baudesson-Stella}, {Becciani}, {Bellazzini}, {Bernet}, {Bertone},
  {Bianchi}, {Blanco-Cuaresma}, {Boch}, {Bombrun}, {Bossini}, {Bouquillon},
  {Bragaglia}, {Bramante}, {Breedt}, {Bressan}, {Brouillet}, {Bucciarelli},
  {Burlacu}, {Busonero}, {Butkevich}, {Buzzi}, {Caffau}, {Cancelliere},
  {C{\'a}novas}, {Cantat-Gaudin}, {Carballo}, {Carlucci}, {Carnerero},
  {Carrasco}, {Casamiquela}, {Castellani}, {Castro-Ginard}, {Castro Sampol},
  {Chaoul}, {Charlot}, {Chemin}, {Chiavassa}, {Cioni}, {Comoretto}, {Cooper},
  {Cornez}, {Cowell}, {Crifo}, {Crosta}, {Crowley}, {Dafonte}, {Dapergolas},
  {David}, {David}, {de Laverny}, {De Luise}, {De March}, {De Ridder}, {de
  Souza}, {de Teodoro}, {de Torres}, {del Peloso}, {del Pozo}, {Delbo},
  {Delgado}, {Delgado}, {Delisle}, {Di Matteo}, {Diakite}, {Diener},
  {Distefano}, {Dolding}, {Eappachen}, {Edvardsson}, {Enke}, {Esquej}, {Fabre},
  {Fabrizio}, {Faigler}, {Fedorets}, {Fernique}, {Fienga}, {Figueras},
  {Fouron}, {Fragkoudi}, {Fraile}, {Franke}, {Gai}, {Garabato},
  {Garcia-Gutierrez}, {Garc{\'\i}a-Torres}, {Garofalo}, {Gavras}, {Gerlach},
  {Geyer}, {Giacobbe}, {Gilmore}, {Girona}, {Giuffrida}, {Gomel}, {Gomez},
  {Gonzalez-Santamaria}, {Gonz{\'a}lez-Vidal}, {Granvik},
  {Guti{\'e}rrez-S{\'a}nchez}, {Guy}, {Hauser}, {Haywood}, {Helmi}, {Hidalgo},
  {Hilger}, {H{\l}adczuk}, {Hobbs}, {Holland}, {Huckle}, {Jasniewicz},
  {Jonker}, {Juaristi Campillo}, {Julbe}, {Karbevska}, {Kervella}, {Khanna},
  {Kochoska}, {Kontizas}, {Kordopatis}, {Korn}, {Kostrzewa-Rutkowska},
  {Kruszy{\'n}ska}, {Lambert}, {Lanza}, {Lasne}, {Le Campion}, {Le Fustec},
  {Lebreton}, {Lebzelter}, {Leccia}, {Leclerc}, {Lecoeur-Taibi}, {Liao},
  {Licata}, {Lindstr{\o}m}, {Lister}, {Livanou}, {Lobel}, {Madrero Pardo},
  {Managau}, {Mann}, {Marchant}, {Marconi}, {Marcos Santos}, {Marinoni},
  {Marocco}, {Marshall}, {Martin Polo}, {Mart{\'\i}n-Fleitas}, {Masip},
  {Massari}, {Mastrobuono-Battisti}, {Mazeh}, {McMillan}, {Messina},
  {Michalik}, {Millar}, {Mints}, {Molina}, {Molinaro}, {Moln{\'a}r},
  {Montegriffo}, {Mor}, {Morbidelli}, {Morel}, {Morris}, {Mulone}, {Munoz},
  {Muraveva}, {Murphy}, {Musella}, {Noval}, {Ord{\'e}novic}, {Orr{\`u}},
  {Osinde}, {Pagani}, {Pagano}, {Palaversa}, {Palicio}, {Panahi}, {Pawlak},
  {Pe{\~n}alosa Esteller}, {Penttil{\"a}}, {Piersimoni}, {Pineau}, {Plachy},
  {Plum}, {Poggio}, {Poretti}, {Poujoulet}, {Pr{\v{s}}a}, {Pulone}, {Racero},
  {Ragaini}, {Rainer}, {Raiteri}, {Rambaux}, {Ramos}, {Ramos-Lerate}, {Re
  Fiorentin}, {Regibo}, {Reyl{\'e}}, {Ripepi}, {Riva}, {Rixon}, {Robichon},
  {Robin}, {Roelens}, {Rohrbasser}, {Romero-G{\'o}mez}, {Rowell}, {Royer},
  {Rybicki}, {Sadowski}, {Sagrist{\`a} Sell{\'e}s}, {Sahlmann}, {Salgado},
  {Salguero}, {Samaras}, {Sanchez Gimenez}, {Sanna}, {Santove{\~n}a},
  {Sarasso}, {Schultheis}, {Sciacca}, {Segol}, {Segovia}, {S{\'e}gransan},
  {Semeux}, {Shahaf}, {Siddiqui}, {Siebert}, {Siltala}, {Slezak}, {Smart},
  {Solano}, {Solitro}, {Souami}, {Souchay}, {Spagna}, {Spoto}, {Steele},
  {Steidelm{\"u}ller}, {Stephenson}, {S{\"u}veges}, {Szabados}, {Szegedi-Elek},
  {Taris}, {Tauran}, {Taylor}, {Teixeira}, {Thuillot}, {Tonello}, {Torra},
  {Torra}, {Turon}, {Unger}, {Vaillant}, {van Dillen}, {Vanel}, {Vecchiato},
  {Viala}, {Vicente}, {Voutsinas}, {Weiler}, {Wevers}, {Wyrzykowski}, {Yoldas},
  {Yvard}, {Zhao}, {Zorec}, {Zucker}, {Zurbach}, \&
  {Zwitter}}]{2021A&A...649A...1G}
{Gaia Collaboration}, {Brown}, A.~G.~A., {Vallenari}, A., {et~al.} 2021, \aap,
  649, A1, \dodoi{10.1051/0004-6361/202039657}

\bibitem[{{Garc{\'{\i}}a} {et~al.}(2014){Garc{\'{\i}}a}, {Ceillier},
  {Salabert}, {Mathur}, {van Saders}, {Pinsonneault}, {Ballot}, {Beck},
  {Bloemen}, {Campante}, {Davies}, {do Nascimento}, {Mathis}, {Metcalfe},
  {Nielsen}, {Su{\'a}rez}, {Chaplin}, {Jim{\'e}nez}, \&
  {Karoff}}]{2014A&A...572A..34G}
{Garc{\'{\i}}a}, R.~A., {Ceillier}, T., {Salabert}, D., {et~al.} 2014, \aap,
  572, A34, \dodoi{10.1051/0004-6361/201423888}

\bibitem[{Gelman {et~al.}(2014)Gelman, Carlin, Stern, Dunson, Vehtari, \&
  Rubin}]{BB13945229}
Gelman, A., Carlin, J.~B., Stern, H.~S., {et~al.} 2014, Bayesian data analysis,
  3rd edn., Texts in statistical science (CRC Press).
\newblock \url{https://ci.nii.ac.jp/ncid/BB13945229}

\bibitem[{{Green} {et~al.}(2019){Green}, {Schlafly}, {Zucker}, {Speagle}, \&
  {Finkbeiner}}]{2019ApJ...887...93G}
{Green}, G.~M., {Schlafly}, E., {Zucker}, C., {Speagle}, J.~S., \&
  {Finkbeiner}, D. 2019, \apj, 887, 93, \dodoi{10.3847/1538-4357/ab5362}

\bibitem[{{Green} {et~al.}(2018){Green}, {Schlafly}, {Finkbeiner}, {Rix},
  {Martin}, {Burgett}, {Draper}, {Flewelling}, {Hodapp}, {Kaiser}, {Kudritzki},
  {Magnier}, {Metcalfe}, {Tonry}, {Wainscoat}, \&
  {Waters}}]{2018MNRAS.478..651G}
{Green}, G.~M., {Schlafly}, E.~F., {Finkbeiner}, D., {et~al.} 2018, \mnras,
  478, 651, \dodoi{10.1093/mnras/sty1008}

\bibitem[{{Hall} {et~al.}(2021){Hall}, {Davies}, {van Saders}, {Nielsen},
  {Lund}, {Chaplin}, {Garc{\'\i}a}, {Amard}, {Breimann}, {Khan}, {See}, \&
  {Tayar}}]{2021NatAs...5..707H}
{Hall}, O.~J., {Davies}, G.~R., {van Saders}, J., {et~al.} 2021, Nature
  Astronomy, 5, 707, \dodoi{10.1038/s41550-021-01335-x}

\bibitem[{{Hogg} {et~al.}(2010){Hogg}, {Myers}, \&
  {Bovy}}]{2010ApJ...725.2166H}
{Hogg}, D.~W., {Myers}, A.~D., \& {Bovy}, J. 2010, \apj, 725, 2166,
  \dodoi{10.1088/0004-637X/725/2/2166}

\bibitem[{{Jackson} \& {Jeffries}(2012)}]{2012MNRAS.423.2966J}
{Jackson}, R.~J., \& {Jeffries}, R.~D. 2012, \mnras, 423, 2966,
  \dodoi{10.1111/j.1365-2966.2012.21119.x}

\bibitem[{{Johnson} {et~al.}(2017){Johnson}, {Petigura}, {Fulton}, {Marcy},
  {Howard}, {Isaacson}, {Hebb}, {Cargile}, {Morton}, {Weiss}, {Winn}, {Rogers},
  {Sinukoff}, \& {Hirsch}}]{2017AJ....154..108J}
{Johnson}, J.~A., {Petigura}, E.~A., {Fulton}, B.~J., {et~al.} 2017, \aj, 154,
  108, \dodoi{10.3847/1538-3881/aa80e7}

\bibitem[{{J{\o}rgensen} \& {Lindegren}(2005)}]{2005A&A...436..127J}
{J{\o}rgensen}, B.~R., \& {Lindegren}, L. 2005, \aap, 436, 127,
  \dodoi{10.1051/0004-6361:20042185}

\bibitem[{{Koch} {et~al.}(2010){Koch}, {Borucki}, {Basri}, {Batalha}, {Brown},
  {Caldwell}, {Christensen-Dalsgaard}, {Cochran}, {DeVore}, {Dunham},
  {Gautier}, {Geary}, {Gilliland}, {Gould}, {Jenkins}, {Kondo}, {Latham},
  {Lissauer}, {Marcy}, {Monet}, {Sasselov}, {Boss}, {Brownlee}, {Caldwell},
  {Dupree}, {Howell}, {Kjeldsen}, {Meibom}, {Morrison}, {Owen}, {Reitsema},
  {Tarter}, {Bryson}, {Dotson}, {Gazis}, {Haas}, {Kolodziejczak}, {Rowe}, {Van
  Cleve}, {Allen}, {Chandrasekaran}, {Clarke}, {Li}, {Quintana}, {Tenenbaum},
  {Twicken}, \& {Wu}}]{2010ApJ...713L..79K}
{Koch}, D.~G., {Borucki}, W.~J., {Basri}, G., {et~al.} 2010, \apjl, 713, L79,
  \dodoi{10.1088/2041-8205/713/2/L79}

\bibitem[{{Lehtinen} {et~al.}(2020){Lehtinen}, {Spada}, {K{\"a}pyl{\"a}},
  {Olspert}, \& {K{\"a}pyl{\"a}}}]{2020NatAs...4..658L}
{Lehtinen}, J.~J., {Spada}, F., {K{\"a}pyl{\"a}}, M.~J., {Olspert}, N., \&
  {K{\"a}pyl{\"a}}, P.~J. 2020, Nature Astronomy, 4, 658,
  \dodoi{10.1038/s41550-020-1039-x}

\bibitem[{{Lindegren} {et~al.}(2021){Lindegren}, {Bastian}, {Biermann},
  {Bombrun}, {de Torres}, {Gerlach}, {Geyer}, {Hern{\'a}ndez}, {Hilger},
  {Hobbs}, {Klioner}, {Lammers}, {McMillan}, {Ramos-Lerate},
  {Steidelm{\"u}ller}, {Stephenson}, \& {van Leeuwen}}]{2021A&A...649A...4L}
{Lindegren}, L., {Bastian}, U., {Biermann}, M., {et~al.} 2021, \aap, 649, A4,
  \dodoi{10.1051/0004-6361/202039653}

\bibitem[{{Masuda}(2022)}]{2022ApJ...933..195M}
{Masuda}, K. 2022, \apj, 933, 195, \dodoi{10.3847/1538-4357/ac7527}

\bibitem[{{Masuda} {et~al.}(2022){Masuda}, {Petigura}, \&
  {Hall}}]{2022MNRAS.510.5623M}
{Masuda}, K., {Petigura}, E.~A., \& {Hall}, O.~J. 2022, \mnras, 510, 5623,
  \dodoi{10.1093/mnras/stab3650}

\bibitem[{{Mazeh} {et~al.}(2015){Mazeh}, {Perets}, {McQuillan}, \&
  {Goldstein}}]{2015ApJ...801....3M}
{Mazeh}, T., {Perets}, H.~B., {McQuillan}, A., \& {Goldstein}, E.~S. 2015,
  \apj, 801, 3, \dodoi{10.1088/0004-637X/801/1/3}

\bibitem[{{McQuillan} {et~al.}(2014){McQuillan}, {Mazeh}, \&
  {Aigrain}}]{2014ApJS..211...24M}
{McQuillan}, A., {Mazeh}, T., \& {Aigrain}, S. 2014, \apjs, 211, 24,
  \dodoi{10.1088/0067-0049/211/2/24}

\bibitem[{{Moe} \& {Kratter}(2021)}]{2021MNRAS.507.3593M}
{Moe}, M., \& {Kratter}, K.~M. 2021, \mnras, 507, 3593,
  \dodoi{10.1093/mnras/stab2328}

\bibitem[{{Morris}(2020)}]{2020ApJ...893...67M}
{Morris}, B.~M. 2020, \apj, 893, 67, \dodoi{10.3847/1538-4357/ab79a0}

\bibitem[{{Morton}(2015)}]{2015ascl.soft03010M}
{Morton}, T.~D. 2015, {isochrones: Stellar model grid package}, Astrophysics
  Source Code Library.
\newblock \doeprint{1503.010}

\bibitem[{{Nielsen} {et~al.}(2013){Nielsen}, {Gizon}, {Schunker}, \&
  {Karoff}}]{2013A&A...557L..10N}
{Nielsen}, M.~B., {Gizon}, L., {Schunker}, H., \& {Karoff}, C. 2013, \aap, 557,
  L10, \dodoi{10.1051/0004-6361/201321912}

\bibitem[{{Paxton} {et~al.}(2011){Paxton}, {Bildsten}, {Dotter}, {Herwig},
  {Lesaffre}, \& {Timmes}}]{2011ApJS..192....3P}
{Paxton}, B., {Bildsten}, L., {Dotter}, A., {et~al.} 2011, \apjs, 192, 3,
  \dodoi{10.1088/0067-0049/192/1/3}

\bibitem[{{Paxton} {et~al.}(2013){Paxton}, {Cantiello}, {Arras}, {Bildsten},
  {Brown}, {Dotter}, {Mankovich}, {Montgomery}, {Stello}, {Timmes}, \&
  {Townsend}}]{2013ApJS..208....4P}
{Paxton}, B., {Cantiello}, M., {Arras}, P., {et~al.} 2013, \apjs, 208, 4,
  \dodoi{10.1088/0067-0049/208/1/4}

\bibitem[{{Paxton} {et~al.}(2015){Paxton}, {Marchant}, {Schwab}, {Bauer},
  {Bildsten}, {Cantiello}, {Dessart}, {Farmer}, {Hu}, {Langer}, {Townsend},
  {Townsley}, \& {Timmes}}]{2015ApJS..220...15P}
{Paxton}, B., {Marchant}, P., {Schwab}, J., {et~al.} 2015, \apjs, 220, 15,
  \dodoi{10.1088/0067-0049/220/1/15}

\bibitem[{{Petigura} {et~al.}(2017){Petigura}, {Howard}, {Marcy}, {Johnson},
  {Isaacson}, {Cargile}, {Hebb}, {Fulton}, {Weiss}, {Morton}, {Winn}, {Rogers},
  {Sinukoff}, {Hirsch}, \& {Crossfield}}]{2017AJ....154..107P}
{Petigura}, E.~A., {Howard}, A.~W., {Marcy}, G.~W., {et~al.} 2017, \aj, 154,
  107, \dodoi{10.3847/1538-3881/aa80de}

\bibitem[{{Petigura} {et~al.}(2022){Petigura}, {Rogers}, {Isaacson}, {Owen},
  {Kraus}, {Winn}, {MacDougall}, {Howard}, {Fulton}, {Kosiarek}, {Weiss},
  {Behmard}, \& {Blunt}}]{2022AJ....163..179P}
{Petigura}, E.~A., {Rogers}, J.~G., {Isaacson}, H., {et~al.} 2022, \aj, 163,
  179, \dodoi{10.3847/1538-3881/ac51e3}

\bibitem[{Phan {et~al.}(2019)Phan, Pradhan, \& Jankowiak}]{phan2019composable}
Phan, D., Pradhan, N., \& Jankowiak, M. 2019, arXiv preprint arXiv:1912.11554

\bibitem[{{Pont} \& {Eyer}(2004)}]{2004MNRAS.351..487P}
{Pont}, F., \& {Eyer}, L. 2004, \mnras, 351, 487,
  \dodoi{10.1111/j.1365-2966.2004.07780.x}

\bibitem[{{Reinhold} \& {Hekker}(2020)}]{2020A&A...635A..43R}
{Reinhold}, T., \& {Hekker}, S. 2020, \aap, 635, A43,
  \dodoi{10.1051/0004-6361/201936887}

\bibitem[{{Reinhold} {et~al.}(2013){Reinhold}, {Reiners}, \&
  {Basri}}]{2013A&A...560A...4R}
{Reinhold}, T., {Reiners}, A., \& {Basri}, G. 2013, \aap, 560, A4,
  \dodoi{10.1051/0004-6361/201321970}

\bibitem[{{Santos} {et~al.}(2021){Santos}, {Breton}, {Mathur}, \&
  {Garc{\'\i}a}}]{2021ApJS..255...17S}
{Santos}, A.~R.~G., {Breton}, S.~N., {Mathur}, S., \& {Garc{\'\i}a}, R.~A.
  2021, \apjs, 255, 17, \dodoi{10.3847/1538-4365/ac033f}

\bibitem[{{Santos} {et~al.}(2019){Santos}, {Garc{\'\i}a}, {Mathur}, {Bugnet},
  {van Saders}, {Metcalfe}, {Simonian}, \&
  {Pinsonneault}}]{2019ApJS..244...21S}
{Santos}, A.~R.~G., {Garc{\'\i}a}, R.~A., {Mathur}, S., {et~al.} 2019, \apjs,
  244, 21, \dodoi{10.3847/1538-4365/ab3b56}

\bibitem[{{See} {et~al.}(2021){See}, {Roquette}, {Amard}, \&
  {Matt}}]{2021ApJ...912..127S}
{See}, V., {Roquette}, J., {Amard}, L., \& {Matt}, S.~P. 2021, \apj, 912, 127,
  \dodoi{10.3847/1538-4357/abed47}

\bibitem[{{Silva Aguirre} {et~al.}(2015){Silva Aguirre}, {Davies}, {Basu},
  {Christensen-Dalsgaard}, {Creevey}, {Metcalfe}, {Bedding}, {Casagrande},
  {Handberg}, {Lund}, {Nissen}, {Chaplin}, {Huber}, {Serenelli}, {Stello}, {Van
  Eylen}, {Campante}, {Elsworth}, {Gilliland}, {Hekker}, {Karoff}, {Kawaler},
  {Kjeldsen}, \& {Lundkvist}}]{2015MNRAS.452.2127S}
{Silva Aguirre}, V., {Davies}, G.~R., {Basu}, S., {et~al.} 2015, \mnras, 452,
  2127, \dodoi{10.1093/mnras/stv1388}

\bibitem[{{Silva Aguirre} {et~al.}(2017){Silva Aguirre}, {Lund}, {Antia},
  {Ball}, {Basu}, {Christensen-Dalsgaard}, {Lebreton}, {Reese}, {Verma},
  {Casagrande}, {Justesen}, {Mosumgaard}, {Chaplin}, {Bedding}, {Davies},
  {Handberg}, {Houdek}, {Huber}, {Kjeldsen}, {Latham}, {White}, {Coelho},
  {Miglio}, \& {Rendle}}]{2017ApJ...835..173S}
{Silva Aguirre}, V., {Lund}, M.~N., {Antia}, H.~M., {et~al.} 2017, \apj, 835,
  173, \dodoi{10.3847/1538-4357/835/2/173}

\bibitem[{{Skrutskie} {et~al.}(2006){Skrutskie}, {Cutri}, {Stiening},
  {Weinberg}, {Schneider}, {Carpenter}, {Beichman}, {Capps}, {Chester},
  {Elias}, {Huchra}, {Liebert}, {Lonsdale}, {Monet}, {Price}, {Seitzer},
  {Jarrett}, {Kirkpatrick}, {Gizis}, {Howard}, {Evans}, {Fowler}, {Fullmer},
  {Hurt}, {Light}, {Kopan}, {Marsh}, {McCallon}, {Tam}, {Van Dyk}, \&
  {Wheelock}}]{2006AJ....131.1163S}
{Skrutskie}, M.~F., {Cutri}, R.~M., {Stiening}, R., {et~al.} 2006, \aj, 131,
  1163, \dodoi{10.1086/498708}

\bibitem[{{Soderblom}(2010)}]{2010ARA&A..48..581S}
{Soderblom}, D.~R. 2010, \araa, 48, 581,
  \dodoi{10.1146/annurev-astro-081309-130806}

\bibitem[{{Takeda} {et~al.}(2007){Takeda}, {Ford}, {Sills}, {Rasio}, {Fischer},
  \& {Valenti}}]{2007ApJS..168..297T}
{Takeda}, G., {Ford}, E.~B., {Sills}, A., {et~al.} 2007, \apjs, 168, 297,
  \dodoi{10.1086/509763}

\bibitem[{{Tayar} {et~al.}(2022){Tayar}, {Claytor}, {Huber}, \& {van
  Saders}}]{2022ApJ...927...31T}
{Tayar}, J., {Claytor}, Z.~R., {Huber}, D., \& {van Saders}, J. 2022, \apj,
  927, 31, \dodoi{10.3847/1538-4357/ac4bbc}

\bibitem[{{Tejada Arevalo} {et~al.}(2021){Tejada Arevalo}, {Winn}, \&
  {Anderson}}]{2021ApJ...919..138T}
{Tejada Arevalo}, R.~A., {Winn}, J.~N., \& {Anderson}, K.~R. 2021, \apj, 919,
  138, \dodoi{10.3847/1538-4357/ac1429}

\bibitem[{{van Saders} {et~al.}(2016){van Saders}, {Ceillier}, {Metcalfe},
  {Silva Aguirre}, {Pinsonneault}, {Garc{\'\i}a}, {Mathur}, \&
  {Davies}}]{2016Natur.529..181V}
{van Saders}, J.~L., {Ceillier}, T., {Metcalfe}, T.~S., {et~al.} 2016, \nat,
  529, 181, \dodoi{10.1038/nature16168}

\bibitem[{{van Saders} {et~al.}(2019){van Saders}, {Pinsonneault}, \&
  {Barbieri}}]{2019ApJ...872..128V}
{van Saders}, J.~L., {Pinsonneault}, M.~H., \& {Barbieri}, M. 2019, \apj, 872,
  128, \dodoi{10.3847/1538-4357/aafafe}

\bibitem[{{Witzke} {et~al.}(2020){Witzke}, {Reinhold}, {Shapiro}, {Krivova}, \&
  {Solanki}}]{2020A&A...634L...9W}
{Witzke}, V., {Reinhold}, T., {Shapiro}, A.~I., {Krivova}, N.~A., \& {Solanki},
  S.~K. 2020, \aap, 634, L9, \dodoi{10.1051/0004-6361/201936608}

\bibitem[{{Wright} {et~al.}(2012){Wright}, {Marcy}, {Howard}, {Johnson},
  {Morton}, \& {Fischer}}]{2012ApJ...753..160W}
{Wright}, J.~T., {Marcy}, G.~W., {Howard}, A.~W., {et~al.} 2012, \apj, 753,
  160, \dodoi{10.1088/0004-637X/753/2/160}

\end{thebibliography}
\bibliographystyle{aasjournal}



\end{document}